\documentclass[11pt]{article}
\pdfoutput=1

\usepackage{amsmath}
\usepackage{amssymb}
\usepackage{graphicx,bbm,mathrsfs}
\usepackage{subfig}
\usepackage{geometry}       
\usepackage{jheppub}

\usepackage{hyperref}
\usepackage{multirow}

\usepackage{float}

\newcommand{\be}{\begin{equation}}  
\newcommand{\ee}{\end{equation}}  
\newcommand{\bea}{\begin{eqnarray}}  
\newcommand{\eea}{\end{eqnarray}}

\newcommand{\specialcell}[2][c]{%
  \begin{tabular}[#1]{@{}c@{}}#2\end{tabular}}

\catcode`\@=11
\font\manfnt=manfnt
\def\Watchout{\@ifnextchar [{\W@tchout}{\W@tchout[1]}}
\def\W@tchout[#1]{{\manfnt\@tempcnta#1\relax%
  \@whilenum\@tempcnta>\z@\do{%
    \char"7F\hskip 0.3em\advance\@tempcnta\m@ne}}}
\let\foo\W@tchout
\def\dubious{\@ifnextchar[{\@dubious}{\@dubious[1]}}

\def\@dubious[#1]{%
  \color{red}\setbox\@tempboxa\hbox{\@W@tchout#1}
  \@tempdima\wd\@tempboxa
  \list{}{\leftmargin\@tempdima}\item[\hbox to 0pt{\hss\@W@tchout#1}]}
\def\@W@tchout#1{\W@tchout[#1]}
\catcode`\@=12

\begin{document}

\begin{flushright}
	LAPTh-005/15
\end{flushright}
\vspace{1cm}

\title{Tasting the $SU(5)$ nature of supersymmetry at the LHC}
\vspace{5em}

\author[a,b]{Sylvain Fichet}
 \affiliation[a]{
ICTP South American Institute for Fundamental Research, Instituto de Fisica Teorica,\\
Sao Paulo State University, Brazil
 	}
 \affiliation[b]{
  	International Institute of Physics, UFRN, Av.~Odilon Gomes de Lima, 1722 - Capim Macio - 59078-400 - Natal-RN, Brazil
 	}

\author[c]{,~Bj\"orn Herrmann}

\author[c]{,~Yannick Stoll}
 \affiliation[c]{
 	LAPTh, Universit\'e de Savoie, CNRS, 9 Chemin de Bellevue, B.P.\ 110, F-74941 Annecy-le-Vieux, France
 	     }

\emailAdd{sylvain.fichet@lpsc.in2p3.fr, herrmann@lapth.cnrs.fr, stoll@lapth.cnrs.fr}


\abstract{
We elaborate on a recently found $SU(5)$ relation confined to the up-(s)quark flavour space, that remains immune to large quantum corrections up to the TeV scale. We investigate the possibilities opened by this new window on the GUT scale in order to find TeV-scale $SU(5)$ tests realizable at the LHC. These $SU(5)$ tests  appear as relations among observables involving  either flavour violation or chirality flip in the up-(s)quark sector. The power of these tests is systematically evaluated using a frequentist, $p$-value based criterion. $SU(5)$ tests in the Heavy supersymmetry (SUSY), Natural supersymmetry and Top-charm supersymmetry spectra are investigated. The latter scenario features light stops and scharms and is well-motivated from various five-dimensional constructions. A variety of $SU(5)$ tests is obtained, involving techniques of top polarimetry, charm-tagging, or Higgs detection from SUSY cascade decays. We find that $O(10)$ to $O(100)$ events are needed to obtain $50\%$ of relative precision at $3\sigma$ significance for all of these tests. In addition, we propose a set of precision measurements in ultraperipheral collisions in order to search for the flavour-changing dipole operators of Heavy supersymmetry. 
}


\maketitle

\newpage



\section{Introduction}
\label{Sec:Intro}

The Standard Model (SM) of particle physics describes to a rather good accuracy all kinds of elementary matter observed up to now -- apart maybe from the elusive dark matter present in our Universe. A fairly fascinating feature of the SM is that these matter fields fit into  complete representations of the $SU(5)$ gauge group \cite{Georgi:1974sy}. More precisely, the fields $\{Q_i,\,U_i,\,E_i\}$ and $\{L_i,\,D_i\}$ can be respectively embedded into three copies of the $\mathbf{10}$ and the $\mathbf{\bar{5}}$ representations of $SU(5)$ \footnote{Note that this is not the case for the Higgs doublet(s), for which a splitting mechanism is required.}. This feature can be interpreted as a hint that the SM gauge group $SU(3) \times SU(2) \times U(1)$ arises as the low-energy limit of a Grand Unified Theory (GUT) containing the $SU(5)$ sub-group at some higher scale (see Refs.\ \cite{SU5GUTS1, SU5GUTS2} for reviews) \footnote{Some early $SU(5)$ realizations are excluded by proton decay \cite{SU5death}.}.

Unravelling whether or not Nature is $SU(5)$-symmetric at short distance constitutes a challenging open problem of particle physics. Many realizations of $SU(5)$-like GUTs are possible, with various low-energy consequences. Moreover, the quantum corrections between  GUT and low-energy scales potentially hide a lot of information. A $SU(5)$ test has thus to face  both challenges of  avoiding a large model-dependence and being doable in the real world.

Besides the equality of the gauge couplings, a consequence of $SU(5)$-like unification of the matter fields is the relation
\begin{equation}
	y_{d} ~=~ y_{\ell}^{t}
	\label{Eq:UnwantedSU5}
\end{equation}
between the Yukawa matrices of the down-type quarks and charged leptons. The fact that we do not observe equal masses for leptons and down-type quarks at low energy may be explained by the renormalization group (RG) flow, which does not respect $SU(5)$ symmetry below the breaking scale. Moreover, Eq.\ \eqref{Eq:UnwantedSU5} is also modified by finite corrections at the GUT scale, coming in when integrating out the heavy degrees of freedom of the theory. It is thus necessary to study to which extent this relation can be fulfilled at both high and low energy, and a quantity of work has been done on the subject \cite{Ananthanarayan:1991xp, Ananthanarayan:1992cd, Anderson:1992ba, Anderson:1993fe, Einhorn:1981sx, Carena:1994bv, Ananthanarayan:1994qt, Rattazzi:1995gk, Blazek:1996wa, Profumo:2003ema, Pallis:2003aw, Gomez:1999dk, Gomez:2002tj, Chattopadhyay:2001va, Badziak:2011bn, Badziak:2012mm, Joshipura:2012sr, Baer:1999mc, Auto:2003ys, Blazek:2002ta, Dermisek:2005sw, Baer:2008jn, Altmannshofer:2008vr, Gogoladze:2010fu, Gogoladze:2011db, Elor:2012ig, Baer:2012jp,Iskrzynski:2014sba}. This is a typical example of $SU(5)$ test where observables are simple but model-dependence is large.

In the hypothesis that supersymmetry (SUSY) is present at a somewhat low scale, new forms of matter exist, that potentially offer new insights on $SU(5)$ symmetry. We will work within this hypothesis throughout this Paper. Supersymmetry is a natural companion of $SU(5)$ unification as the simplest supersymmetric models (such as MSSM, NMSSM, \dots) lead to gauge coupling unification with a good precision. This happens when no new degree of freedom shows up between the electroweak (EW) and the GUT scales, but also if new spacetime structure appears at some intermediate scale, or if additional fields in $SU(5)$-like representations are present. Apart from that, SUSY also naturally stabilizes the EW scale and provides candidates for cold dark matter in our Universe.

In the $SU(5)$ framework, the two Higgs supermultiplets of the MSSM, denoted $H_{d}$ and $H_{u}$, need to be embedded in a $5$ and a $\bar{5}$ representation, denoted $\mathcal{H}_{d}$ and $\mathcal{H}_{u}$, respectively. Interactions between matter and Higgs fields are then given by the superpotential
\begin{equation}
	W = \lambda_{d}^{ij} \mathcal{H}_{d} 10_i \bar{5}_j + \lambda_{u}^{ij} \mathcal{H}_{u} 10_i 10_j \,.
	\label{Eq:Superpotential}
\end{equation}
After $SU(5)$ breaking, assuming that unwanted Higgs triplets are heavy due to some mechanism, the superpotential reads 
\begin{equation}
	W = y_d^{ij} H_d Q_i D_j + y_{\ell}^{ij} H_d L_i E_j + y_u^{ij} H_u Q_i U_j \,.
\end{equation}
At the GUT scale,  Eq.\ \eqref{Eq:UnwantedSU5} is still verified, with $y_u = y_{\ell}^t = \lambda_u$. Furthermore, supersymmetry needs to be broken, such that the Lagrangian also contains the SUSY-breaking scalar trilinear terms,
\begin{equation}
	\mathcal{L}_{\rm soft} ~\supset~ a_{u} h_{u} \tilde{q} \tilde{u} + a_{d} h_{d} \tilde{q} \tilde{d} 
		+ a_{\ell} h_{d} \tilde{\ell} \tilde{e} \,,
	\label{eq:L_soft}
\end{equation}
and SUSY-breaking scalar masses,
\begin{equation}
	\mathcal{L}_{\rm soft} ~\supset~ -\tilde{q}^*M^2_Q \tilde{q}- \tilde{u}^*M^2_U \tilde{u}-\tilde{d}^*M^2_D \tilde{d}
	-\tilde{e}^*M^2_E \tilde{e}- \tilde{\ell}^*M^2_L \tilde{\ell} \,,
	\label{eq:L_soft_masses}
\end{equation}
where $\tilde{q}$, $\tilde{u}$, $\tilde{d}$, $\tilde{\ell}$, and $\tilde{e}$ are the supersymmetric partners of the SM fermions. 
With the assumption that the source of SUSY breaking is $SU(5)$-symmetric, SUSY-breaking terms can be parametrized by a singlet spurion and the aforementioned $SU(5)$ relation is also verified for the corresponding trilinear terms, 
\begin{equation}
	a_{d} \approx a_{\ell}^{t}\,.
	\label{Eq:TrilinearDown}
\end{equation}
It would be again tempting to use these relations in order to test the $SU(5)$ GUT hypothesis. However, 
contrary to SUSY-preserving parameters, it turns out that the RG evolution of Eq.\ \eqref{Eq:TrilinearDown} is much more model-dependent than the one of Eq.\ \eqref{Eq:UnwantedSU5}. It depends on the other parameters of the model, on the hypothesis about SUSY breaking, on the hypothesis of having no new degrees of freedom, and so on. It may even be dominated by the SUSY breaking hidden sector dynamics. Therefore this relation can hardly constitute a generic test of the $SU(5)$ GUT hypothesis.
The scalar masses satisfy the $SU(5)$ relations
\begin{equation}
    M^2_Q \approx M^2_U \approx M^2_E \,, \qquad
    M^2_L \approx M^2_D  	\,.
	\label{Eq:softmasses_relations}
\end{equation}
The same arguments from model-dependence apply  to these relations as for the one of Eq. \eqref{Eq:TrilinearDown}.

Testing the $SU(5)$ hypothesis in a way as model-independent as possible, even with low-energy supersymmetry, seems therefore to be a rather arduous task. However it turns out that an additional $SU(5)$ relation, first identified in Ref.\ \cite{Fichet:2014vha}, opens intriguing possibilities.  Building on this observation, the primary goal of this work is to find realistic  tests of  the $SU(5)$ hypothesis, and to evaluate their potential for the Large Hadron Collider (LHC) era.  Let us stress that our study opens a new possibility with respect to those relying on the quark-lepton complementarity of Eqs.\ \eqref{Eq:TrilinearDown} and \eqref{Eq:softmasses_relations} (see, e.g., Refs. \cite{SU5relations1,SU5relations2,SU5relations3}). Such correlations, although certainly interesting in specific models, can be hardly used as a generic test of the $SU(5)$ hypothesis, as the RG corrections received by quark and leptons are fundamentally different. 

The paper is organized as follows. In Sec.\ \ref{Sec:GUTrelation}, we review the sector of up-type squarks and discuss the $SU(5)$ property which is to be exploited in the subsequent Sections. The persistence of this property is closely examined and quantified in Sec.\ \ref{Sec:Numerics}. The relation is found to be exact up to $O(1\%)$ relative discrepancies. Sec.\ \ref{se:EFT_MIA} comprises the strategy and tools needed to setup $SU(5)$ tests at low energy. It includes in particular the up-squark effective Lagrangian, and a systematic frequentist method to evaluate the power of the proposed $SU(5)$ tests.

We then investigate various SUSY scenarios. In Sec.\ \ref{se:Heavy_SUSY}, we consider a scenario where all sparticles are too heavy to be produced at the LHC. A $SU(5)$ test is possible through the SUSY-induced flavour-changing dipole operators. Beyond standard LHC searches, we propose a precision search for dipole operators in ultraperipheral collisions. As an aside it is pointed out that EW fluxes in p-p, p-Pb, Pb-Pb collisions are expected to be coherent, and thus drastically enhanced. 

In Sec.\ \ref{se:natural} we consider the natural SUSY case. Depending on the stop/wino ordering, a test on stop flavour-violation involving charm-tagging and a test involving top polarimetry are proposed. The latter is favoured in presence of a light mainly right-handed stop. In Sec.\ \ref{se:Top_charm}, both stops and scharms are assumed to be directly observable at the LHC. This top-charm SUSY scenario is found to be motivated in various $5d$ constructions. Tests relying on Higgs detection in cascade decays are presented. Summary and  conclusions are given in Sec.\ \ref{Sec:Conclusions}.


\section{The sector of up-type squarks in \texorpdfstring{$SU(5)$}{Lg} theories \label{Sec:GUTrelation}}

In this Section, we review the sector of up-type squarks in the light of the $SU(5)$ hypothesis, and discuss a $SU(5)$ relation which has, to our knowledge, only been reported in the earlier work of Ref.\ \cite{Fichet:2014vha}. 
Starting from the superpotential given in Eq.\ \eqref{Eq:Superpotential}, the $10_i 10_j$ term is symmetric, such that only the symmetric part of the $\lambda_{u}^{ij}$ is selected. This enforces a symmetric top Yukawa coupling at the GUT scale,
\begin{equation}
	y_{u} = y_{u}^{t} \,.
	\label{Eq:YukSymGUT}
\end{equation}
Contrary to the aforementioned $SU(5)$ relation of Eq.\ \eqref{Eq:UnwantedSU5} which is potentially modified by model-dependent GUT threshold corrections or higher dimensional operators, Eq.\ \eqref{Eq:YukSymGUT} remains exact whatever extra corrections are included. More precisely, for any operator of the form $\mathcal{O}_{ij} 10_i 10_j$, the antisymmetric part of $\mathcal{O}_{ij}$ always vanishes in the contraction with $10_i 10_j$. Let us emphasize that $y_{u}$ is symmetric, but generally not hermitian.

For a non-SUSY theory, Eq.\ \eqref{Eq:YukSymGUT} does not seem particularly exploitable. In this case the only physical parameters are mass eigenvalues and CKM angles, and this does not seem to be enough to find out whether or not $y_u$ is symmetric. This situation changes once we consider SUSY. More precisely, thanks to the required SUSY breaking, more degrees of freedom of the Yukawa matrices become physical. This is due to the presence of the scalar couplings, which in general are not flavour singlets. Just like the $SU(5)$ relation Eq.\ \eqref{Eq:TrilinearDown} for the SUSY-breaking terms comes along with Eq.\ \eqref{Eq:UnwantedSU5} in the sector of slepton and down squarks, we have a relation between the trilinear SUSY-breaking terms in the up-squark sector associated to Eq.\ \eqref{Eq:YukSymGUT},
\begin{equation}
	a_u = a_u^t \, .
	\label{Eq:TrilinearUp}
\end{equation}
As we will see in the following, this property provides a potential way to test the $SU(5)$ GUT hypothesis. We remind that we work in the assumption that the SUSY breaking source does not break $SU(5)$, i.e.\ the SUSY breaking is parametrized by a spurion $F$-term which is singlet under $SU(5)$. Going beyond this simplest possibility, one observes that a spurion $F$-term in the $24$ and $75$ of $SU(5)$ would lead respectively to $a_u=-a_u^t$, and $a_u=a_u^t$. This comes from the tensor product $10 \times 10 = \bar{5}_{\rm sym}+\overline{45}_{\rm antisym}+\overline{50}_{\rm sym}$. We do not consider these possibilities here.

A crucial point is that Eq.\ \eqref{Eq:TrilinearUp} survives when rotating the supermultiplets into the Yukawa matrices eigenvalue basis (generally referred to as the super-CKM basis). Generally, the singular-value decomposition of the complex Yukawa matrices is $y_f = U_f^{\dagger} \hat{y}_f W_f$, where $U_f$ and $W_f$ are unitary matrices and $\hat{y}_f$ is the diagonal matrix containing the (positive) eigenvalues. The rotated  trilinear terms are therefore $\hat{a}_f = W_f^{\dagger} a_f U_f$. The fact that the Yukawa coupling $y_u$ is symmetric implies $U_u^* = W_u$, such that we can write $y_u = W_u^{t} \hat{y}_u W_u$. Further, we obtain $\hat{a}_u = W_u^{\dagger} a_u W_u^*$, which is a symmetric matrix if $a_u$ is. One can remark that both properties Eqs.\ \eqref{Eq:YukSymGUT} and \eqref{Eq:TrilinearUp} are necessary to keep $a_u$ symmetric in the super-CKM basis \footnote{More generally, $a_u$ remains symmetric for any rotation of $y_u$ of the form $y_f=U_f^{\dagger}\hat{y}_f U_f$, i.e.\ preserving its symmetry.}. 

 Let us emphasize that the relations \eqref{Eq:YukSymGUT}, \eqref{Eq:TrilinearUp} of our interest are confined to the flavour space of up-(s)quarks, and are therefore expected to be more stable against quantum corrections than the quark-lepton relations \eqref{Eq:TrilinearDown} and \eqref{Eq:softmasses_relations}. Moreover, let us note that there is no connection between the $SU(5)$ relation $a_u=a_u^t$ and potentially dangerous FCNCs in the down-sector. In our later analyses of SUSY scenarios, we assume some up squarks to be observed, 
which means that constraints from down-sector FCNCs are assumed to be fulfilled by construction.

What we have found up to now is summarized by the implication
\begin{equation}
	\Big\{ \textrm{$SU(5)$-type SUSY GUT} \Big\} 
		~\Longrightarrow~ 
	\Big\{ y_u=y_u^t\,,~	a_u=a_u^t \textrm{ at the GUT scale} \Big\} \,. 
	\label{Eq:Implication1}
\end{equation}
The next step is to see how this translates into physical properties, which can be tested at the TeV scale. Since the property given in Eq.\ \eqref{Eq:Implication1} is confined to up-type squarks, we only need to scrutinize this sector. The object comprising all necessary information for our purpose is the up-squark mass matrix contained in the Lagrangian 
\begin{equation} 
	\mathcal{L} ~\supset~ - \tilde{u}^\dagger \mathcal{M}^2_{\tilde{u}} \tilde{u} \,, 
	\label{Eq:MassTerm}
\end{equation}
where $\tilde{u} = (\tilde{u}_L,\tilde{c}_L,\tilde{t}_L,\tilde{u}_R,\tilde{c}_R,\tilde{t}_R)^t$ contains the six up-type squarks. The $6 \times 6$ mass matrix $\mathcal{M}^2_{\tilde{u}}$ is hermitian by construction. 

In the broken electroweak phase and within the super-CKM basis, the up-squark mass matrix $\mathcal{M}^2_{\tilde{u}}$ takes the form
\begin{equation}
 	\mathcal{M}^2_{\tilde{u}} = 
	\begin{pmatrix}
		\hat{M}_Q^2 + O(v_u^2)\mathbf{1}_3~~~ & \frac{v_u}{\sqrt{2}}\, \hat{a}_u\, (1+c_{\alpha}\frac{h}{v_u}+\ldots) + O(v_u M_{\rm SUSY}) \mathbf{1}_3 \\
 		\frac{v_u}{\sqrt{2}}\, \hat{a}_u^\dagger\,(1+c_{\alpha}\frac{h}{v_u}+\ldots) + O(v_u M_{\rm SUSY}) \mathbf{1}_3 
 		& \hat{M}_U^2 +O(v_u^2)\mathbf{1}_3 \\
	\end{pmatrix} .
	\label{Eq:MassMatrix}
\end{equation}
Here, the parameters $\hat{M}_Q^2$ and $\hat{M}_U^2$ are the scalar masses  of the left- and right-handed up-squarks in the super-CKM basis, and $M_{\rm SUSY}$ is the typical SUSY scale. The $H_u$ Higgs field of the MSSM is decomposed into its vacuum expectation value $v_u$ and its light component $h$ (the SM-like Higgs boson) with fraction $c_\alpha$. The heavy neutral component $H$ is not shown.  More detailed expressions can be found in any SUSY review (see, e.g., Ref.\ \cite{MSSM}) and are not needed for the rest of this work.

We can see that the property $ \hat a_u = \hat a_u^t $ implies that the off-diagonal block of $\mathcal{M}^2_{\tilde{u}}$  is symmetric. 
This property will be at the center of our attention in this work. Finding evidence for a symmetric off-diagonal block of $\mathcal{M}^2_{\tilde{u}}$ would constitute a rather striking hint in favour of the $SU(5)$ hypothesis. Finding evidence against a symmetric off-diagonal block in $\mathcal{M}^2_{\tilde{u}}$ would prove the $SU(5)$ hypothesis to be wrong. From this matrix we see that the physics needed to test this $SU(5)$ relation will have potentially to do with  up-squark flavour violation, chiral (LR) symmetry, and will possibly involve the Higgs boson.

While the property $ \hat a_u = \hat a_u^t $ is exactly valid at the GUT scale, it could be lost in the renormalization group (RG) evolution and not observable at ${\cal O}({\rm TeV})$ scales, just as it is the case for the $SU(5)$ relations Eqs.\ \eqref{Eq:TrilinearDown} and \eqref{Eq:softmasses_relations}. In the following, we are going to quantify how much information about the property $ \hat a_u = \hat a_u^t $  disappears through the RG flow between the GUT and the TeV scales.

Let us first consider the relevant RG equations \cite{MSSM}.  We have to study the beta functions of the Yukawa and trilinear matrices, $y_u$ and $a_u$, since these beta functions conserve the property $\hat a_u=\hat a_u^t$ at any scale if they are symmetric. The one-loop MSSM beta function of the up-type Yukawa matrix is given by
\begin{equation}
	16\pi^2 \beta_{y_u} = y_u \left[ 3\, \textrm{Tr} \big\{ y_u^{\dag} y_u^{} \big\} + 3\, y_u^{\dag} y_u^{} + y_d^{\dag} y_d^{} - \frac{16}{3}\, g^2_3 - 3\, g^2_2 - \frac{13}{15}\, g^2_1 \right] \,. 		
	\label{Eq:BetaYu}
\end{equation}
In this expression, the gauge and the trace terms are flavour singlets, and $y_u y_u^{\dag} y_u$ is symmetric by hypothesis. The only non-symmetric contribution is the term $y_u y_d^{\dag} y_d$. In SUSY, the relative magnitude of the up- and down-type Yukawa couplings depends on $\tan\beta = \frac{v_u}{v_d}$, such that the contribution of this term to the asymmetry grows with $\tan\beta$. Given the observed low-energy quark masses, one expects $y_d^{\dag} y_d$ to compete with $y_u^{\dag} y_u$ only at very high $\tan\beta$. Moreover, the $y_u y_d^{\dag} y_d$ contribution is suppressed by the elements of the CKM matrix, as $y_d$ and $y_u$ would commute if they were simultaneously diagonalizable. One can thus expect that $y_u$ stays symmetric to a very good approximation at low energy. 

The one-loop MSSM beta-function of $a_u$ reads
\begin{align}
	16\pi^2 \beta_{a_u} =
	& ~a_u \left[ 3\, \textrm{Tr} \big\{ y_u^{\dag} y_u \big\} + 5\, y_u^{\dag} y_u + y_d^{\dag} y_d 
		- \frac{16}{3}\, g^2_3 - 3\, g^2_2 - \frac{13}{15}\, g^2_1 \right] \nonumber \\
	&+ y_u \left[ 6\, \textrm{Tr} \big\{ a_u y_u^{\dag} \big\} + 4\, y_u^{\dag} a_u+ 2\, y_d^{\dag} a_d
		+\frac{32}{3}\, M^2_3+6\, M^2_2 + \frac{26}{15}\, M^2_1\, \right].
	\label{Eq:BetaAu}
\end{align}
Again, the gauge and trace terms are flavour singlets. But the other terms are generally not symmetric, because $a_u$ generally does not commute with $y_u^{\dag} y_u$, nor does $a_d$ with $y_u y_d^{\dag}$. Thus, $a_u$ stays symmetric to a good precision at low energy only if the RG flow is dominated by gauge contributions. Therefore having a symmetric $a_u$ at low energy seems not to be such a generic feature. In practice, however, the beta-function is often dominated by the large gluino mass contribution. Moreover, as the $M^2_3$ contribution is positive, it decreases $a_u$ with the energy, such that the non-symmetric terms become smaller and smaller in the beta function. Therefore, one can expect that although $\beta_{a_u}$ is generally non-symmetric, its asymmetry at the TeV scale remains rather small in many concrete cases (see Sec.\ \ref{Sec:Numerics}).

Overall, we expect $y_u$ and $a_u$ to remain symmetric to a good approximation at scales that can be tested, e.g., at the Large Hadron Collider (LHC). Somehow, this $SU(5)$-type relation persists because it occurs inside the flavour space of the up-(s)quark sector, while the other $SU(5)$-type relations Eqs.\ \eqref{Eq:UnwantedSU5} and \eqref{Eq:TrilinearDown} are relating different sectors, and are thus more sensitive to quantum corrections. 

The theoretical, model-dependent deviations from $\hat a_u=\hat a_u^t$ induced by the RG flow must be considered as an irreducible uncertainty. Quantifying this uncertainty by $\epsilon^{\rm irr}$, the implication Eq.\ \eqref{Eq:Implication1} becomes therefore
\begin{equation}
	\Big\{ \textrm{$SU(5)$-type SUSY GUT} \Big\} 
		\Longrightarrow  \Big\{ \hat a_u=\hat a_u^t\,\left( 1 + O(\epsilon^{\rm irr}) \right) 
		 \textrm{ at the TeV scale} \Big\} \,. 
	\label{Eq:Implication3}
\end{equation}
We numerically estimate $\epsilon^{\rm irr}$ in Sec.\ \ref{Sec:Numerics} for typical scenarios within the MSSM.

Beyond the MSSM, one can also expect that $y_u$ and $a_u$ be symmetric at low-energy to a good precision. The condition is that the non-symmetric terms in Eqs.\ \eqref{Eq:BetaYu} and \eqref{Eq:BetaAu} do not dominate, and that the possible hidden sector contributions do not contribute either to $\epsilon^{\rm irr}$. For example, a hidden SUSY breaking sector would need to be flavour singlet to not spoil the symmetry. From now on, let us simply keep in mind that having $\epsilon^{\rm irr} \ll 1$ is not absolutely generic, although it is present in many classic cases. In the rest of this paper we work in the super-CKM basis unless stated otherwise, and will not write the hat on $a_u$ anymore. The low-energy $SU(5)$ relation will be commonly denoted by $a_u\approx a_u^t.$

Before coming to low-energy observables and to concrete numerical examples, let us set the notations of the up-squark parameters. The complete low-energy up-squark mass matrix in the super-CKM basis is denoted
\begin{equation}
	\mathcal{M}^2_{\tilde{u}} ~=~ \begin{pmatrix}
	m^2_{11} &m^2_{12} &m^2_{13} &m^2_{14} &m^2_{15} &m^2_{16} & \\
	 &m^2_{22} &m^2_{23} &m^2_{24} &m^2_{25} &m^2_{26} & \\
	  & &m^2_{33} &m^2_{34} &m^2_{35} &m^2_{36} & \\
		& & &m^2_{44} &m^2_{45} &m^2_{46} & \\
		& & & &m^2_{55} &m^2_{56} & \\
			& & & & &m^2_{66} & \\
		\end{pmatrix}\,.
	\label{eq:LE_MassMatrix}
\end{equation}
We will denote the masses of the up squarks by $m_{\tilde{u}_i}$ ($i=1,\dots,6$) with the convention $m_{\tilde{u}_1} < m_{\tilde{u}_2} < \dots < m_{\tilde{u}_6}$. 
At the TeV scale, the squared mass matrix ${\cal M}_{\tilde{u}}^2$ defined in Eq.\ \eqref{Eq:MassMatrix} is diagonalized according to
\begin{equation}
	{\rm diag}\left( m_{\tilde{u}_1}^2, \dots, m_{\tilde{u}_6}^2 \right) ~=~
		{ R}_{\tilde{u}}^{} {\cal M}_{\tilde{u}}^2 { R}_{\tilde{u}}^{\dag} ,
\end{equation}
where ${ R}_{\tilde{u}}$ is the $6 \times 6$ rotation matrix containing the flavour decomposition of each squark mass eigenstate $\tilde{u}_i$.


\section{Persistence of the $SU(5)$ relation at low-energy} 
\label{Sec:Numerics}

In the previous Section, we saw that the $SU(5)$ relation $a_u\approx a_u^t$ remains true at low scale up to a relative discrepancy  measured by the parameter $\epsilon^{\rm irr}$. This discrepancy is induced by the RG flow, and a qualitative study of the MSSM RG equations  suggests that $\epsilon^{\rm irr}$ is  parametrically small in most of the parameter space. In this Section we numerically evaluate the magnitude of $\epsilon^{\rm irr}$.   It is mandatory to know this magnitude as it constitutes an irreducible bound on the precision at which one can potentially test  $a_u\approx a_u^t$. A too large $\epsilon^{\rm irr}$ would imply that this relation is spoiled by large model-dependent quantum corrections.

Here we present the results for a reference scenario of the MSSM. 
The  scenario is defined at the GUT scale by the parameter values given in Table \ref{Num:Tab:BenchmarkPoint}. The  SUSY-breaking parameters of Eqs.\ \eqref{eq:L_soft} and \eqref{eq:L_soft_masses} satisfy $M^2_{\bold{10}} = M_Q^2 = M_U^2 = M_E^2$, $M^2_{\bar{\bold{5}}} = M_D^2 = M_L^2$, and $a_d = a_{\ell}^t$ at  $Q = M_{\rm GUT}$. The only $CP$-violating phase is the one of the CKM matrix.

\begin{table}
	\begin{center}
				\begin{tabular}{|c|c|c|} 
				  \hline
				  \multicolumn{3}{|c|}{$\left(M^2_{\bold{10}}\right)_{ij}$}      \\
				  \hline
				                                            $(1000)^2$& $0$          & $0$       \\
				  \hline
				                           			        $0$         & $(3270)^2$ & $(557)^2$\\ 
				  \hline
				                            		       $0$         & $(557)^2$  & $(1550)^2$\\ 
				  \hline  
			\end{tabular}
			\hspace{0.5cm}
			\begin{tabular}{|c|c|c|} 
				  \hline
				\multicolumn{3}{|c|}{$\left(M^2_{\bold{\bar{5}}}\right)_{ij}$}          \\
				  \hline
				                                            $(860)^2$ & $0$         & $0$         \\
				  \hline
				                           			        $0$         & $(374)^2$ & $0$         \\ 
				  \hline
				                            		        $0$         & $0$         & $(412)^2$ \\ 
				  \hline  
			\end{tabular} 
			\vspace{0.5cm} \\
			\begin{tabular}{|c|c|c|} 			
				  \hline
				  \multicolumn{3}{|c|}{$(a_u)_{ij} $}       \\
				  \hline
				    $~~~0~~~$         & $0$         & $0$         \\
				  \hline
				    $~~~0~~~$         & $0$         & $(a_u)_{23}$\\ 
				  \hline
				    $~~~0~~~$         & $(a_u)_{32}$& $-1500$      \\ 
				  \hline  
			\end{tabular} 
			\hspace{0.5cm}
			\begin{tabular}{|c|c|c|} 
				  \hline
				  \multicolumn{3}{|c|}{$(a_d)_{ij} $}     \\
				  \hline
				    $~~~0~~~$          & $0$         & $0$         \\
				  \hline
				    $~~~0~~~$          & $0$         & $(a_d)_{23}$\\ 
				  \hline
				    $~~~0~~~$          & $(a_d)_{32}$& $400$      \\ 
				  \hline  
			\end{tabular}
			\hspace{0.5cm}
		\begin{tabular}{|l|} 
		 	\hline
		 	 $M_{1/2}=1000$ \\
		 	\hline 
			 $M_{H_{u,d}}^2=(1000)^2$  \\
			\hline 
			 $\mathrm{sign}(\mu)=+1$ \\
		 	\hline
		\end{tabular}
	\end{center}
\caption{Supersymmetry-breaking parameters at $Q=M_{\rm GUT}$ of the MSSM reference scenario used in our numerical analysis. Masses and trilinear couplings are given in GeV. Two values for $\tan\beta$ are considered, $\tan\beta = 10$ and $40$.}
\label{Num:Tab:BenchmarkPoint}
\end{table}

We compute the SUSY spectrum in the $\overline{\textrm{DR}}$ scheme at the scale $Q=1$ TeV according to the SPA convention \cite{AguilarSaavedra:2005pw} using two-loop RGEs. We use the spectrum calculator {\tt SPhenoMSSM} obtained from {\tt SARAH} \cite{Staub:2013tta} and {\tt SPheno} \cite{Porod:2003um}. 
To quantify the discrepancy between $a_u$ and $a_u^t$, it is convenient to use a quantity normalized with respect to the SUSY scale.  Therefore instead of using $\epsilon^{\rm irr}$, we introduce the asymmetry normalized to ${\rm tr}(\mathcal{M}^2_{\tilde{u}}) $, which is arguably representative of the SUSY scale. More precisely, we define
\begin{equation}
	\mathcal{A}_{ij} ~=~ 
	\frac{\left|(a_{u})_{ij}-(a_{u})_{ji}\right|}{{\rm Tr}\big\{\mathcal{M}^2_{\tilde{u}}\big\}^{1/2}}\bigg|_{Q=1\,\textrm{TeV}}\,, 
	\qquad i\neq j \,,
\end{equation}
which is related to $\epsilon^{\rm irr}$ by $\mathcal{A}_{ij} = \epsilon^{\rm irr}\, |(a_u)_{ij}|/{\rm Tr}\{\mathcal{M}^2_{\tilde{u}}\}^{1/2}$. Note that the asymmetry $\mathcal{A}_{ij}$ being dimensionless, it is independent of the overall scale chosen for our reference scenario.

There are in principle three asymmetries to be investigated: $\mathcal{A}_{12}$, $\mathcal{A}_{13}$, and $\mathcal{A}_{23}$. However, the $1-2$ and $1-3$ mixing terms appear to be too constrained by the requirements that no tachyons should be present in the low-scale spectrum and that loop corrections to the SM Yukawas should not be too large. Moreover, the largest asymmetry is expected to come from the $2-3$ sector, so that $\mathcal{A}_{12}, \mathcal{A}_{13} < \mathcal{A}_{23}$. We focus therefore on the asymmetry $\mathcal{A}_{23}$.

As discussed in Sec.\ \ref{Sec:GUTrelation}, the main dependence of $\mathcal{A}_{23}$ is with respect to the parameters $(a_u)_{23,32}$ and $(a_d)_{23,32}$. We vary these parameters assuming that the $SU(5)$ hypothesis is true, i.e.\ $(a_u)_{23} = (a_u)_{32}$ at $Q = M_{\rm GUT}$. We also take $(a_d)_{23} = (a_d)_{32}$ for simplicity.

\begin{table}
	\begin{center}
	\begin{tabular}{|c|c|}
		\hline
		$\Delta M_{Bs} $&$ (17.76 \pm 3.37)$ $\mathrm{ps}^{-1}$ \\
		\hline
	 	$\epsilon_K/\epsilon_K^{\textrm{SM}} $&$ (1 \pm 0.25)$ \\
		\hline
	 	$\mathrm{BR}(B^0_s\rightarrow \mu \mu) $&$ (3.12 \pm 2.08)\times 10^{-9}$ \\
		\hline
	 	$\mathrm{BR}(b \rightarrow s \gamma) $&$ (355 \pm 68)\times 10^{-6} $ \\
		\hline	
	 	$\mathrm{BR}(\tau \rightarrow \mu \gamma)$&$ < 4.5\times 10^{-8}$ \\
		\hline
	\end{tabular}
	\end{center}
	\caption{Low-energy flavour data  taken into account in the numerical analysis.}
	\label{Num:Tab:ListConstraints}
\end{table}

For a complete analysis of the reference scenario,
we take into account constraints from flavour physics at low-energy, among which the $B_s$ mesons oscillation frequency $\Delta M_{Bs}$ and the ratio of the $K$-meson mixing parameter normalized to the Standard Model value $\epsilon_K/\epsilon_K^{\textrm{SM}}$. The list of constraints is given in Table \ref{Num:Tab:ListConstraints}. For each constraint, 
experimental and theoretical errors are combined in quadrature. The central values and experimental errors can be found in Refs.\ \cite{Amhis:2012bh}, while the theoretical errors are taken from Refs.\ \cite{Misiak:2006zs, Ball:2006xx, Brod:2010mj}. In our reference scenario, the SM Higgs mass is found to be about $m_{h^0} = 124$ GeV, which is consistent with the latest ATLAS result, $m_{h^0} = 125.36 \pm 0.41$ (sys+stat) GeV \cite{Aad:2014aba}, combined in quadrature with  a theoretical error estimate of $\sim 3$ GeV.

\begin{figure}[t]
	\centering
    \includegraphics[scale=0.60]{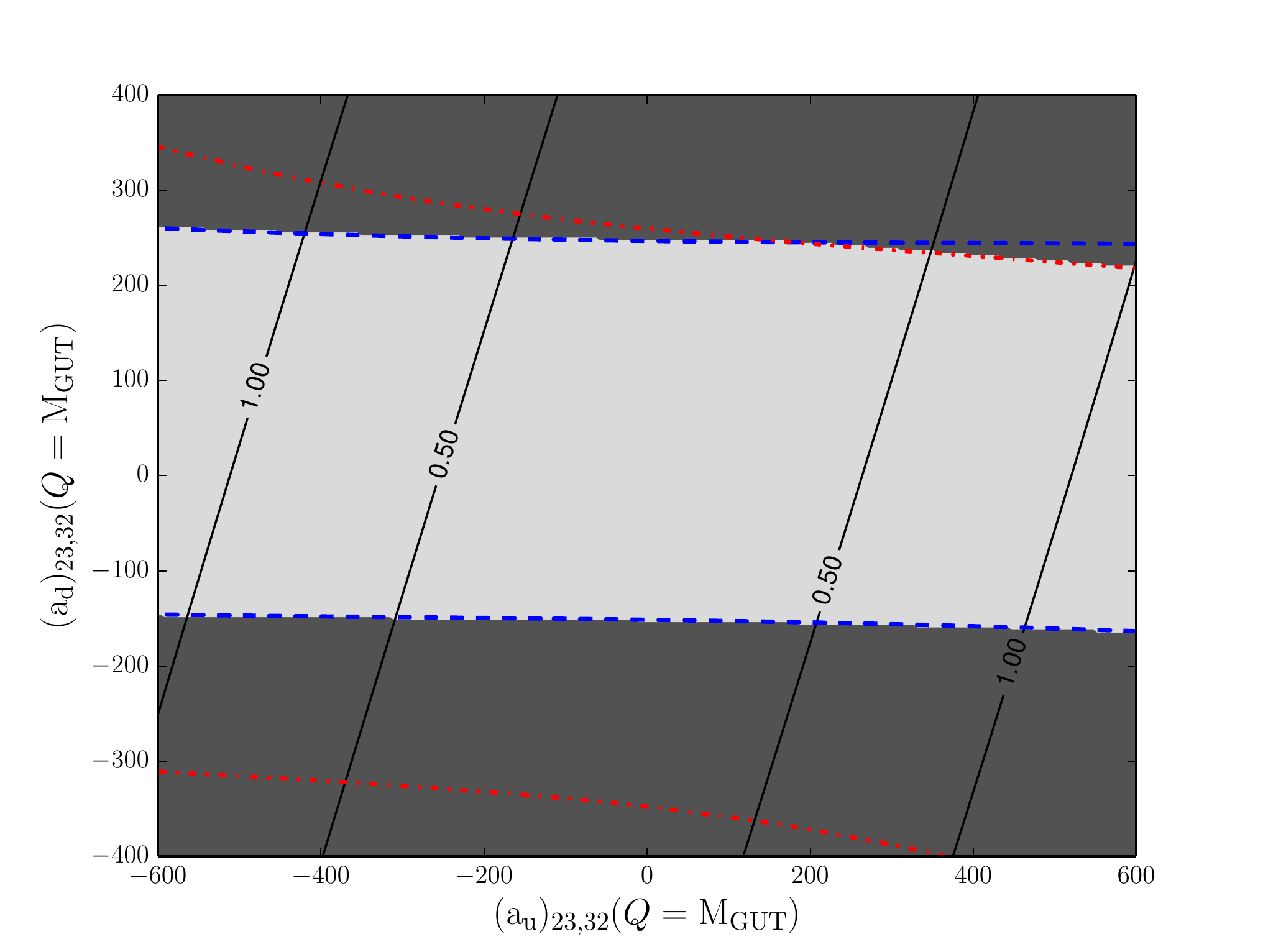}
	\caption{The asymmetry $\mathcal{A}_{2,3}$ (black solid line) together with the 2$\sigma$  exclusion bands from $\Delta M_\mathrm{Bs}$ (blue dashed lines) and $\mathrm{BR}(B^0_s\rightarrow \mu \mu)$ (red dashed-dotted lines) evaluated in  the reference scenario of Tab.\ \ref{Num:Tab:ListConstraints} for the case of low $\tan\beta = 10$. The grey area represents the allowed zone once all constraints are taken into account.}
	\label{Num:Pic:LowTanB}
\end{figure}		

\begin{figure}[t]
	\centering
	\includegraphics[scale=0.60]{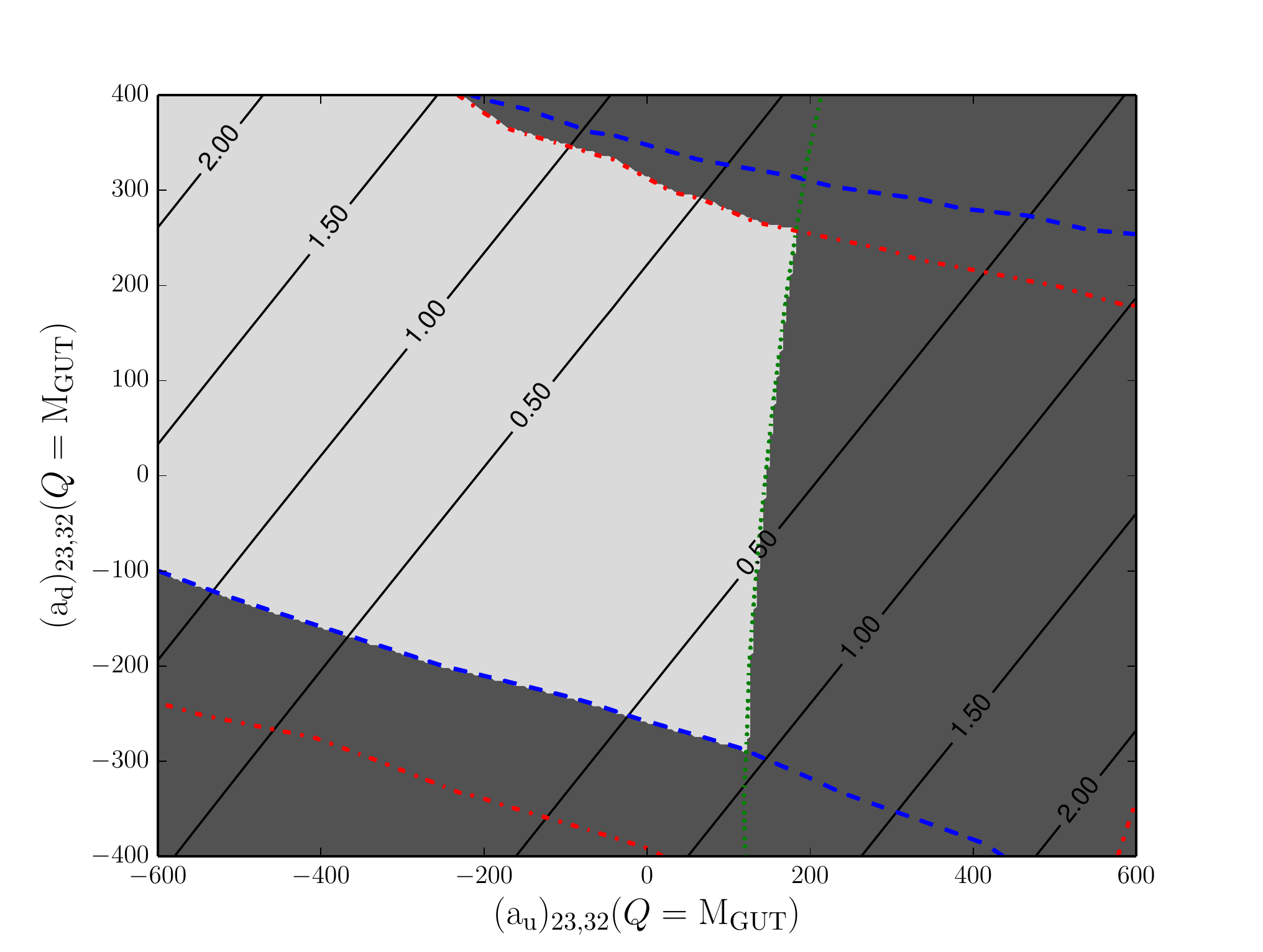}
	\caption{Same as Fig.\ \ref{Num:Pic:LowTanB} for the case of high $\tan\beta=40$. The green dotted line represents the 2$\sigma$ lower bound on $\mathrm{BR}(b \rightarrow s \gamma)$.} 
	\label{Num:Pic:HighTanB}
\end{figure}

We evaluate the asymmetry $\mathcal{A}_{23}$ at the TeV scale for a scan in the parameters $(a_u)_{23}=(a_u)_{32}$ and $(a_d)_{23}=(a_d)_{32}$ at the GUT scale. 
 The numerical results are presented in Figs.\ \ref{Num:Pic:LowTanB} and \ref{Num:Pic:HighTanB} for $\tan\beta = 10$ and $\tan\beta = 40$, respectively. The thick black isolines indicate the asymmetry $\mathcal{A}_{23}$ at the scale $Q=1$ TeV in percent. Note that the bounds $|(a_u)_{23}| < 600$ GeV and $|(a_d)_{23}| < 400$ GeV are taken into account in order to avoid tachyons in the resulting mass spectrum. The leading constraints are $\mathrm{BR}(B^0_s\rightarrow \mu \mu)$, $\Delta M_{Bs}$ (see Fig.\ \ref{Num:Pic:LowTanB}), and $\mathrm{BR}(b \rightarrow s \gamma)$ (at high $\tan\beta$, see Fig.\ \ref{Num:Pic:HighTanB}). We find $\epsilon_K/\epsilon_K^{\textrm{SM}} \sim 1$ due to the fact that we have chosen real SUSY-breaking parameters. We have numerically verified that this observable constitutes a stringent constraint on $CP$-violating phases. Finally, let us note that for both low and high $\tan\beta$, $\mathrm{BR}(\tau \rightarrow \mu \gamma)$ never exceeds the upper experimental value. Moreover, the value of the Higgs mass also does not strongly depend on the variation of $(a_u)_{23,32}$ or $(a_d)_{23,32}$.

Coming back to our main center of interest, we can see in Figs.\ \ref{Num:Pic:LowTanB} and \ref{Num:Pic:HighTanB} that the asymmetry $\mathcal{A}_{23}$ does not exceed a few percent. As expected from RGE considerations, we can conclude that none of the asymmetries $\mathcal{A}_{ij}$ exceeds this value on a large part of the MSSM parameter space. In the following, we find that, during the LHC era, such a level of precision will be most probably difficult to reach. We will therefore not mention $\epsilon^{\rm irr}$ or $\mathcal{A}_{ij}$ for the rest of this Paper.

\section{Strategies and tools for testing the $SU(5)$ hypothesis at the TeV scale}
\label{se:EFT_MIA}

Any strategy that can be set up to test the $SU(5)$ relation $a_u \approx a_u^t$ necessarily relies on a comparison involving at least two up-squarks. Apart from this relation, the squark matrix is in general arbitrary, so that each of the six up-type squarks can take any mass. Some of the squarks can be light enough to be produced on-shell at the LHC, while others may be too heavy such that they appear only virtually in intermediate processes.  

As a result, a panel of possibilities for $SU(5)$ tests will appear, depending on the exact features of the up-squark spectrum. It can be convenient to split the possibilities of $SU(5)$ tests into three categories, depending on whether the tests involve only virtual, both virtual and real, or only real up-type squarks. In the following, we outline the tools that are relevant for each of these cases. Note that the $SU(5)$ tests on virtual up-type squarks will necessarily involve one-loop processes, because the interaction terms probed in the tests are $R$-parity conserving. In contrast, the $SU(5)$ tests on real squarks can in principle rely on tree-level processes only. 
 
Dealing with the eigenvalues and rotation matrices of the full $6 \times 6$ mass matrix is in general rather technical, and may constitute an obstacle on the quest for simple $SU(5)$ relations among observables. However, depending on the pattern of $\mathcal{M}^2_{\tilde u}$, two expansions can be used in order to simplify the problem: the mass insertion approximation (MIA) and the effective field theory (EFT) expansion. 

The MIA is an expansion with respect to small eigenvalues-splittings, and is in particular valid in presence of small off-diagonal elements. In contrast, the EFT expansion can be applied if a large hierarchy is present in $\mathcal{M}^2_{\tilde{u}}$. The heavy up-type squarks with mass matrix $M^2$ are integrated out, the expansion parameter is $E^2 \hat{M}^{-2}$, and $M^2$ can have arbitrary off-diagonal entries. The two expansions, which we detail in the following, are therefore complementary.

It is clear that the feasibility of the $SU(5)$ tests we will setup depends crucially on the amount of data available --
whatever they involve real or virtual up-squarks. This feasibility needs to be quantified using appropriate statistical tools. Whenever a $SU(5)$ test can be obtained through a definite relation among observables, we will use a frequentist $p$-value in order to evaluate to which precision this relation can be tested, for a given significance and amount of data. We will systematically report the expected precision associated with this test.

\subsection{Tree-level up-squark effective theory}

The pattern of the squark masses  can in principle exhibit some hierarchy, and this actually occurs in many classes of new physics models. In such a situation, the physics of the light squarks can be conveniently captured into a low-energy effective Lagrangian, where heavy squarks are integrated out. This has, e.g., be discussed in Ref.\ \cite{Fichet:2014vha} for the situation where only two squarks are light and the remaining four can be integrated out. The higher-dimensional operators present in this tree-level Lagrangian will serve as a basis for the $SU(5)$ tests based on \textit{both virtual and real squarks}.

Following Ref.\ \cite{Fichet:2014vha}, the up-squark mass term of Eq.\ \eqref{Eq:MassTerm} can be reorganized such that  
\begin{equation}
	\mathcal{L} ~\supset~ \tilde{u}^\dagger \mathcal{M}^2_{\tilde{u}} \tilde{u} ~\equiv~ 
	\Phi^\dagger  \mathcal{M}^2 \Phi ~=~
	\left(\hat{\phi}^\dagger, \phi^\dagger \right)
	\begin{pmatrix} \hat{M}^2 & \tilde{M}^2 \\ \tilde{M}^{2\dagger} &  M^2 \end{pmatrix}
	\begin{pmatrix} \hat{\phi} \\ \phi \end{pmatrix} \,,
\end{equation} 
where $\hat{\phi}$ contains the heavy states and $\phi$ the light ones. The relevant piece of the corresponding Lagrangian has the general form
\begin{equation}
	\mathcal{L} ~\supset~ \big| D\Phi \big|^2 - \Phi^\dagger \mathcal{M}^2 \Phi + \Big( \mathcal{O} \phi +\hat{\mathcal{O}} \hat{\phi} + {\rm h.c.} \Big) \,,
\end{equation}
where the operators $\mathcal{O}$ and $\hat{\mathcal{O}}$ represent the interactions with other fields, that are potentially exploited to probe the up-squark sector.

Assuming that the matrix norm  of $\hat{M}^2$ is large with respect to the energy $E^2$ at which the theory is probed, i.e. $||\hat{M}^{2}||=(\sum_{i,j} |\hat{M}^{2}_{i,j}|^2)^{1/2}\gg E^2$,
 the heavy squarks $\hat{\phi}$ can be integrated out. We are  thus left with the low-energy effective Lagrangian of light squarks, 
\begin{align}
		\mathcal{L}_{\rm eff} ~=~ &\big| D\phi \big|^2 
		- \phi^\dagger \left[ M^2  
		- \tilde{M}^{2\dagger} \hat{M}^{-2} \tilde{M}^{2}
		- \frac{1}{2} \left\{ \tilde{M}^{2\dagger}  \hat{M}^{-4}  \tilde{M}^2, M^2 \right\} \right] \phi \nonumber \\  
		& + \left[ \mathcal{O} - \hat{\mathcal{O}} \big( \hat{M}^{-2} - \hat{M}^{-4} D^2 \big) \tilde{M}^2 
		 - \frac{\mathcal{O}}{2} \tilde{M}^{2\dagger}  \hat{M}^{-4} \tilde{M}^2 \right] \phi + {\rm h.c.}		
		 \,\,.  
	\label{eq:Leff}	
\end{align}    
We keep only the leading and the subleading terms of the $E^2 \hat{M}^{-2}$ expansion relevant to build $SU(5)$ tests. This effective Lagrangian contains in principle higher dimensional couplings and derivative terms, which are either subleading or irrelevant for the observables we are going to consider, and are thus neglected. Eq.\ \eqref{eq:Leff} has been obtained using the field redefinition $\Phi\rightarrow (\mathbf{1}-\frac{1}{2} \tilde{M}^{2\dagger} \hat{M}^{-4} \tilde{M}^2) \Phi$ in order to canonically normalize the light squarks kinetic terms. The $\{\,,\}$ denotes the anti-commutator. 

From Eq.\ \eqref{eq:Leff} we see that flavour-violating couplings of the light squarks enter at first order and are controled by  $\hat{M}^{-2} \tilde{M}^2$. The flavour-conserving couplings will instead be modified at the second order. The mass matrix $M^2$, associated to the light squarks, receives a correction independent of $M^2$ at first order, and corrections proportional to $M^2$ at second order. The imprint of the heavy up-squarks in the Lagrangian of Eq.\ \eqref{eq:Leff} appears as corrections to the masses and couplings of the light up-squark states. Physically, these corrections have to be understood both as tree-level exchange of heavy up-squarks, and as the first terms of the expansion with respect to the small parameters that describe mixing of heavy and light squarks. While in Ref.\ \cite{Fichet:2014vha} it has been applied to the special case of two light squarks, this framework is more general and can be used for any number of states that are accessible at the LHC, while the others are integrated out. Finally, we remind that, although this effective theory approach might bear some resemblance with the mass insertion approximation, the two approaches are fundamentally different.

\subsection{One-loop effective operators}

When all the up-squarks are heavier than the typical LHC scale, they can only appear off-shell. Potential tests of the $SU(5)$ hypothesis are thus based on \textit{virtual squarks} only. All squarks can be integrated out, and the resulting low-energy Lagrangian then contains the SM plus possible other light SUSY particles. Using other light SUSY particles to  test $a_u\approx a_u^t$ does not seem to provide attractive possibilities. In particular gauginos are flavor-blind, and bottom squarks depend on their own flavour structure. 

Here, we focus instead on the case where the whole SUSY spectrum is heavy. The resulting low-energy Lagrangian then has the form 
\begin{equation}
	\mathcal{L}_{\rm eff} ~=~ \mathcal{L}_{\rm SM} + \mathcal{L}^{(5)}+\mathcal{L}^{(6)}+O(1/M_{\rm SUSY}^3)\,. 
\end{equation}
The effective operators with dimension larger than $6$ are not relevant for our present study.
  The dimension-six Lagrangian takes the form
\begin{equation}
	\mathcal{L}^{(6)} ~=~ \sum_i\frac{\alpha_i}{M^2_{\rm SUSY}} \mathcal{O}_i ,
\end{equation}
where the $\alpha_i$ are dimensionless constants accompanying the operators $\mathcal{O}_i$. This Lagrangian contains 59 operators, which have been first fully classified in Ref.\ \cite{Grzadkowski:2010es}.

In SUSY models with conserved $R$-parity,  all $CP$-even operators are generated at loop level. In general, they receive contributions from a large number of loop diagrams, and we shall identify which among these operators can be potentially relevant for a $SU(5)$ test.
We are thus looking for one-loop diagrams involving
\begin{itemize}
	\item at least one $(a_{u})_{ij} h_{u} \tilde{q}_i \tilde{u}_j$ vertex, as it encloses the $SU(5)$ relation Eq.\ \eqref{Eq:TrilinearUp} that we want to test,
	\item fermions on the outgoing legs in order to access information on the flavour structure of the $(a_{u})_{ij}$ coupling.
\end{itemize}
These requirements naturally lead to the sector of up-type flavour-changing dipole operators of the form
\begin{equation}
	\mathcal{O}^{V}_{u,ij} ~=~ \bar q_i \sigma^{\mu\nu} u_j H_u V_{\mu\nu} \Big|_{i\neq j}
\end{equation}
where $V=(G,W,B)$ and $\sigma^{\mu\nu}=\frac{i}{2}[\gamma^\mu,\gamma^\nu]$.
The key point to obtain a test of the $SU(5)$ hypothesis, is to be able to distinguish between $\mathcal{O}^{V}_{u,ij}$ and $\mathcal{O}^{V}_{u,ji}$. The generation of these operators and the possibilities they offer to build a $SU(5)$ test will be discussed in Sec.\ \ref{se:Heavy_SUSY}.

\subsection{Mass-insertion approximation}
\label{Sec:MIA}

If the relative difference between the eigenvalues of the mass matrix $\mathcal{M}_{\tilde u}^2$ is small, an expansion can be made
by taking this difference as the expansion parameter (see, e.g., Refs.\ \cite{old_MIA, Raz:2002zx, Hiller:2008sv}). This is the so-called mass-insertion approximation (MIA) \footnote{Note that the MIA is sometimes considered as an expansion with respect to small off-diagonal elements of $\mathcal{M}_{\tilde u}^2$. This is a more restrictive expansion, which is in general not equivalent to the MIA as we define it here \cite{Raz:2002zx, Hiller:2008sv}.}. In the super-CKM basis, the mass matrix is written in terms of the mass insertions $\delta^{\alpha\beta}_u$ ($\alpha,\beta=L,R$) as follows,
\begin{equation}
	\mathcal{M}_{\tilde u}^2 ~\approx~ m_{\tilde q}^2 \Biggr[ \mathbf{1} +
	\begin{pmatrix} \delta^{LL}_u & \delta^{LR}_u \\ \delta^{RL}_u & \delta^{RR}_u \end{pmatrix} \Biggr] ,
	\label{Eq:MIA}
\end{equation}
where $m_{\tilde q}^2$ is an \textit{average} squark mass, that can, e.g., conveniently be chosen to be $m_{\tilde q}^2= {\rm Tr}\{\mathcal{M}_{\tilde u}^2\}$. Note that, here, the $\delta^{\alpha\beta}_u$ are $3 \times 3$ matrices in flavour space, and we have $\delta^{RL}_u = (\delta_u^{LR})^{\dagger}$ since the mass matrix $\mathcal{M}_{\tilde u}^2$ is hermitian. The $SU(5)$ relation $a_u\approx a_u^t$ translates on the chirality-flipping, flavour-violating mass insertion matrices as 
\begin{equation}
	\delta_u^{LR}=(\delta_u^{LR})^t\,,
\end{equation}
or equivalently, in terms of the generation-mixing entries,
\begin{equation}
	(\delta^{LR}_u)_{12} = (\delta^{LR}_u)_{21}, \quad 
	(\delta^{LR}_u)_{31} = (\delta^{LR}_u)_{13}, \quad 	
	(\delta^{LR}_u)_{32} = (\delta^{LR}_u)_{23}.
\end{equation}
The MIA also applies when only a subset of the $\mathcal{M}_{\tilde u}^2$ eigenstates is nearly-degenerate. 

The MIA is commonly used in the broken electroweak phase with $\langle H\rangle\equiv v/\sqrt{2}\neq 0$. However, the effect of electroweak breaking in  the mass matrix is small by assumption, as it participates in splitting the eigenvalues. This is in particular verified whenever $M_{\rm SUSY}\gg v/\sqrt{2}$. Therefore electroweak breaking does not have influence on the mean squark mass $m_{\tilde q}$ by assumption. As a result, the MIA can also be used in the unbroken electroweak phase. In this case, the VEVs have to be replaced by the original Higgs fields. Electroweak symmetry breaking appears in the squark mass matrices exclusively through the chirality-flippling mass insertions $\delta^{LR}$, $\delta^{RL}$. More precisely, the latter are proportional to the VEV according to $\delta^{LR,RL} \propto v/M_{\rm SUSY}$, while the chirality-conserving ones are not. Thus, the use of the parameters $\delta^{LL}$ and $\delta^{RR}$ remains unchanged. In contrast, for the chirality-flipping, flavour-violating mass insertions, which we are interested in, we have to replace
\begin{equation}
	(\delta^{LR}_u)_{ij} ~\rightarrow~ \frac{\sqrt{2}\,H_u}{v_u} (\delta^{LR}_u)_{ij} \,. 
\end{equation}
for $i \neq j$. This is also a way to recover the physical Higgs boson in the complete mass matrix Eq.~\eqref{Eq:MassMatrix}.

\subsection{Expected precision}
\label{Sec:ExpPrec}

When the $SU(5)$ test consists of a simple relation, for example among event rates at the LHC, one can use a frequentist $p$-value to 
evaluate the potential of the test. Assume that there is a relation $R(\mathcal{O}_i)$ among observables quantities $\mathcal{O}_i$, that satisfies 
\begin{equation} 
	{\rm{E}}\big[ R(\mathcal{O}_i) \big] ~=~ 0\,
	\label{eq:r}
\end{equation}
whenever the $SU(5)$ hypothesis is verified. Here, $\textrm{E}$ denotes the expectation operator.  We use a $p$-value test to quantify
whether or not the \textit{observed} value of $R$, denoted $\hat r$, is compatible with zero. This test is not trivial to perform in general, but simplifies whenever the probability density function (PDF) of $\hat r$ can be approximated by a half-normal law $\hat r \sim 2\mathcal{N}(r,\sigma)\theta(r)$ (see e.g.\ Ref.\ \cite{Cowan:2010js}), where $\theta$ is the unit step function.

In this case, the compatibility of $\hat r$ with the $r=0$ hypothesis can be expressed  in terms of a statistical significance as
\begin{equation}
	Z ~=~ \frac{|\hat r  |}{\sigma} \,. 
	\label{eq:Z}
\end{equation}
Here, $Z = \Phi^{-1}(1-p)$ translates the $p$-value into a significance given in terms of standard Gaussian deviations. $\Phi$ is the standard Gaussian repartition function.  

For a given relation satisfying Eq.\ \eqref{eq:r} and an hypothesized amount of data, one can evaluate the expected value of $\sigma$.  For a given significance $Z$, the quantity 
\begin{equation} 
	P_Z ~=~ Z \sigma 
	\label{eq:PZ} 
\end{equation}
represents the \textit{expected precision} to which the relation Eq.\ \eqref{eq:r} can be tested. Throughout this paper we will systematically report the value of $P_Z$ when a relation like Eq.\ \eqref{eq:r} is available.

As an example, consider the relation $ a X_1 = b X_2$. The $X_i$ are assumed to be normal variables, centered on $\mu_i$ and with variance $\sigma^2_i$. Defining $R = | a X_1 - b X_2 |$, the PDF of $R$ takes the form $f_R \sim 2\mathcal{N}(r,\sigma^2)\theta(r)$ where $r = a \mu_1 - b \mu_2$, and the variance is given by $\sigma^2=a^2 \sigma^2_1+b^2 \sigma^2_2$. In order to compute the expected significance defined in Eq.\ \eqref{eq:PZ}, we use this value of $\sigma$ assuming that the $SU(5)$ hypothesis is true.  Note that this implies $r=0$ by construction.

In practice it is useful to normalize $R$ such that $\textrm{E}[R] \in [0,1]$. The expected precision can then be expressed in percent. Concretely, having $P_Z = 20\%$ means that a violation of the relation $\textrm{E}[R] = 0$ by $20\%$ or more (i.e.\ $\textrm{E}[R] > 20\%$) can be assessed with a significance of $Z\sigma$. The use of a normalized quantity is more convenient to interface the $SU(5)$ test with other studies -- either experimental or theoretical. Normalizing $R$ implies that one frequently encounters the random variable $|A-B|/(A+B)$, where $A$ and $B$ are potentially correlated. We will use the notation $\textrm{E}[A] = \mu_A$, $\textrm{E}[B] = \mu_B$,  ${\textrm V}[A] = \sigma_A^2$, $\textrm{V}[B] = \sigma_B^2$, $\textrm{Cov}[A,B]=\rho \sigma_A \sigma_B$. The following relations are useful in order to derive the expected precision,
\begin{align}
	\textrm{E}\big[A/B\big] ~&\approx~ \textrm{E}[A]/\textrm{E}[B]\,, \\
	\textrm{V}\big[A/B\big] ~&\approx~ \frac{1}{\mu_B^4}\bigg(\mu_A^2 \sigma_B^2+\mu_B^2 \sigma_A^2 - 2\rho \sigma_A \sigma_B \mu_A\mu_B \bigg)\,,\\
	\textrm{V}\big[(A-C)/(A+C)\big] ~&\approx~ 
		4\frac{\mu_A^2 \sigma^2_C+\mu_C^2 \sigma^2_A}{(\mu_A+\mu_C)^4} 
		~\approx~ 4{\textrm V}\big[(A)/(A+C)\big]~~~ \textrm{with}~ \textrm{Cov}[A,C]=0\,.
\end{align}
These approximations are in particular valid in the Gaussian limit, where a ratio of random variables is described by the Fieller-Hinkley distribution \cite{Hinkley}. For a large number of events $N_A \equiv \mu_A \approx \sigma^2_A$ and similarly for $C$, one has for example
\begin{equation}
	\textrm{V}\big[|N_A-N_C|/(N_A+N_C)\big] ~=~ \frac{4N_A N_C}{(N_A + N_C)^3} \,.
\end{equation}

\section{Heavy supersymmetry }
\label{se:Heavy_SUSY}

We can now apply the strategy and tools presented above in various SUSY scenarios. In this Section we assume that the SUSY scale is large with respect to the scales probed by the LHC. This of course increases the fine-tuning of the electroweak scale which is a primary motivation for low-energy SUSY \footnote{On the other hand the SUSY scale can still be far from the Planck scale, so that it still improves a lot the gauge-hierarchy problem. To go further into naturalness and fine-tuning arguments, these notions  need to be properly defined and quantified, see Ref.\ \cite{Fichet:2012sn}.}. In such a situation it is appropriate to integrate the whole SUSY sector including the charged Higgses. As discussed in Sec.~\ref{se:EFT_MIA}, the dimension-six operators that one should scrutinize to test the $SU(5)$ hypothesis are the up-sector dipoles
\begin{equation}
	\mathcal{L}^{(6)}_{1\rm-loop} ~\supset~ \frac{\alpha_{ij}^G}{M_{\rm SUSY}^2}\, \bar q_i \sigma^{\mu\nu} T^a u_j  H_u G^a_{\mu\nu} + 	\frac{\alpha_{ij}^W}{M_{\rm SUSY}^2}\, \bar q_i \sigma^{\mu\nu} T^I u_j H_u W^I_{\mu\nu}+\frac{\alpha_{ij}^B}{M_{\rm SUSY}^2}\, \bar q_i \sigma^{\mu\nu} u_j H_u B_{\mu\nu}\,, \label{eq:dipoles}
\end{equation}
where $i \neq j$. Our ability to build a $SU(5)$-test relies crucially on distinguishing the chiralities (i.e.\ between $L$ and $R$) at some level. The only quark for which this can be done is the top quark, as the lighter quarks hadronize too fast. Our test has therefore to rely on \textit{top polarimetry}, which itself relies on the $V-A$ structure of the leading top decay $t\rightarrow bW$ (see e.g.\ Ref.\ \cite{Top_Polarization}).

We therefore focus only on the dipole operators involving a top quark,
\begin{equation}
\begin{split}
	\mathcal{L}^{(6)}_{1\rm-loop} ~\supset~ &\frac{\alpha_{3i}^G}{M_{\rm SUSY}^2}\, \bar t_L \sigma^{\mu\nu} T^a u_{R\,i} H_u G^a_{\mu\nu} +	   \frac{\alpha_{i3}^G}{M_{\rm SUSY}^2}\, \bar u_{L\,i} \sigma^{\mu\nu} T^a t_{R} H_u G^a_{\mu\nu}
	 \\
	 &+\frac{\alpha_{3i}^W}{2\,M_{\rm SUSY}^2}\, \bar t_L \sigma^{\mu\nu}  u_{R\,i} H_u W^3_{\mu\nu}	  
	 + \frac{\alpha_{i3}^W}{2\,M_{\rm SUSY}^2}\, \bar u_{L\,i} \sigma^{\mu\nu} T^I t_{R} H_u W^3_{\mu\nu} 
	  \\
	 &+ \frac{\alpha_{3i}^B}{M_{\rm SUSY}^2}\, \bar t_L \sigma^{\mu\nu}  u_{R\,i} H_u B_{\mu\nu}  
	 + \frac{\alpha_{i3}^B}{M_{\rm SUSY}^2}\, \bar u_{L\,i} \sigma^{\mu\nu}  t_{R} H_u B_{\mu\nu} 
	 	  \,, \label{eq:top_dipoles}
\end{split}
\end{equation}
where $i=1,2$.

The operators given in Eq.\ \eqref{eq:dipoles} are generated at one-loop by penguin diagrams. They receive their main contributions from the chargino--down-squark, charged Higgs--down-quark and gluino--up-squark amplitudes. The latter contribution is expected to dominate because it is enhanced by the QCD coupling. Moreover, contrary to the well-known case of down quarks, higher-loop corrections in the up-sector from flavour-changing self-energies \cite{Hofer:2009xb, Crivellin:2009ar} are $\tan\beta$-suppressed, so that they cannot become large. We therefore end up with dipole operators mainly generated by the \textit{gluino loop}. 

From now on we assume that the up-squark mass matrix is degenerate enough so that the MIA applies. The expression of the electroweak and chromo-dipoles operators are then given by
\begin{align}
\frac{\alpha_{3i}^G}{M_{\rm SUSY}^2} ~&=~
	\frac{g_s^3}{16\pi^2\, m_{\tilde q}^2}\left(
	7\sqrt{2} \,\frac{ m_{\tilde g}\, \delta^{LR}_{3i}\, F_s^{(1)}(x_g)}{v_u\,240}-5\,\frac{y_t\, \delta^{LL}_{3i}\, F_s^{(2)}(x_g)}{36}
	\right) \,, \\
	\frac{\alpha_{3i}^W}{M_{\rm SUSY}^2} &= \frac{g\,g_s^2}{16\pi^2\, m_{\tilde q}^2}\left(
	-\sqrt{2}\frac{m_{\tilde g} \delta^{LR}_{3i} F_{EW}^{(1)}(x_g)}{v_u\,30}+\,\frac{y_t \delta^{LL}_{3i} F_{EW}^{(2)}(x_g)}{9}
	\right) \,, \\
	\frac{\alpha_{3i}^B}{M_{\rm SUSY}^2} &= \frac{g'\,g_s^2}{16\pi^2\, m_{\tilde q}^2}\left(
	-\sqrt{2}\frac{m_{\tilde g} \delta^{LR}_{3i} F_{EW}^{(1)}(x_g)}{v_u\,180}+\,\frac{y_t \delta^{LL}_{3i} F_{EW}^{(2)}(x_g)}{54}
	\right) \,,
	\label{eq:dipole_coef}
\end{align}
and the $\alpha_{i3}$ coefficients are obtained by replacing $\delta_{3i}^{LR} \rightarrow \delta_{i3}^{LR}$, $\delta_{3i}^{LL} \rightarrow \delta_{i3}^{RR}$. Here, we have defined $x_g = m_{\tilde g}/m_{\tilde q}$, the form factors satisfy $F(1)=1$, $F(x)\rightarrow 0$ for $x\rightarrow \infty$, and are given in App.\ \ref{App:dipole_ff}. Moreover, $m_{\tilde q}$ denotes the average squark mass introduced in Sec.\ \ref{Sec:MIA}. The above expressions have been deduced using loop functions from Refs.\ \cite{Altmannshofer:2010pqa, Gabbiani:1996hi} and by making use of Sec.\ \ref{se:EFT_MIA}. The coefficients of the EW operators in the broken phase,
\begin{equation}
	\mathcal{L}^{(6)}_{1\rm-loop} ~\supset~ \frac{\alpha_{3i}^\gamma}{M_{\rm SUSY}^2}\, \bar t_L \sigma^{\mu\nu} 
		u_{R\,i} H_u F_{\mu\nu} + \frac{\alpha_{3i}^Z}{M_{\rm SUSY}^2}\, \bar t_L \sigma^{\mu\nu} u_{R\,i} H_u Z_{\mu\nu} \,,
\end{equation}
are obtained by replacing $\alpha^\gamma_{ij} \rightarrow s_w \alpha^W_{ij}/2+c_w \alpha^B_{ij}$ and $\alpha^Z_{ij} \rightarrow c_w \alpha^W_{ij}/2-s_w \alpha^B_{ij}$. Here $s_w$ and $c_w$ are the sine and cosine of the weak angle.

As already pointed out above, for the purpose of obtaining a $SU(5)$ test, the contributions that we are interested in are the chirality-flipping ones, $\delta^{LR}$. 
The chirality-conserving contributions from $\delta^{LL}$ and $\delta^{RR}$ are suppressed by a factor $y_t v_u/m_{\tilde g} = m_t\, s_\beta/m_{\tilde g} $) with respect to the contributions from $\delta^{LR}$. This factor is compensated if one considers that $\delta^{LR}\propto v a_u /m_{\tilde q}^2$, as it is usually the case in SUSY models with soft breaking. However, the magnitude of the $a_u$-term can be larger than $m_{\tilde q}$ (for example from naive dimensional analysis \cite{Brummer:2011cp}). One should therefore let $\delta^{LR} \sim \mathcal{O}(1)$. 

Considering usual models, one should also notice that
\begin{equation}  
	(\delta^{LL}_{3i}) ~\approx~ (\delta^{RR}_{3i}) 
\end{equation}
at low-energy to a very good aproximation.
We can now identify the consequences of our $SU(5)$ relation $a_u\approx a_u^t$ at the LHC.
This relation implies the following equalities between the coefficients of the dipoles in Eq.\ \eqref{eq:top_dipoles},
\begin{equation}
	\alpha_{3i}^G ~\approx~ \alpha_{i3}^{G}\,,\qquad 
	\alpha_{3i}^W ~\approx~ \alpha_{i3}^{W}\,,\qquad 
	\alpha_{3i}^B ~\approx~ \alpha_{i3}^{B}\,. 
\end{equation}
Concretely, the dipole operators induce both new top decays and top production modes at the LHC. These processes will be discussed in the Subsecs.\ \ref{se:dipoles_LHC} and \ref{se:dipoles_UPC} of the present Paper.

\subsection{$SU(5)$ test through single top polarimetry}
\label{se:top_polarization}

We have seen above that measuring the top spin is a necessary ingredient to build a $SU(5)$ test that distinguishes between the $\mathcal{O}^V_{3i}$ and $\mathcal{O}^V_{i3}$ dipole operators. The expected precision associated with such polarization-based tests can be evaluated in a generic way. The top spin has to be measured with  some polarization-sensitive observable $z$ with distribution $f_Z$. For the $t\rightarrow bW$ decay, $z$ can be for example  the angle between the top and the lepton from the $W$-decay, or the $b$-quark energy in case of a boosted top \cite{Top_Polarization}. The usual way to proceed to get information on the top spin is to split the phase space into two subsets $\mathcal{D}=\mathcal{D}_++\mathcal{D}_-$. The total sample of $N$ events is then split into two subsamples $N_\pm ~=~ N|_{z\in \mathcal{D}_\pm} $ satisfying 
\begin{equation}
	N ~=~ N_++N_-\,,\quad 
	{\rm E}[N_\pm] ~=~ {\rm E}[N] \int_{\mathcal{D}_\pm} \!\! f_Z(z) \,{\rm d}z\,.
\end{equation}
The choice of $\mathcal{D}_\pm$ is in general a freedom of the analysis, although it is preferable to have $N_+\sim N_-$ to minimize the statistical error.  In what follows, we will systematically choose the subsets of phase space $\mathcal{D}_\pm$  such that ${\rm E}[N_+] = {\rm E}[N_-]$ if the $SU(5)$ hypothesis is satisfied. 

The power of the $SU(5)$ test depends on the power of the spin-analyzing variable, which is parametrized by $\kappa\in[0,1]$. Here we assume that the $z$-dependent distribution used to analyze the spin has the form $f_Z\propto(1+\kappa \,P_t\,z)$, where  $z\in[-1,1]$ and $P_t=\pm 1$ is the top polarization \cite{Top_Polarization}. A similar approach can be done for any other distributions. The lepton angle in $t \rightarrow b W$ has a maximal spin-analyzing power $\kappa=1$ because of the $V-A$ structure of electroweak interactions. In that case $z$ is the top-lepton angle cosine in the top rest frame. Note that $\kappa \leq 1$ from unitarity.  For the subdomains of $z$ we choose   $\mathcal{D}_+=[0,1]$, $\mathcal{D}_-=[-1,0[$. Note that this matches the usual definition of a forward-backward asymmetry.

In order to quantify the precision at which the $SU(5)$ relation can be tested, we introduce the normalized asymmetry
\begin{equation}
	R ~=~ \frac{2}{\kappa}\frac{|N_+-N_-|}{N_++N_-}\,,
\end{equation}
which satisfies
\begin{equation}
	{\rm E}[R] ~=~ \frac{\left||\alpha_{3i}|^2-|\alpha_{i3}|^2\right|}{|\alpha_{3i}|^2+|\alpha_{i3}|^2}\,.
\end{equation}
Assuming $B$ background events for the total signal, the expected precision defined in Sec.\ \ref{se:EFT_MIA} reads 
\begin{equation}
	P_Z ~=~ Z \,\frac{2}{\kappa} \frac{(N+B)^{1/2}}{N}\,.
\end{equation}

\begin{figure}
  	\centering
	\includegraphics[width=8cm]{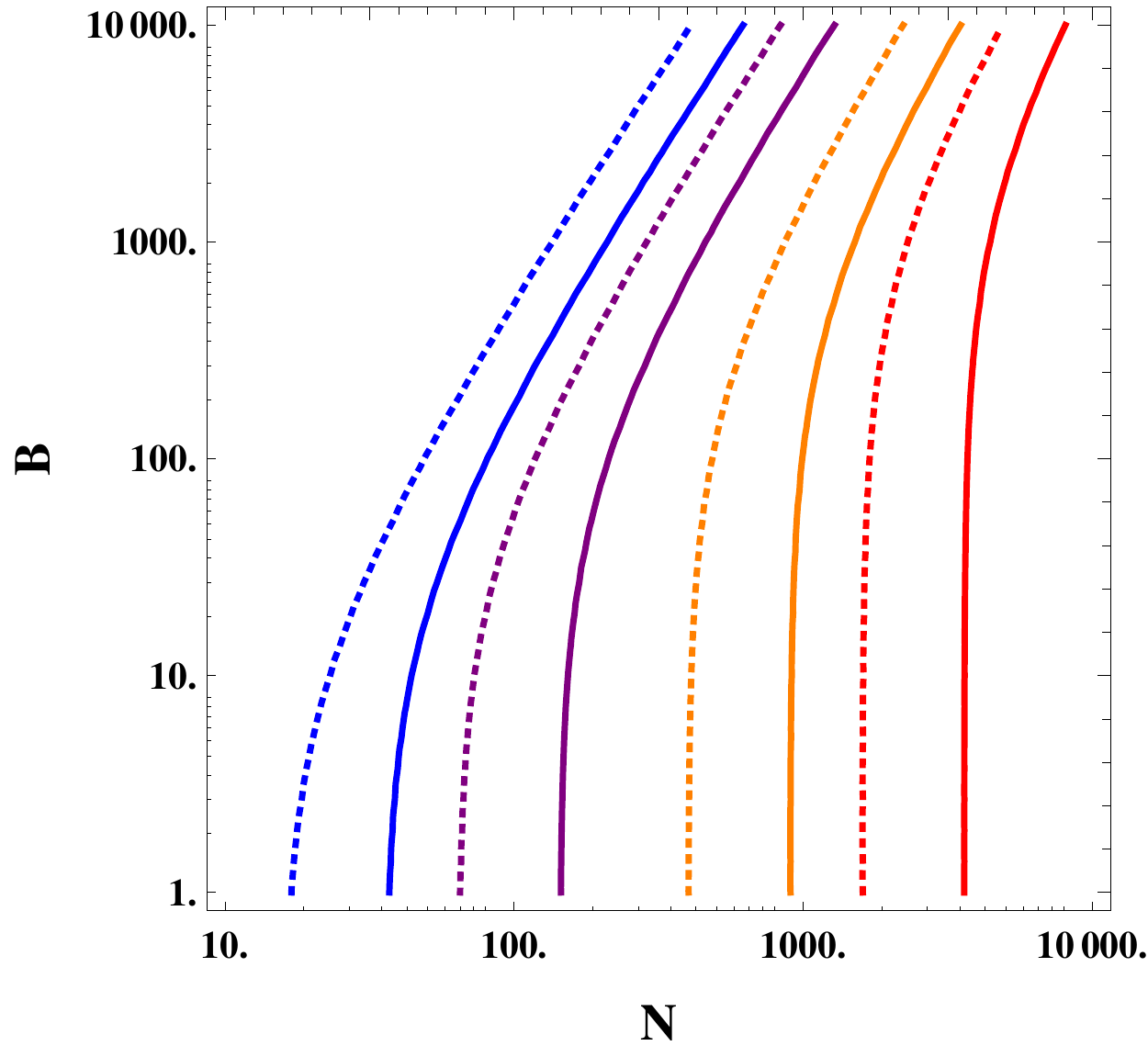}
   	\caption{Expected precision $P_Z$ for top-polarization based $SU(5)$-tests as a function of the total number of signal and background events $N$ and $B$. The spin-analyzing power is fixed to $\kappa=1$. Plain (dotted) lines denote $2\sigma$ ($3\sigma$) significance, respectively. Blue, purple, orange, and red lines respectively show $P_{2,3} = 100\%,~50\%,~20\%,~10\%$ isolines of expected precision.}
     \label{fig:NB_top}
\end{figure}

Note that $P_Z \rightarrow \infty$ for $\kappa \rightarrow 0$, as no spin information is available in this limit. The expected precision is shown in Fig.\ \ref{fig:NB_top} as a function of the background and total event numbers $B$ and $N$. The $SU(5)$ hypothesis starts to be testable when $P_Z$ is lower than $100\%$. 

For $\kappa=1$ and $B=0$, we see that the hypothesis starts to be testable for $N \gtrsim 16$ at $2\sigma$ and for $N \gtrsim 36$ at $3\sigma$ significance. About $144$, $900$, and $3600$ events are needed to test the relation at respectively $50\%$, $20\%$, $10\%$ precision with $3\sigma$ significance. Clearly, and contrary to a mere signal discovery, a substantial amount of signal events is necessary in order to test the $SU(5)$ hypothesis.

\subsection{Existing LHC searches}
\label{se:dipoles_LHC}

Let us now discuss the various LHC processes that we expect to be relevant to perform the $SU(5)$ test described previously. The dipole-induced anomalous top couplings induce the two-body decays
\begin{equation} 
	t\rightarrow q\, \gamma\,, \qquad 
	t \rightarrow q\, Z\,, \qquad 
	t \rightarrow q\, g\,.
\end{equation}
Three and four-body decays also exist, they are obtained from the ones above by adding an extra Higgs in the final state and replacing $Z$ by $W^+ W^-$ \footnote{As the dipole operators of Eq.\ \eqref{eq:top_dipoles} involve a Higgs field, the effective vertices that they produce can be either taken with $\langle H\rangle = v/\sqrt{2}$, i.e.\ the $t u V$ vertices, or with an extra Higgs boson, i.e.\ the $t u V h$ vertices. The Higgs vertices can be overlooked if one directly introduces the effective operators in the broken EW phase, i.e.\ with $\langle H\rangle =v/\sqrt{2}$, as customarily done in the anomalous-top coupling literature.}. Among the two-body decays,  $t \rightarrow q Z$ and $t \rightarrow q \gamma$ carry information about the dipoles, that can be accessed through polarization of the outgoing gauge bosons.
 The $t \rightarrow q \gamma$ process has been searched for at the Tevatron and LHC \cite{Yazgan:2013pxa,CMS_tqZ_decay}, the leading CMS bound is provided in Table \ref{tab:LHC_bounds}.

We now turn to dipole-induced top production. All the LHC processes we can consider have the particularity of featuring a \textit{single top}. For a proton-proton collider, the main partonic processes one can think about are  $g\,q\rightarrow t$, $g\,q\rightarrow t Z $, $g\,q\rightarrow t \gamma $, $g\,g\rightarrow t\,q $. Apart from the latter, various LHC searches have already been performed in these channels, see \cite{CMS_tqZ,CMS:tgam,ATLAS_tqg, Agram:2013koa, Abramowicz:2011tv}. The sensitivities translated on dipole coefficients are given in Tab. \ref{tab:LHC_bounds}. 
The leading bounds on $\alpha^{\gamma}_{3i,i3}$, $\alpha^{Z}_{3i,i3}$, $\alpha^{G}_{3i,i3}$ come from CMS and ATLAS searches for rare top decays, top-$\gamma$ production and single top production. 

\begin{table}

\begin{center}
\begin{tabular}{|c|c|c|}
\hline
Dipole coefficients combinations & $95\%$ CL limits & SUSY\\
\hline 
$\left(|\alpha_{31}^\gamma|^2+|\alpha_{13}^\gamma|^2\right)^{1/2}M_{\rm SUSY}^{-2}$ & 
$<0.19 \textrm{ TeV}^{-2}$ \cite{CMS:tgam} (CMS) & $ 7\cdot 10^{-5}$ TeV$^{-2}$ \\ 
\hline 
$\left(|\alpha_{32}^\gamma|^2+|\alpha_{23}^\gamma|^2\right)^{1/2}M_{\rm SUSY}^{-2}$ & 
$<0.65 \textrm{ TeV}^{-2}$ \cite{CMS:tgam} (CMS) & $ 7\cdot 10^{-5}$ TeV$^{-2}$ \\ 
\hline 
$\left(|\alpha_{31}^Z|^2+|\alpha_{13}^Z|^2\right)^{1/2}M_{\rm SUSY}^{-2}$ & 
$<0.68 \textrm{ TeV}^{-2}$ \cite{CMS_tqZ} (CMS) & $ 1\cdot10^{-4}$ TeV$^{-2}$  \\ 
\hline 
$\left(|\alpha_{32}^Z|^2+|\alpha_{23}^Z|^2\right)^{1/2}M_{\rm SUSY}^{-2}$ & 
$<3.44 \textrm{ TeV}^{-2}$ \cite{CMS_tqZ} (CMS) & $ 1\cdot10^{-4}$ TeV$^{-2}$  \\ 
\hline 
$\left(|\alpha_{31}^G|^2+|\alpha_{13}^G|^2\right)^{1/2}M_{\rm SUSY}^{-2}$ & 
$<0.029 \textrm{ TeV}^{-2}$ \cite{ATLAS_tqg} (ATLAS) & $ 3\cdot10^{-4}$ TeV$^{-2}$
\\
\hline
$\left(|\alpha_{32}^G|^2+|\alpha_{23}^G|^2\right)^{1/2}M_{\rm SUSY}^{-2}$ & 
$<0.063 \textrm{ TeV}^{-2}$ \cite{ATLAS_tqg} (ATLAS) & $ 3\cdot10^{-4}$ TeV$^{-2}$
\\
\hline
\end{tabular} 
\end{center}
\caption{Leading experimental limits on dipole operators. All the limits are given at $95\%$ confidence level. The SUSY values are given for $M_{\rm SUSY}=1$ TeV, $\delta\sim 1$, and $\tan\beta=5$.} 
\label{tab:LHC_bounds}
\end{table}

We observe that all these bounds seem to be far from the  sensitivity required to observe the dipoles from SUSY, which are loop-generated. The typical order of magnitude for the SUSY dipole operators (Eq. \eqref{eq:dipole_coef}) $\alpha^{G}_{3i, i3}/M_{\rm SUSY}^2$ ($\alpha^{W}_{3i, i3}/M_{\rm SUSY}^2$) is 
indicated in the last column of Table \ref{tab:LHC_bounds}. It might be in fact possible that none of the searches above will ever be able to reach the SUSY dipoles at the LHC, even with $3000$ fb$^{-1}$ of integrated luminosity.

For illustration, let us consider the ATLAS $gq \rightarrow t$ search discussed in Ref.\ \cite{ATLAS_tqg}. Knowing the expected background $B\approx 10^{5}$ and the $95\%$ confidence level bound on $|\alpha^G_{3i}|^2+|\alpha^G_{i3}|^2$ for $L=14.2 $ fb$^{-1}$ one can readily estimate the event rate by interpreting the significance as $Z=S/\sqrt{B+S}$ \footnote{The ATLAS statistical analysis is much more evolved, but we expect this estimation to be enough for qualitative considerations.}. We obtain $\sigma\approx 52 (|\alpha^G_{3i}|^2+|\alpha^G_{i3}|^2)/M_{\rm SUSY}^4$ pb$^{-1}$. Using the typical SUSY-induced values of $\alpha^G_{3i}$ shown above and extrapolating to $L=3000$ fb$^{-1}$, we find that only $\sim 28$ signal events would be collected in this analysis. The background would need to be drastically reduced in order to obtain a $O(1)$ significance. A mere discovery of the SUSY dipoles using these analyses being apparently difficult, applying top polarimetry is even more compromised, as a substantial amount of events is necessary, as shown in Fig.\ \ref{fig:NB_top}.

\subsection{Proposal for ultrapheripheral searches}
\label{se:dipoles_UPC}

Given that the backgrounds are rather large in the LHC analysis described above, we  now propose  an alternative type of measurement relying on
ultraperipheral collisions (UPCs). These are the quasi-diffractive processes that occur when the incoming protons or heavy-ions interact with an impact parameter larger than the sum of their radius (see Ref.\ \cite{Bertulani:2005ru} for a review). In the ultraperipheral regime the remnant of these nuclei are only slightly deflected and proceed in forward direction. The protons have a good chance to remain intact and can be tagged in the forward detectors planned for both ATLAS \cite{atlas_FW} and CMS \cite{cms_FW}. Tagging both protons in fully exclusive processes provides a powerful background rejection, because the complete kinematics of the events are known (see e.g.\ Refs.\ \cite{Heinemeyer:2007tu,Tasevsky:2014cpa,Royon:2014kta, Fichet:2014uka}). Heavy-ion collisions have no pile-up, and do not need forward tagging. For both proton and heavy ions, UPCs are therefore an attractive tool for precision physics.

As an accelerated charged object, the proton radiates short pulses of electromagnetic field. The associated photon flux $f_\gamma$ emitted by the proton is proportional to the squared electric charge, $f_\gamma \propto e^2 \sim 0.1$. Recent prospects for photon fusion in UPCs can be found in Ref.\ \cite{d'Enterria:2013yra}. In the case of heavy ions, a striking feature is that all charges inside the nucleus radiate coherently, such that the photon flux scales as $f_\gamma\propto Z^2 e^2$. This phenomenon is known and exploited since a long time (see, e.g., Ref.\ \cite{Baur:1999fz} and references therein). The enhancement factor is, e.g., $82^4\sim 4.5\cdot 10^7$ for Pb-Pb collisions. This enhancement has to be contrasted with the low expected heavy-ion run luminosity, typically $1$ nb$^{-1}$ per year. This type of experiment is useful if the large photon flux can compensate the small luminosity. 

We now point out that the same reasoning can qualitatively apply to electroweak charges. The $W$- and $Z$-boson fluxes can be treated similarly to the photon \cite{Kane:1984bb}. In the UPC regime, contrary to a seemingly widespread belief, electroweak charges radiate coherently just like the electric ones.  Although the $W$ and $Z$ boson masses are responsible for the ``short range'' of the EW interaction in a non-relativistic view, in relativistic QFT these masses have to be put in balance with the centre-of-momentum energy of the di-boson collision. The maximum energy from a flux is estimated to be roughly $\gamma/R$. For pp collisions, the maximum energy of the vector collision system is $\sqrt{\omega}_{\rm max}\sim 4.5 $ TeV \cite{d'Enterria:2013yra}. This is large with respect to $m_W$ or $m_Z$, so that the flux of transverse $W$ or $Z$ is essentially identical to the one of the photon. Also, by the equivalence theorem, the flux of longitudinal $W$- and $Z$-bosons can be replaced by the flux of corresponding Goldstone bosons. For heavy-ion collisions, we have $\sqrt{\omega}_{\rm max} \sim 260 $ GeV and $\sim 160 $ GeV respectively for p-Pb and Pb-Pb, such that $\sqrt{\omega}_{\rm max}=O(m_{W,Z})$. In this case the EW fluxes are thus expected to be suppressed in comparison with the photon flux case. Understanding more precisely the features of these EW fluxes deserves a dedicated analysis. 
Note that with lead ions in initial states, the emission of $W^+$, $W^-$ transforms $Pb$ into  $Bi$ and $Tl$ ions respectively. 
Such isotopes are unstable, but are long-lived with respect to the LHC scale.
 
Here we simply focus on the effect of the total EW charge, i.e.\ the overall prefactor of the $W$- and $Z$-fluxes. For coherent fluxes $f_{W^{\pm}}$ and $f_Z$, we have $f_{W^+} \propto 2 g^2\sim 0.9$, $f_{W^-} \propto g^2/2\sim 0.2 $, and $f_{Z}\propto g^2/c^2_w (1/4 -s^2_w+2s^4_w) \sim 0.07$. For the proton, the EW fluxes are therefore slightly enhanced for $W$-bosons because of the magnitudes of the involved couplings. For the heavy-ion EW fluxes, in contrast to the electromagnetic case, the neutrons inside the nucleus are charged under the weak interaction. We obtain
\begin{align}
	f_{W^+} ~&\propto~ \frac{g^2 (Z+A)^2}{2}\,, \\
	f_{W^-} ~&\propto~ \frac{g^2 (2A-Z)^2}{2}\,, \\
	f_Z ~&\propto~ \frac{g^2}{c^2_w}\left[\frac{(2Z-A)^2}{4}-Z(2Z-A)s^2_w+2Z^2s^4_w \right]\,.
\end{align}
For Pb-Pb collisions, we have therefore an enhancement of $4.1 \cdot 10^6$ for $f_Z^2$ and $9.4 \cdot 10^9$ for $f_{W^{+}}f_{W^{-}}$ with respect to the p-p case. For $WW$ the enhancement factor is $200$ times larger than the enhancement for $\gamma\gamma$. Taking into account the gauge couplings, the factor from $f_{W^{+}}f_{W^{-}}$ is $10^3$ times larger than the factor from $f_\gamma^2$. Provided that the  $W$ fluxes are not strongly suppressed, processes involving intermediate $W$-bosons in heavy-ion collisions at the LHC are thus rather attractive.

\begin{figure}
  	\centering
	\includegraphics[width=6cm]{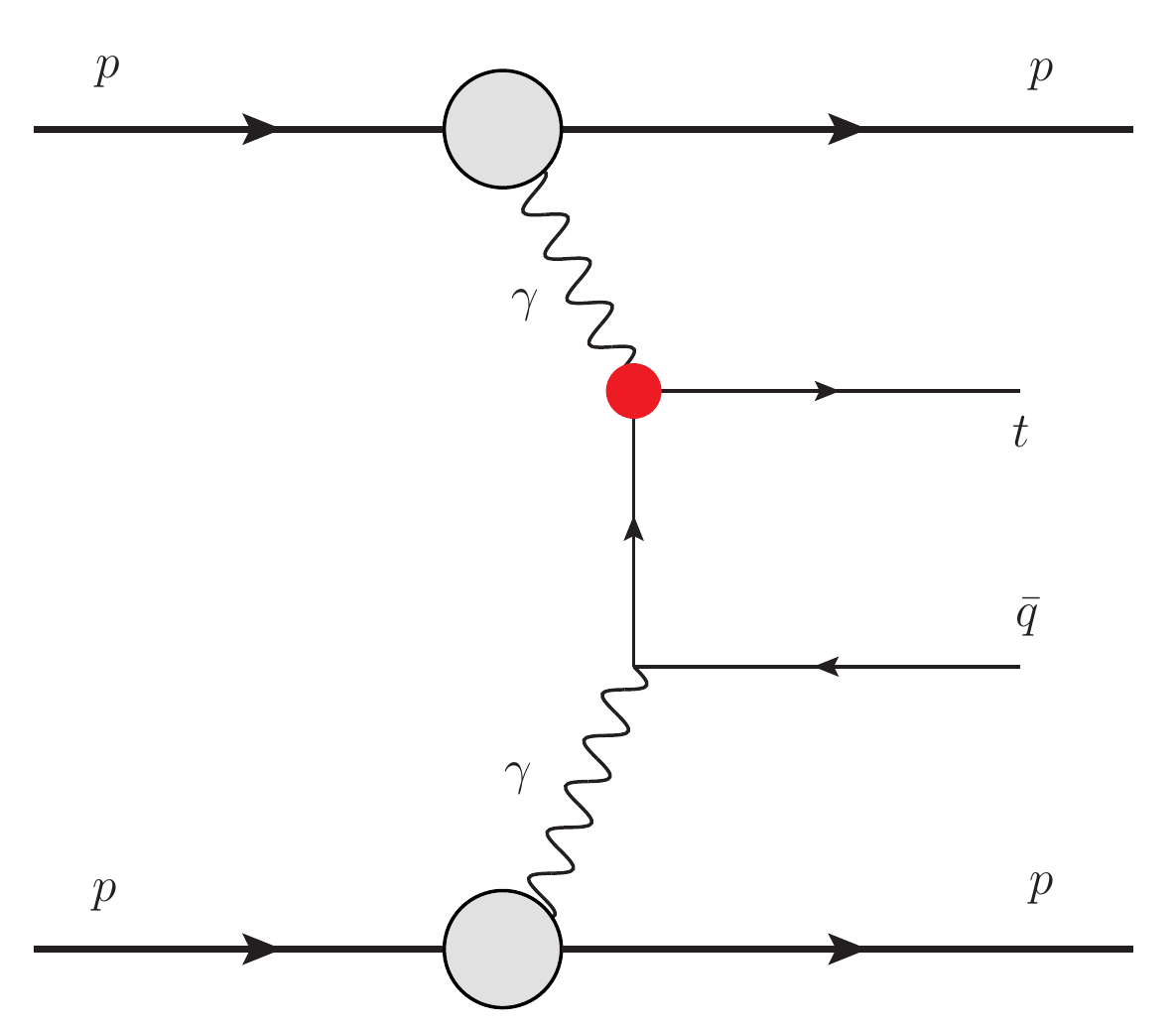}
	\includegraphics[width=6cm]{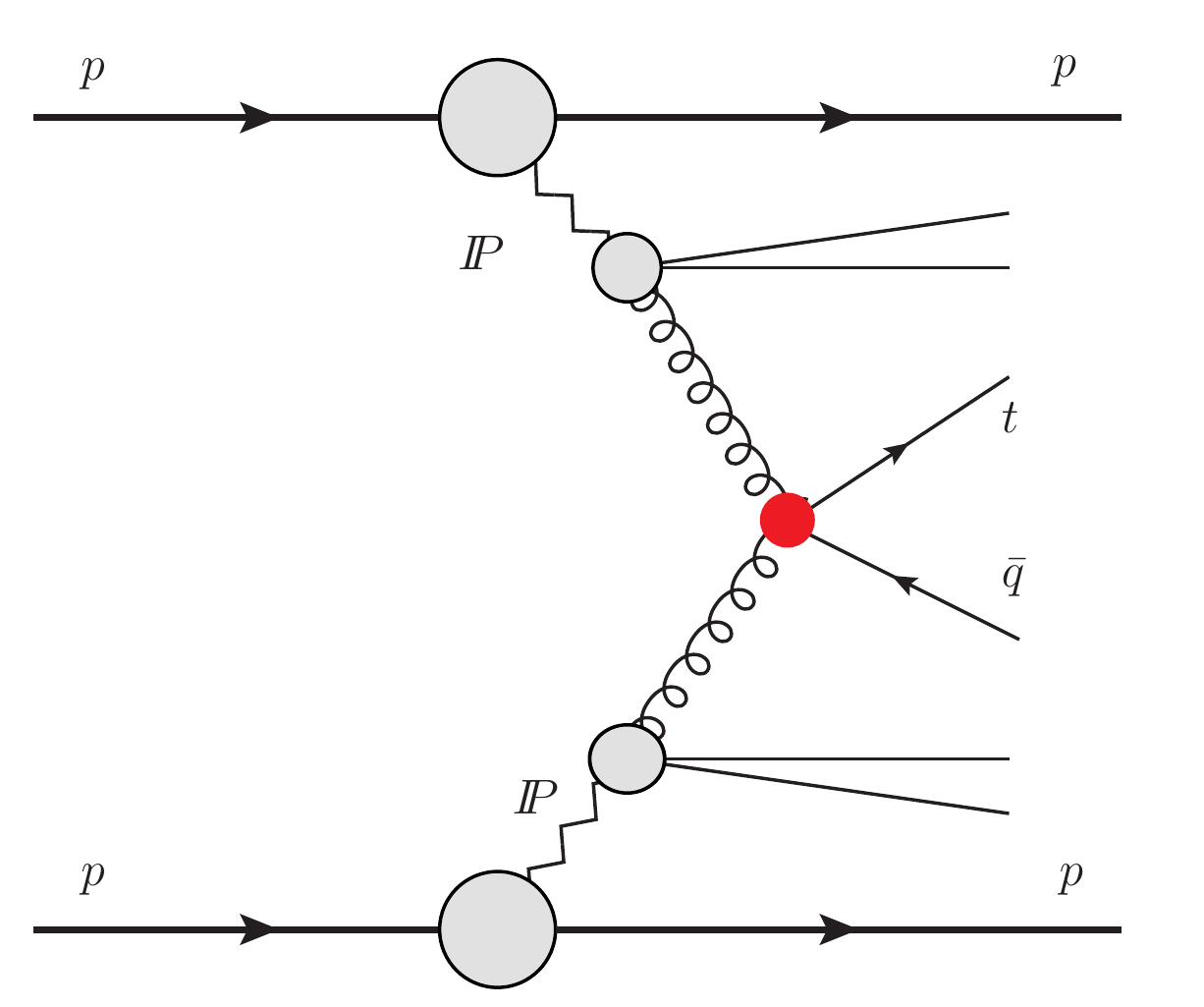}\\
	\includegraphics[width=6cm]{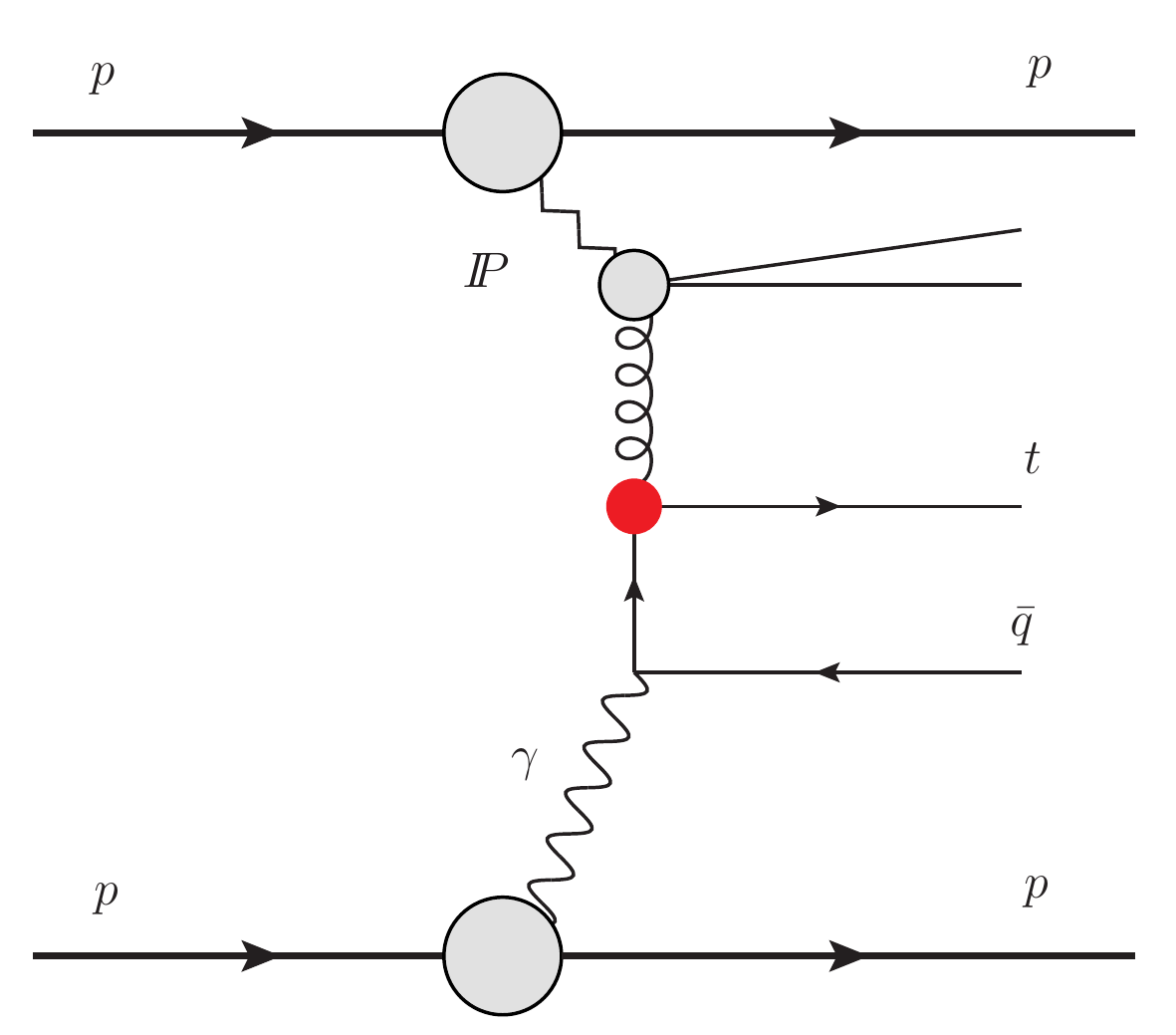} \\
	\includegraphics[width=6cm]{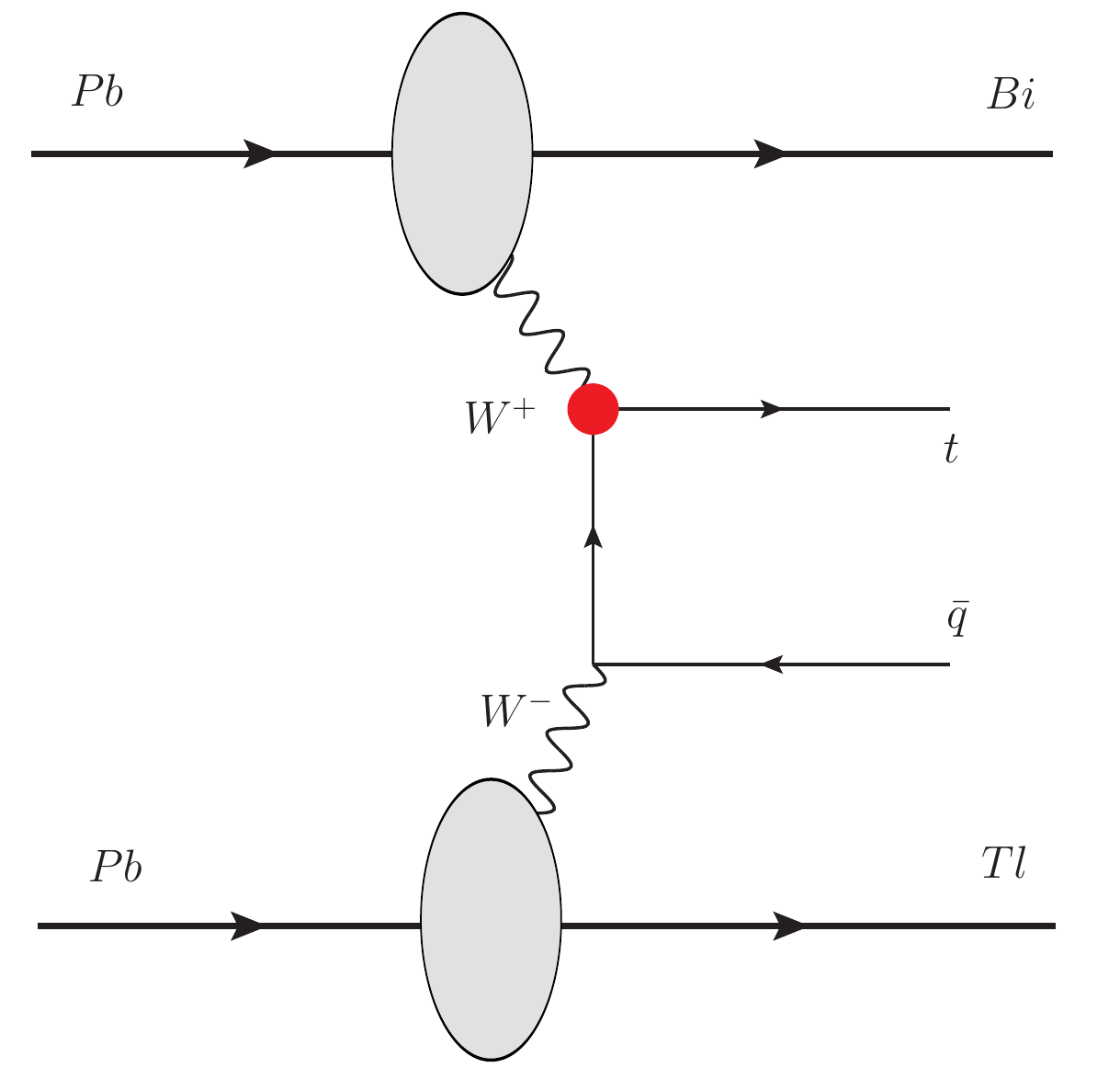}
	\includegraphics[width=6cm]{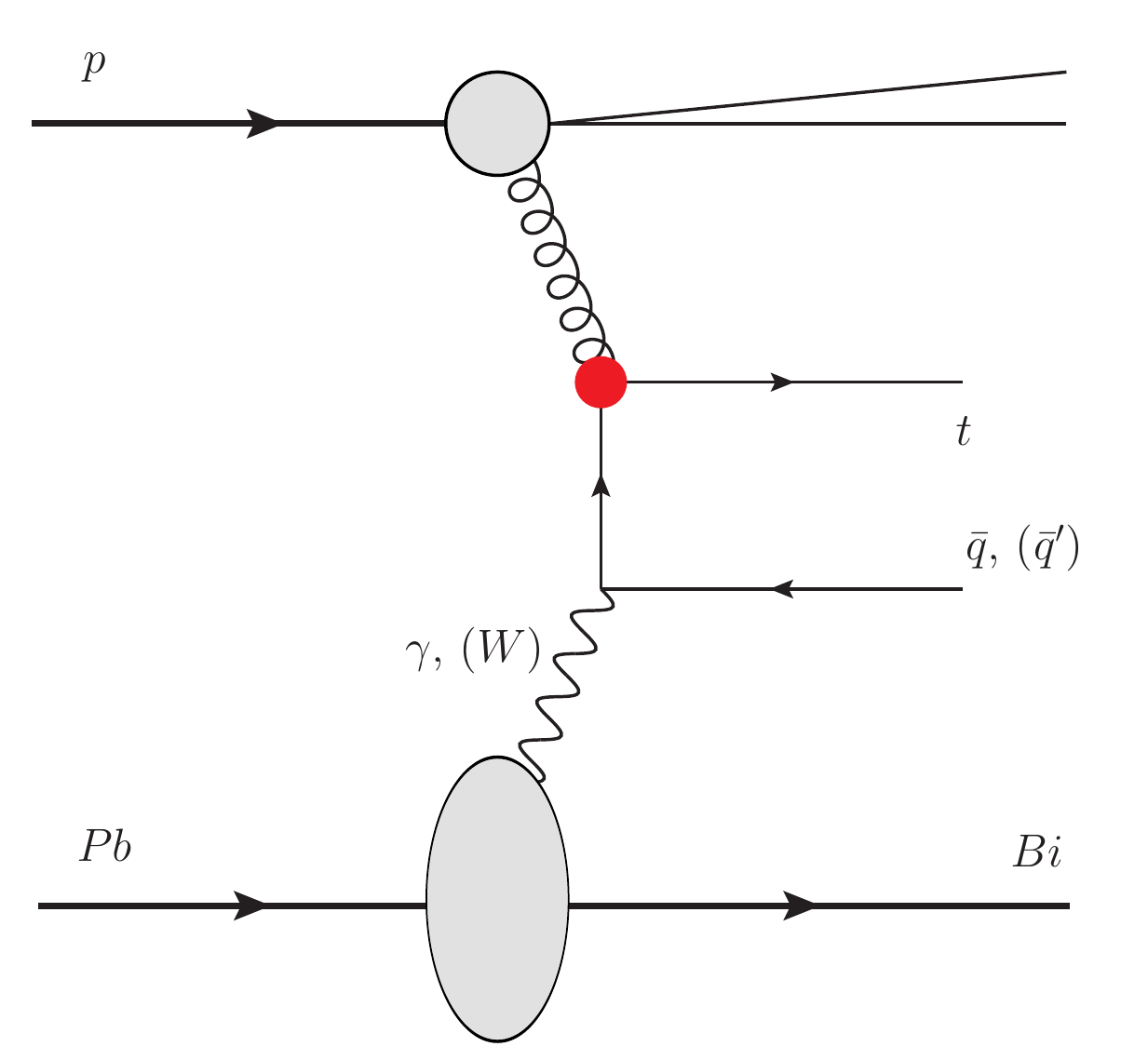}
   	\caption{Ultraperipheral processes potentially relevant to search for the dipole operators $\mathcal{O}^{G,\gamma,W}$ at high precision. The processes shown are the ones relevant for $SU(5)$ tests. The  vertex marked in red denotes the SUSY dipole. }
   	\label{fig:UPCs}
\end{figure}

The proton cannot emit a single gluon and stay intact because of color conservation. For UPCs however, an intriguing possibility is that the gluon comes from a Pomeron emitted by the proton \cite{IS}. In that case the proton remains intact and can be tagged. However, the Pomeron itself cannot remain intact, such that the process is not fully exclusive. Taking all features into account, we now propose a selection of processes that deserves particular attention in order to probe the dipole operators $\mathcal{O}^{G,\gamma,W}$ with high sensitivity. The processes are pictured in Fig.\ \ref{fig:UPCs}:
\begin{itemize}
\item {\it $\gamma\,\gamma \rightarrow t\, q$ in p-p collisions (Central exclusive production)}\\
The two protons can be tagged and the reaction is exclusive. However, because of the $e^4$ suppression for photon fluxes, the cross-section might be too small to detect the $\mathcal{O}^\gamma$ dipoles.
\item {\it $g\,g\rightarrow t\,q $ through Pomeron fusion in p-p collisions (Double Pomeron exchange)}\\
The two intact protons can be tagged. However, this process is not fully exclusive as the Pomerons break up. A detailed study using diffractive PDFs would be necessary in order to evaluate the potential of this channel. This process is sensitive to the $\mathcal{O}^{G,\gamma}$ dipoles. 
\item {\it $\gamma\,g \rightarrow t\, q$ in p-p collisions (Diffractive photoproduction)}\\
The proton radiating the photon can be tagged. One may also require a second intact proton. The remnant of the subsequent Pomeron would then be observed in one hemisphere only. This process is sensitive to $\mathcal{O}^G$, and may be an interesting trade-off between the two above processes. 
\item {\it $WW\rightarrow t\,q$ in Pb-Pb collisions (Central exclusive production)}\\
This configuration fully relies on the huge enhancement factor from the coherent action of the EW charges of the nucleons. A computation of the $W$-boson fluxes is required. This process is sensitive to the $\mathcal{O}^W$ dipoles. 
\item {\it $g\gamma\rightarrow t\,q$, $gW\rightarrow t\,q'$ in p-Pb collisions (Single diffraction dissociation)} \\
No proton tagging is necessary in this case, so that ones lets the gluon be emitted by the proton  in order to take profit of the large gluon PDF. The luminosity is higher than in the Pb-Pb case, about $200$ nb$^{-1}$ per year. One takes profit of the $\gamma$- or $W$-flux enhancement from the heavy ion. This process is sensitive to the $\mathcal{O}^G$ dipoles. Therefore this configuration might be a good compromise between p-p and Pb-Pb collisions.
\end{itemize}

All these processes feature a top-quark plus a jet in the final state. Would the expert reader obtain an estimate for one of these processes, the potential to test the $SU(5)$ hypothesis through top polarimetry could be directly drawn from Fig.\ \ref{fig:NB_top}. Identifying the jet charge -- if possible -- would be useful to reject fake QCD events \cite{Krohn:2012fg}. Finally, we remark that all these processes are relevant for generic dipole operators searches beyond SUSY. Also, the enhancement of $W$-boson fluxes from the total EW charge has implications for the searches of other anomalous couplings like triple and quartic gauge interactions. 

\section{Natural supersymmetry }
\label{se:natural}

We now investigate $SU(5)$ tests in the framework of ``natural'' (or effective) supersymmetry. The mass spectrum of this scenario features a first and second generation of up-squarks that are considerably heavier than the squarks of the third generation. This pattern is motivated by both naturalness considerations \cite{Cohen:1996vb} and by ultraviolet (UV) constructions like partially-supersymmetric composite models \cite{Redi:2010yv} or supergravity contributions generically present in five-dimensional models \cite{Dudas:2012mv}. 

Clearly, in natural SUSY, the up-squarks effective theory (see Sec.\ \ref{se:EFT_MIA}) will consist of two mostly stop-like squarks. Their mixing is not constrained and can potentially be large, which is often needed to satisfy the constraint coming from the observed Higgs-boson mass. The effective operators that appear when integrating out the heavy up-squarks can potentially induce flavour-changing stop decays, and violation of unitarity relations in the stop chiral decays \cite{Fichet:2014vha}. 

All the information about the $SU(5)$ relation $a_u\approx a_u^t$ is enclosed in the higher-dimensional operators of the effective Lagrangian for the stops. Potential $SU(5)$ tests have to rely on the effect of these effective operators in the stop sector. Therefore these tests involve both \textit{real} squarks (the stops) and \textit{virtual} squarks. We assume that both stops ($\tilde{t}_1$ and $\tilde{t}_2$) are produced at the Large Hadron Collider (LHC). The total production cross-sections of  stop pairs at next-to-next-to-leading order (NNLO) are $\sigma_{\tilde t \tilde t} = 89~{\rm fb}^{-1} \pm 10\%  $, $8.6~{\rm fb}^{-1} \pm 15\%$, $0.71~{\rm fb}^{-1} \pm 20\%$, and $0.08\pm 25\%$ fb$^{-1}$ for stop masses of $m_{\tilde{t}} = 700$, 1000, 1400, and 1800 GeV, respectively \cite{SUSYXSWG}.

In the following, we develop $SU(5)$ tests that already have been sketched in Ref.\ \cite{Fichet:2014vha}. We refer to this paper for certain details of the derivation. We evaluate the potential of these tests using the frequentist treatment described in Sec.\ \ref{se:EFT_MIA}. We consider two cases concerning the mass ordering between the stops ($m_{\tilde{t}_{1,2}}$), the bino ($m_{\tilde{B}}$), and the wino ($m_{\tilde{W}}$). We assume that the lightest neutralino is mostly bino-like, $\tilde{\chi}^0_1 \sim \tilde{B}$, and the second-lightest neutralino is mostly wino-like, $\tilde{\chi}^0_2 \sim \tilde{W}$, as it is typically the case in scenarios with gaugino mass unification at the GUT scale.

\subsection{The $m_{\tilde{t}_{1,2}} > m_{\tilde{W}} > m_{\tilde{B}}$ case}

This mass ordering allows the stops to decay either into the lightest neutralino $\tilde{\chi}^0_1 \approx \tilde{B}$ or into the second-lightest neutralino $\tilde{\chi}^0_2 \approx \tilde{W}$. In order to build a $SU(5)$ test, we are interested in the flavour-changing decays 
\begin{equation}
	\tilde{t} ~\rightarrow~ \tilde{W}\, u/c ~\rightarrow~ \tilde{B}\,Z/h \, u/c
	\qquad {\rm and} \qquad
	\tilde{t} ~\rightarrow~ \tilde{B}\, u/c\,.
\end{equation}
We assume that the stop masses are unknown, and that only the event rates of the stop decays into binos and winos, respectively denoted by $N_Y$ and $N_L$, are experimentally accessible. Moreover, we assume that a certain fraction, denoted by $N_Y^c$ and $N_L^c$, of these events can be charm-tagged. The remaining events $N_Y^{\not c}=N_Y-N_Y^c$ and $N_L^{\not c}=N_L-N_L^c$ then contain both up-quark events and miss-tagged charm jets. 

Assuming the same charm-tagging efficiency $\epsilon_c$ for $N_Y$ and $N_L$ \footnote{The formula can be generalized in a straightforward manner if the efficiencies are different.}, we have
\begin{equation} 
	{\rm E}[N_Y^c] ~=~ \epsilon_c \gamma_Y {\rm E}[N_Y]\,,\qquad {\rm E}[N_L^c] ~=~ \epsilon_c\gamma_L {\rm E}[N_L]\,,
\end{equation}
where $\gamma_{Y,L}$ are the actual fractions of charm events. As shown in Ref.\ \cite{Fichet:2014vha}, the $SU(5)$ hypothesis implies $\gamma_{Y} = \gamma_{L} \equiv \gamma$. The $SU(5)$ test is therefore
\begin{equation}
	\frac{N_Y^c}{N_L^c} ~=~ \frac{N_Y^{\not c}}{N_L^{\not c}}\,.
	\label{eq:charm_test}
\end{equation}
Let us remark that the large QCD error on the underlying cross-sections roughly cancels out in the above ratios of event rates. Moreover, no information on the stop mixing angle nor the stop masses is necessary to perform the test.

\begin{figure}
	\centering
	\includegraphics[width=8cm]{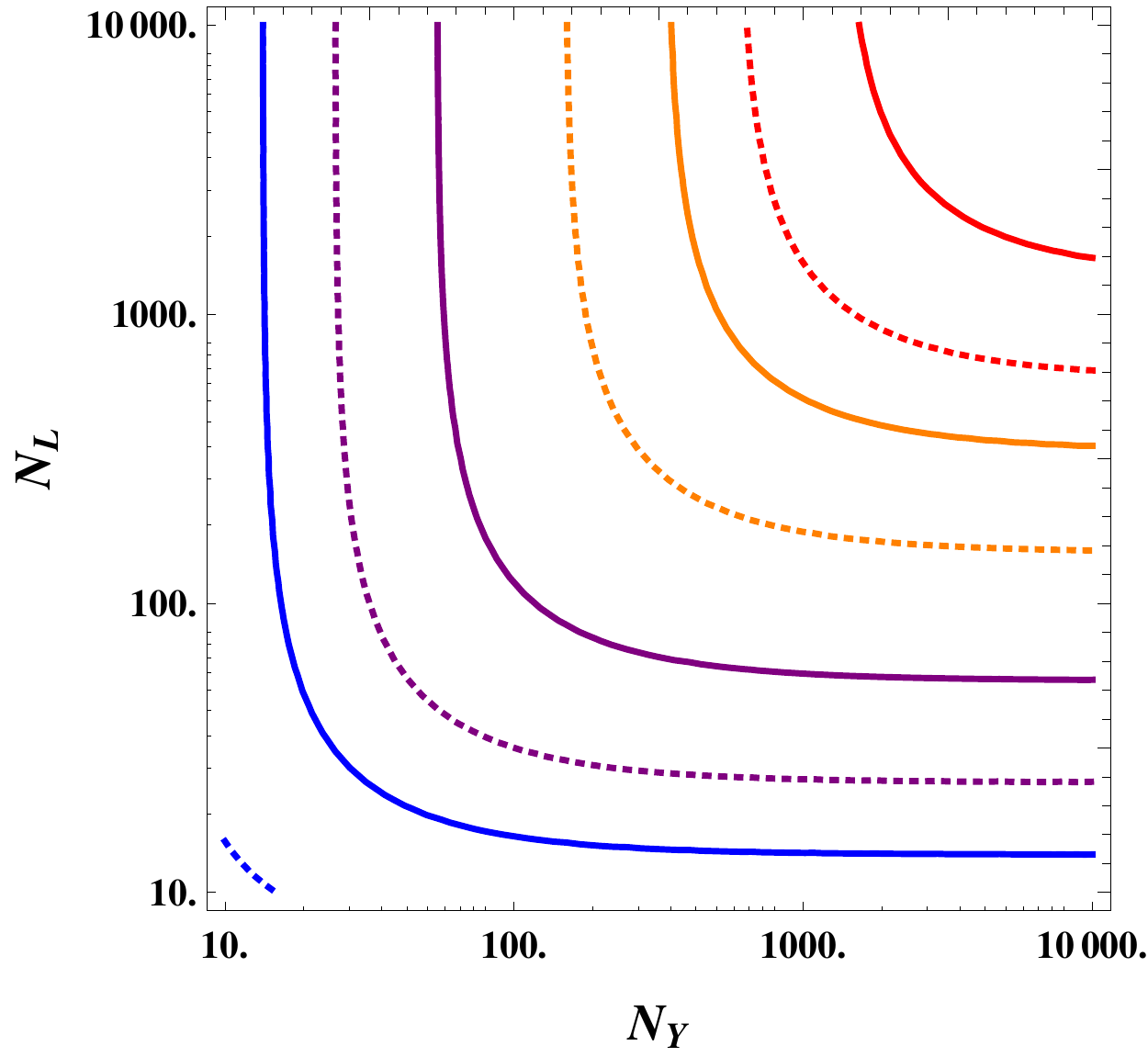}
   	\caption{Expected precision $P_Z$ for the $SU(5)$ test on flavour-changing stop decays Eq.\ \eqref{eq:charm_test}. $N_Y$ and $N_L$ are the numbers of observed stop decays to bino and wino, respectively. The charm fraction is fixed to $\gamma = 0.5$ and the charm-tagging efficiency to $\epsilon_c = 0.5$. Plain (dotted) lines denote $2\sigma$ ($3\sigma$) significance respectively. Blue, purple, range and red lines respectively show $P_{2,3} = (100\%,50\%,20\%,10\%)$ isolines of the expected precision.}
     \label{fig:charm_tag} 
\end{figure}

Following Sec.\ \ref{Sec:ExpPrec}, we define the normalized test quantity as
\begin{equation}
	R ~=~ {\bigg|\frac{N_Y^c}{N_L^c}-\frac{N_Y^{\not c}}{N_L^{\not c}} \bigg| } \bigg/
		  {\bigg(\frac{N_Y^c}{N_L^c}+\frac{N_Y^{\not c}}{N_L^{\not c}}\bigg)}\,,
\end{equation}
such that
\begin{equation}
	{\rm E}[R] ~=~ \frac{|\gamma_Y-\gamma_L|}{\gamma_Y+\gamma_L} \,.
\end{equation}

The expected precision (see Sec.\ \ref{Sec:ExpPrec}) associated with this test is found to be 
\begin{equation}
	P_Z ~=~ \frac{Z}{2\epsilon_c^{1/2}} \bigg(\frac{1}{N_Y}+\frac{1}{N_L} \bigg)^{\! 1/2}
		\bigg( \frac{1}{\gamma}+\frac{1}{\gamma-1}-\frac{(1-\epsilon_c)\gamma}{(\gamma-1)^2} \bigg)^{\! 1/2}\,.
\end{equation}
Note that $P_Z\rightarrow \infty$ in the limits $\gamma\rightarrow 0$ (no charm jets expected), $\gamma\rightarrow 1$ (no up jets expected), or $\epsilon_c\rightarrow 0$ (no charm-tagging). The expected precision is shown in Fig.\ \ref{fig:charm_tag}, where the actual charm fraction is set to $\gamma = 0.5$, and the charm-tagging efficiency is set to $\epsilon_c = 0.5$. We see that $N_Y = N_L \gtrsim 27$ events are sufficient to start to probe the relation at 3$\sigma$ significance, i.e.\ to have $P_3 < 100\%$. 

Testing the relation with $50\%$, $20\%$, or $10\%$ precision at $3\sigma$ requires respectively $N_Y \sim N_L \sim 110 $, 675, or 2700 events. For comparison, assuming stop flavour-violating branching ratios of $0.05$ and $300$ fb$^{-1}$ of integrated luminosity, we expect about $1340$, $130$, and $11$ events for stop masses of $m_{\tilde{t}}=700$, $1000$, and $1400$ GeV, respectively.

\subsection{The $m_{\tilde{W}} > m_{\tilde{t}_{1,2}} > m_{\tilde{B}}$ case}

For this mass ordering, the stops decay only into the bino according to $\tilde{t}_{1,2} \rightarrow t_{L,R} \tilde{B}$. Performing top polarimetry on a decaying stop potentially gives access to the stop mixing angle \cite{Top_polarization_stop}. Here we point out that the same kind of procedure also provides a $SU(5)$-test.

In the language of the up-squark effective Lagrangian of Sec.\ \ref{se:EFT_MIA}, the operator coupling to $(\tilde{t}_L,\tilde{t}_R)$ is $\mathcal{O}\propto \tilde{B}(t_L,-4 t_R)$ \cite{MSSM}. The factor $-4$ comes from the different hypercharges of the left and right top components. In the stop effective Lagrangian, the effective mass term mixes $\tilde{t}_L$ and $\tilde{t}_R$. The mass eigenstates $\tilde{t}_a$, $\tilde{t}_b$ are obtained by 
\begin{equation}
	\begin{pmatrix}
 	   \tilde{t}_a \\ \tilde{t}_b
 	\end{pmatrix}
 	=
	\begin{pmatrix}
 	   c_\theta & s_\theta \\
 	  -s_\theta & c_\theta
 	\end{pmatrix}
	\begin{pmatrix}
 	   \tilde{t}_L \\ \tilde{t}_R
 	\end{pmatrix} \, ,
\end{equation}
where $s_\theta$ and $c_\theta$ are sine and cosine of the stop mixing angle $\theta_t$. No implicit assumption is made about the mass ordering of $\tilde{t}_a$, $\tilde{t}_b$, i.e.\ we do not know in principle whether $\tilde{t}_a = \tilde{t}_1$, $\tilde{t}_b = \tilde{t}_2$  or vice versa \footnote{The stop rotation matrix is real as we consider real SUSY breaking terms.}. The stop angle can be large and is important for phenomenology.
 
Using the effective Lagrangian of the stop, it can be shown \cite{Fichet:2014vha} that the top-stop-bino coupling is distorted with a structure
\begin{equation}
	\tilde{B}(t_L,-4 t_R)\,\mathcal{K}  \,R(\theta_t)\,
	\begin{pmatrix}
		\tilde{t}_a \\ \tilde{t}_b
	\end{pmatrix} \,,\qquad \qquad
	\mathcal{K} ~=~ 
	\begin{pmatrix}
		1~ & ~x \\ x~ & ~1
	\end{pmatrix}
	\label{eq:K_x}
\end{equation}
if the $SU(5)$ hypothesis is true. The parameter $x$ encloses the effect of effective operators and therefore $x \ll 1$. The matrix $\mathcal{K}$ is instead non-symmetric if the $SU(5)$ hypothesis is not satisfied. At leading order in the EFT expansion, i.e.\ neglecting effective operators, we have $\mathcal{K} = \mathbf{1}_2$.

As in Sec.\ \ref{se:Heavy_SUSY}, let us assume that the spin of the tops is analyzed through distributions of the form $(1 + \kappa P_t z)$ with $z \in[-1,1]$. The decays of the stops $\tilde{t}_{a}$ and $\tilde{t}_{b}$, leading to the event rates $N_a$ and $N_b$, are then splitted over the domains $\mathcal{D}_- = [-1,0]$ and $\mathcal{D}_+ = [0,1]$, such that $N_a = N_{a+} + N_{a-}$ and $N_b = N_{b+} + N_{b-}$. The expected event rates depend on $x$ and $\theta_t$, and are given in App.\ \ref{app:stop_decay}. 

The stop angle is obtained up to $O(x)$ corrections using the event rate expressions with $x=0$ (see App.\ \ref{app:stop_decay}), i.e.\ at leading order in the EFT expansion. Note that only one stop is necessary for that purpose. The stop mixing angle is for example obtained from the $\tilde{t}_a$ forward-backward asymmetry, that satisfies
\begin{equation} 
	{\rm E}[A_a] ~=~ \frac{\kappa}{2}\frac{15-17c_{2\theta}}{17-15c_{2\theta}}\,. 
\end{equation}	
The $x$ parameter is obtained using the full event rates expressions of App.\ \ref{app:stop_decay}, i.e.\ at next-to-leading order in the EFT expansion. The measurement of $N_{a,b\pm}$ provides two inequivalent ways of knowing $x$. The event rates $N_{a,b\pm}$ satisfy therefore one non-trivial relation if the $SU(5)$ hypothesis is verified. This $SU(5)$ relation can be put in the normalized form
\begin{equation}
	R ~=~ \bigg|\frac{(A_a-A_b)(1606+450c_{4\theta}) +(1028\kappa -2040 (A_a+A_b))c_{2\theta}) }
		{(765+255c_{4\theta})\kappa+2040(A_a-A_b)c_{2\theta}-(1606+450c_{4\theta})(A_a+A_b)
		}\bigg|\,,
		\label{eq:R_stop_decay}
\end{equation}
where we have introduced the usual asymmetries 
\begin{equation}
	A_a ~=~ \frac{|N_{a+}-N_{a-}|}{N_{a+}+N_{a-}} \quad{\rm and}\quad
	A_b ~=~ \frac{|N_{b+}-N_{b-}|}{N_{b+}+N_{b-}}\,.
\end{equation}
One has ${\rm E}[R]=0$ if the $SU(5)$ hypothesis is satisfied.

\begin{figure}[t]
\begin{picture}(400,150)
\put(80,0){		\includegraphics[trim=0cm 0cm 0cm 0cm, clip=true,width=9cm]{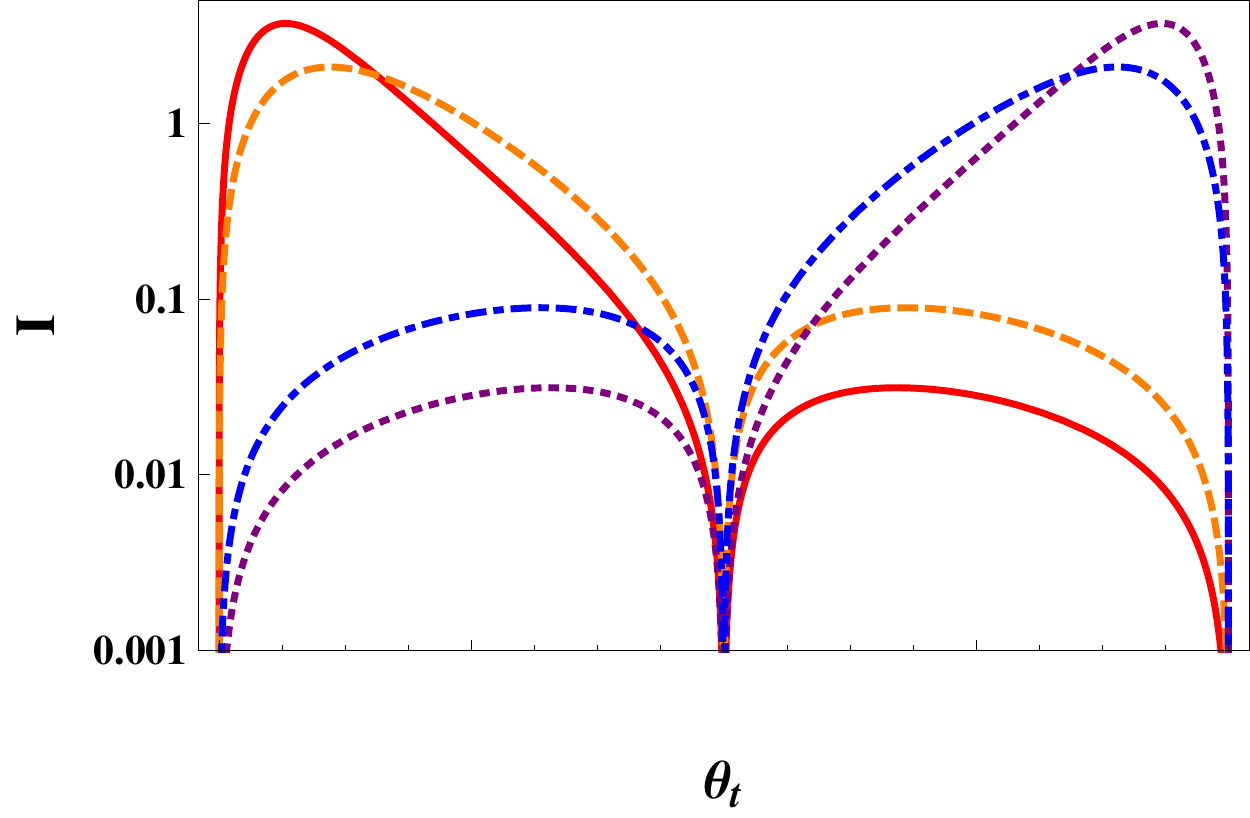}}
\put(127,24){$\scriptstyle{\mathbf{0}}$}
\put(172,24){$\scriptstyle{\mathbf{\pi/8}}$}
\put(223,24){$\scriptstyle{\mathbf{\pi/4}}$}
\put(272,24){$\scriptstyle{\mathbf{3\pi/8}}$}
\put(324,24){$\scriptstyle{\mathbf{\pi/2}}$}
\end{picture}
\vspace{0.cm}
   	\caption{Information content  of the event rates $N_{a,b\pm}$ with respect to the $x$ parameter (see Eq.~\eqref{eq:K_x}). Plain, dashed, dotted and dot-dashed lines correspond to $I[N_{a+}]$, $I[N_{a-}]$, $I[N_{b+}]$, and $I[N_{b-}]$, respectively. The spin-analyzing power is set to $\kappa=1$.}
   	\label{fig:info_stop}   
	\end{figure}
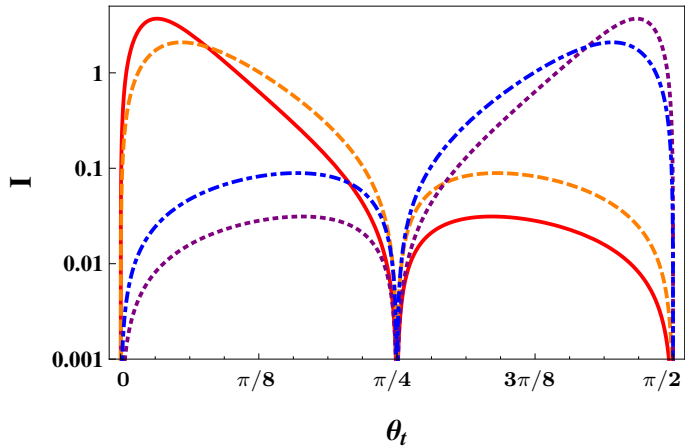

The information about $x$ carried by the expected event rates is conveniently measured by $I[N_{a,b\pm}] =| \partial \log N_{1,2\pm}/\partial x|$. It turns out that this information depends crucially on the stop mixing angle. This can be seen in Fig.\ \ref{fig:info_stop}, where we show $I[N_{a,b\pm}]$ as a function of the latter. The information becomes small for $\theta_t\sim 0$ (no stop mixing) and vanishes exactly for $\theta_t = \pi/4$ (maximal stop mixing). In between these two limit cases one has $I[N_{a\pm}] = O(1)$, $I[N_{b\pm}]=O(0.1)$, and the reverse for the interval $[\pi/4,\pi/2]$.  

\begin{figure}[t]
  	\centering
	\includegraphics[width=8cm]{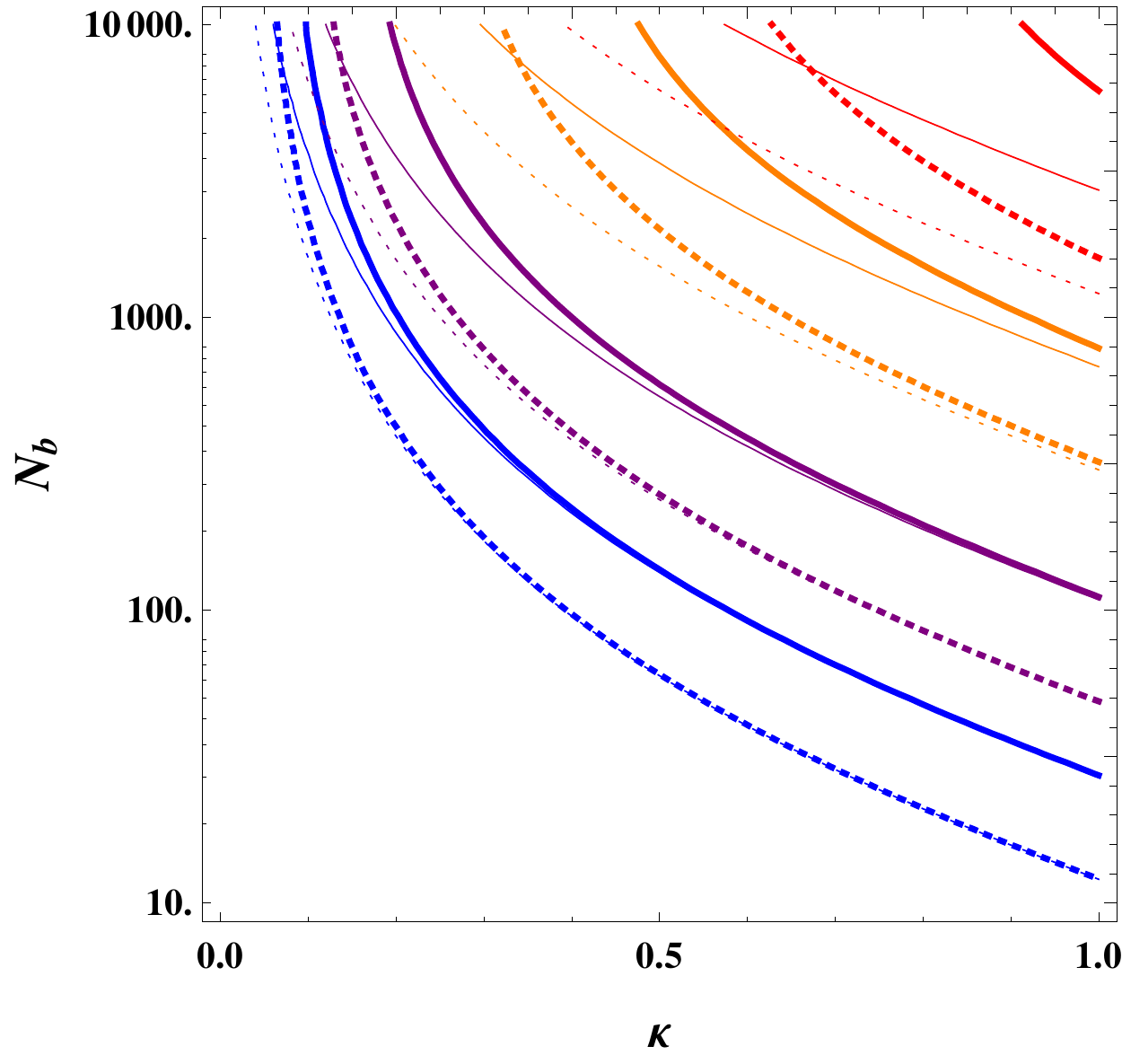}
   	\caption{Expected precision $P_Z$ for the $SU(5)$ test on polarization of stop decays Eq.\ \eqref{eq:R_stop_decay} as a function of the spin-analyzer efficiency $\kappa$ and of the amount of observed $\tilde{t}_b$ decays $N_b$. The number of $\tilde{t}_a$ decays is fixed to $N_a=20$ (thick lines) and $N_a=10000$ (thin lines). The stop mixing angle is fixed to $\theta_t=0.4$. Plain (dotted) lines denote $2\sigma$ ($3\sigma$) significance respectively.  Blue, purple, range and red lines respectively show $P_{2,3}=(100\%,50\%,20\%,10\%)$ isolines of expected precision.}
        \label{fig:PZ_stop_decay}
\end{figure}

The associated expected precision is found to be approximatively 
\begin{equation}
	P_Z \approx Z\frac{(17+15c_{2\theta}) }{255\sqrt{2} \,\kappa\, (3+c_{ 4\theta})}
	\bigg( \frac{3212-739\kappa^2-1020(\kappa^2-4)c_{2\theta} + (900-289\kappa^2)c_{4\theta}}{N_b} \bigg)^{1/2} \,
\end{equation}
in the $\theta_t \in [0, \pi/4]$ case.
We have $P_Z \rightarrow \infty$ for $\kappa \rightarrow 0$ as the test relies on top polarimetry. For $\theta_t \in [0, \pi/4]$, the event rates $N_{b\pm}$ are less sensitive to $x$ than the event rates $N_{a\pm}$. The expected precision depends thus mainly on the amount of $\tilde{t}_b$ produced. For this reason we drop the $1/N_a$ term is the equation above. The exact formula is given in App.\ \ref{app:stop_decay}.

As for $\theta_t \in [0, \pi/4]$ more $\tilde{t}_b$ than $\tilde{t}_a$ are needed, scenarios where the $\tilde{t}_b$ is the lightest are more interesting. For these values of the mixing angle, this lightest stop is mainly right-handed.
The same reasoning can be applied for $\theta_t\in[\pi/4,\pi/2]$. This time, a larger amount of $\tilde{t}_a$ is necessary. A spectrum where $\tilde{t}_a$ is the lightest stop is therefore more favorable for the $SU(5)$ test. Again, 
this lightest stop would be mainly right-handed. We conclude that scenarios with a light mainly right-handed stop are always more appropriate to carry out this $SU(5)$ test, for any value of the stop angle.

The expected precision is shown in Fig.\ \ref{fig:PZ_stop_decay} for $\theta_t=0.4$, as well as for $N_a=20$ and $1000$. We see that with a spin-analyzer of efficiency $\kappa=0.5$ and $N_a=20$, $N_b \gtrsim 137$ events are needed to probe the relation at 3$\sigma$ significance, i.e.\ to have $P_3<1$. Testing the relation with $50\%$ or $20\%$ precision at $3\sigma$ requires $N_b \sim 589$ or $7560$. For comparison, for $300$ fb$^{-1}$ of integrated luminosity, one expects about $26700$, $2580$, $213$, and $24$ events for respectively $m_{\tilde{t}}=700$, $1000$, $1400$, and $1800$ GeV.

\section{Top-charm supersymmetry}
\label{se:Top_charm}

Let us now focus on classes of models that feature a heavy first generation of up-type squarks, i.e.\ with only the stop and scharm states potentially accessible at the LHC. In the super-CKM basis, this means assuming up-squark mass terms of the form
\begin{equation}
	m_Q^2 ~=~ m_U^2 ~=~ 
	\begin{pmatrix} \Lambda^2 & 0 & 0 \\ 0 & 0 & 0 \\ 0 & 0 & 0 \end{pmatrix} 
	+ {\cal O}(M^2_{\rm SUSY}) 
	  \begin{pmatrix} 0 & 0 & 0 \\ 0 & \lambda_{22} & \lambda_{32} \\ 0 & \lambda_{23} & \lambda_{33} \end{pmatrix}
	\label{eq:2gen_matrix}
\end{equation}
where the $\lambda_{ij}$ form in general a hermitian matrix of $\mathcal{O}(1)$ parameters and $\Lambda \gg M_{\rm SUSY}$. 

Such a framework is immune against $D^0-\bar D^0$ mixing induced from the up-squark sector \cite{Altmannshofer:2009ne} because the first up-squark generation is heavy. We assume that the down-type squarks are either aligned or heavy in order to avoid FCNC induced by the down sector. Let us recall that, in the $SU(5)$-like GUT context, slepton mass matrices are related to the squark mass matrices. 

Phenomenologically, this top-charm SUSY framework constitutes an ideal playground to gain knowledge about how to test the $SU(5)$-like GUT hypothesis of our interest. It is also useful in order to carry out stop searches at the LHC taking into account flavour violation \cite{QFV1, QFV2}. In particular, large top-charm mixing is found to both improve naturalness and to relax the constraints on stop masses \cite{Blanke:2013uia}.

\subsection{Extradimensional realizations}
\label{se:5Dtopcharm}

Here we provide some elements of model-building to demonstrate how the top-charm SUSY structure can appear in presence of  a compact extra-dimension. We focus on the features that make the mass matrix Eq.\ \eqref{eq:2gen_matrix} appear. The discussion below is not meant to be exhaustive, nor fully detailed. Although not all details are discussed, care has been taken that the elements proposed are compatible with the observed SM flavour structure and that $a_u$-terms are potentially generated.   The reader with no interest in model-building can safely skip this Subsection. 
    
\begin{itemize}
\item \textit{A flat $O$(TeV) extra-dimension}\\
Consider a flat extra-dimension $y\in[0,\pi R]$, with $1/R$ of order of few TeVs. Let the Higgs $H_u$ and the first generation $10_1$ come from hypermultiplets propagating in the bulk, while the two other generations are brane-localized. Supersymmetry is broken by twisted boundary conditions (BCs), i.e.\ Scherk-Schwarz breaking with maximal twist (see Refs.\ \cite{Antoniadis:1990ew,ADPQ,DPQ,Delgado:1999sv, Barbieri:2000vh,AHNSW,DGJQ,DDGQ,Weiner,Barbieri:2002uk,
Barbieri:2002sw,BMP,MP,Delgado:2001si,Quiros:2003gg,Diego:2005mu,Diego:2006py, Dimopoulos:2014aua} and the review \cite{von Gersdorff:2007kk}). The first generation up-squarks get a tree-level mass of $1/2R$ and come within a SUSY $N=2$ hypermultiplet. The masses of the second and third generations are generated at one-loop \cite{ADPQ},
\begin{equation}
	m_Q^2 ~=~ m_U^2 ~\approx~ 
	\textrm{diag}\bigg[\frac{1}{4R^2}, \frac{7\zeta(3)}{16\pi^4R^2} C,\frac{7\zeta(3)}{16\pi^4R^2} C\bigg],
\end{equation}
where $C$ is a model-dependent factor containing group theory invariants. The hierarchy of Eq.\ \eqref{eq:2gen_matrix} is generated due to this loop suppression.

\item \textit{A flat $O(M_{\rm GUT})$ extra-dimension with gauge-Higgs unification}\\
Let us consider now $1/R\sim M_{\rm GUT}$. The MSSM Higgses come as zero modes of the scalar component of the $\mathcal{N}=2$ vector multiplet. Matter fields are in the bulk with some exponential profile controlled by their bulk mass $a_i/\pi R$. The Yukawa couplings come from the overlap with matter in the bulk (see Refs.\  \cite{Burdman:2002se, Hebecker:2008rk, Brummer:2009ug}). Supersymmetry is broken by radion mediation parameterized by the radion F-term $F_T$ (i.e.\ a Scherk-Schwarz with small twist) and by a spurion $F_X$ on the $y=0$ brane. The $10_1$ is localized towards the $y=0$ brane using a $a_1>0$ bulk mass, $10_2$ is localized towards the $y=\pi R$ brane using $a_2<0$, and the $10_3$ has flat profile, $a_3=0$. A correct Yukawa hierarchy is obtained in that way. For the up-squark masses one obtains \footnote{We do not write the cross-term $F_T F_X$ for simplicity.}
\begin{equation}
	m_Q^2 ~=~ m_U^2 ~\approx~ 
	\textrm{diag}\bigg[ \bigg(\frac{F_X}{M^2_*}\bigg)^2\frac{a_1}{\pi R M_*},~0~,~\bigg(\frac{F_T}{2R}\bigg)^2
	+\bigg(\frac{F_X}{M^2_*}\frac{1}{\pi R M_*}\bigg)^2\bigg],
\end{equation}
where the second-generation term is exponentially suppressed with respect to $(F_T/2R)^2$. The five-dimensional cutoff $M_*$ is such that $\pi R M_*$ is of order of few units.  It is natural to assume $F_X/M_*\sim F_T/2R$. In order to reproduce the up-squark Yukawa, one has $a_1\sim 10$, so that there is one order of magnitude between the first and third generation. The gaugino having a flat profile, its mass is of same order as the third generation of the soft masses $M_{1/2}^2\approx m^2_{Q,U\,3}$. Below the GUT scale, the theory reduces to the MSSM. The RG flow of the up-squark soft masses is dominated by the gluino, whose contribution is universal.  The low-energy matrix is roughly given by $m^2_{Q,U\,3}\approx m^2_{Q,U\,3}+ K_3 M^2_{1/2}$ with $K_3=(4.5-6.5)$  \cite{MSSM}. One has therefore a low-energy matrix of the form  Eq.\ \eqref{eq:2gen_matrix}.

\item \textit{A $O(M_{\rm GUT})$ weakly warped extra-dimension  }\\
Assume a warped extra-dimension with $AdS$ curvature $k$, with first KK excitations  $m_{\rm KK}\sim \pi k \epsilon$ at the GUT scale,  and the warp factor $\epsilon \approx 1/20$. The MSSM Higgses (possibly containing pNGBs) are localized on the IR brane. The bulk masses naturally generate the SM flavour hierarchy by taking $O(1)$ values \cite{HGUT, Brummer:2011cp}. Assume that an IR spurion $F_X$ breaks the SM flavour symmetry, giving a soft mass $O((F_X/M_*)^2)$ only to the first generation of squarks and to nothing else. Radius stabilization and a vanishing cosmological constant \cite{Luty:1999cz, Luty:2000ec, Brummer:2011cp} typically implies that radion mediation is suppressed by a warped factor with respect to the IR brane breaking $F_T/2R\sim \epsilon F_X/M_*$. We therefore have
\begin{equation}
	m_Q^2 ~=~ m_U^2 ~\approx~ \bigg(\frac{F_X}{M_*}\bigg)^2 \textrm{diag} [ 1, \epsilon^2, \epsilon^2 ],
\end{equation}
The gaugino mass is generated by the radion, $M_{1/2} \sim F_T/2R $. The low-energy mass matrix has therefore the structure given in Eq.\ \eqref{eq:2gen_matrix}.

\item \textit{A $O$(TeV) partly-supersymmetric warped extra-dimension}\\
Finally, let us consider a warped extra-dimension with first KK excitations $m_{\rm KK}\sim \pi k \epsilon$ about $ 100$ TeV and the warp factor about $\epsilon=10^{-15}$. The Higgs lies on the IR brane. Supersymmetry is broken by boundary conditions. Fermion component of the hypermultiplet are given $(\pm,\pm)$ BCs, while scalar components have twisted BCs $(\pm,\mp)$ or $(\mp,\pm)$.  This setup has a natural holographic interpration has a $\mathcal{N}=2$ strongly interacting CFT coupled to a non-SUSY elementary sector \cite{Gherghetta:2003he, Gherghetta:2004sq, Redi:2010yv}. Bulk masses of $10_i$ are set in order to reproduce the Yukawa hierarchy, with $c_1>c_2>c_3\sim 1/2$. Assuming left-handed zero modes for the quarks, one sets $(\mp,\pm)$  BCs to the 1st generation and $(\pm,\mp)$ to the two other generations. From the latter BCs  a light mode emerges, with $m\approx (4c^2-1)^{1/2} k \epsilon^{c+1/2}$ for $c>1/2$ and $m\approx \sqrt{2}/|\log\epsilon|^{1/2}k \epsilon$ for $c=1/2$.  This mode does not exist for  the first generation, whose first mode is given by $m_{\rm KK}=(c_1/2+1) \pi k \epsilon$. The bulk masses being set in order to reproduce the up quark masses pattern, one ends up with \footnote{One uses $c_1\approx0.7$, $c_2\approx0.6$, $c_1 \approx 0.5$. One has in particular $|\log\epsilon|(2c_2-1)\epsilon^{2c_2-1}/(1-\epsilon^{2c_2-1})=y_c/y_t $.}
\begin{equation}
	m_Q^2 ~=~ m_U^2 ~\approx~ (k\epsilon)^2~\textrm{diag} \bigg[ 18,~4\cdot 10^{-4},~0.05 \bigg]\,.
\end{equation}
Notice one obtains an inverted hierarchy where the 2nd generation is the lightest. Extra source of SUSY breaking like the universal gravity contributions discovered in Ref.\ \cite{Dudas:2012mv} could be useful to lift the scharm masses and to generate $a$-terms.  The phenomenology of this last model is rather intriguing as the light stops and scharms are accompanied by their complex conjugate, as a remnant of the broken $\mathcal{N}=2$ hypermultiplet. We stress that this model would deserve a dedicated study.

\end{itemize}

\subsection{Effective Lagrangian}

The effective Lagrangian derived in Sec.\ \ref{se:EFT_MIA} applies directly to the top-charm SUSY spectrum, identifying the first generation up-squarks as the heavy field $\hat \phi=(u_L,u_R)$, and the second and third generation up-squarks as the light field $\phi=(c_L,t_L, c_R, t_R)$. The blocks of the mass matrix $\mathcal{M}^2$ have then the form 
\begin{equation}
	\hat{M}^2 = \begin{pmatrix} m_{11}^2 & m_{14}^2 \\  & m_{44}^2 \end{pmatrix}\,,\quad 
	\tilde{M}^2 = 
		\begin{pmatrix}
	 	m_{12 }^2  & m_{13 }^2 & m_{15 }^2  & m_{16 }^2 \\   m_{42 }^2  & m_{43 }^2 & m_{45 }^2  & m_{46 }^2
		\end{pmatrix}\,,\quad
	M^2 = \begin{pmatrix}
		m_{22 }^2  & m_{23 }^2 & m_{25 }^2  & m_{26 }^2 \\
        	   	   & m_{33 }^2 & m_{35 }^2  & m_{36 }^2 \\
				   &           & m_{55 }^2  & m_{56 }^2 \\
				   &           &            & m_{66 }^2 
	   \end{pmatrix}
\end{equation}

Furthermore, the low-energy $SU(5)$ relation $a_u\approx a_u^t$ of our interest translates into 
\begin{equation}
	m_{24}^2 ~\approx~ m_{15}^2 \qquad {\rm and} \qquad m_{16}^2 ~\approx~ m_{34}^2.
\end{equation} 
Using also the $SU(5)$ relation $M^2_Q\approx M^2_U$ (valid only for the two first generations), we have in addition $m_{12}^2 \approx m_{45}^2$ and $m_{13}^2\approx m_{46}^2$.  It is therefore natural to scrutinize the effects of the virtual first generation up-squarks on the light top-charm squarks. The $\hat{\mathcal{O}} \hat{M}^{-2} \tilde{M}^{2} \phi$ term of the effective Lagrangian Eq.\ \eqref{eq:Leff} induces flavour-changing decays of the light top-charm squarks into $u\tilde{B}$ and $u\tilde{W}$. Distinguishing between the initial $\tilde{c}$ and $\tilde{t}$ seems difficult, and we therefore do not pursue this direction. Note that a distorsion of the flavour-conserving couplings coming from the $\mathcal{O}(\mathbf{1}-\tilde{M}^{2\dagger} \hat{M}^{-4} \tilde{M}^{2\dagger}/2)\phi$ term in Eq.\ \eqref{eq:Leff} is also present.

 However, unlike for the Natural SUSY case, looking at the higher-dimensional operators is not the only possibility available, because a $SU(5)$ information from $a_u\approx a_u^t$ also remains at leading order in the low-energy mass matrix $M^2$. 
  This is the relation of the top-charm sector $m^2_{26}=m^2_{35}$. We have therefore the possibility of testing the $SU(5)$ hypothesis using only the \textit{real} up-squarks. From now on we thus focus only on the top-charm sector.

\subsection{$SU(5)$ test through Higgs production}

Let us first consider a case where all stop and scharm masses are nearly degenerate. This possibility happens in particular in low-energy GUTs, where no large stop mixing is needed in order to have the correct Higgs mass, see e.g.\ Ref.\ \cite{Dimopoulos:2014aua}. Nearly-degenerate squarks imply that the mass insertion approximation (MIA) is valid for the sector of stops and scharms, and $m_{\tilde{t}_L} \approx m_{\tilde{t}_R} \approx m_{\tilde{c}_L} \approx m_{\tilde{c}_R}\equiv m_{\tilde{q}}$. These states should be produced in an equally abundant way at the LHC, as their production occurs mainly through gluon fusion.

The off-diagonal elements of the up-type trilinear matrix are identified with mass insertions, $(\delta^{LR}_u)_{ij} = v_u (a_u)_{ij}/(\sqrt{2} m_{\tilde{q}}^2)$. The $SU(5)$ hypothesis then implies
\begin{equation}
	(\delta^{LR}_u)_{23} ~=~ (\delta^{LR}_u)_{32} \,. 
\end{equation}
To experimentally test this relation, one may scrutinize the composition of the stop and scharm eigenstates.  At first view, even in the MIA, such an analysis seems difficult because of the presence of additional mass-insertions $\delta^{LL}_{23}$, $\delta^{RR}_{23}$, and $\delta^{LR}_{22,33}$. To overcome this issue, one should note the fundamental difference between $\delta^{LL,RR}_{23}$ and $\delta^{LR}_{23}$. The former relates to a truly bilinear term, i.e.\ the scalar masses, while the latter is induced by a trilinear term, the squark-Higgs scalar coupling. This fact is somehow hidden if one lets the Higgs be on its VEV. The physical Higgs exclusively couples to the LR components of the squark eigenstates. This can be seen from the complete low-energy mass matrix in Eq.\ \eqref{Eq:MassMatrix}. In order to set up a $SU(5)$ test, one may therefore use the Higgs as a probe of the squark eigenstates. Detecting a Higgs gives a direct access to the coupling $(\delta^{LR}_u)_{23}$, i.e.\ to $(a_u)_{23}$.

\begin{figure}
\begin{picture}(400,270)
\put(100,170){\includegraphics[width=8cm]{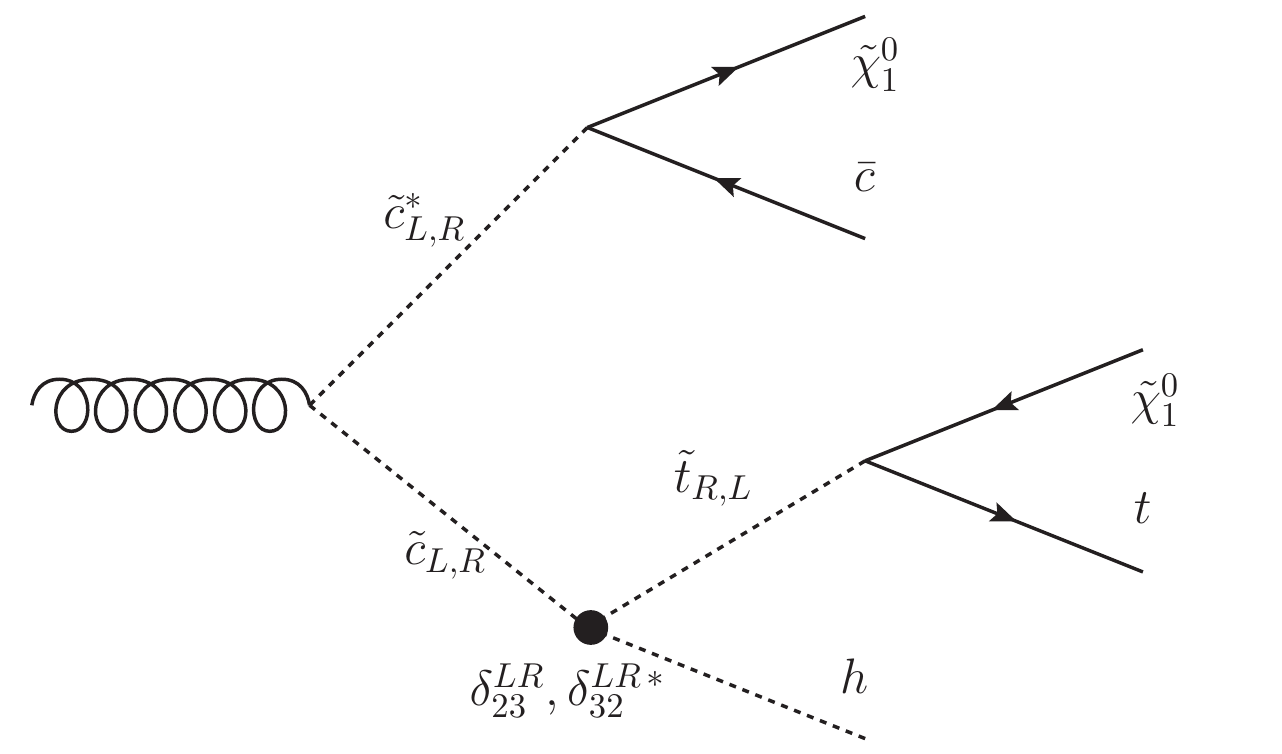}}
\put(220,0){\includegraphics[width=8cm]{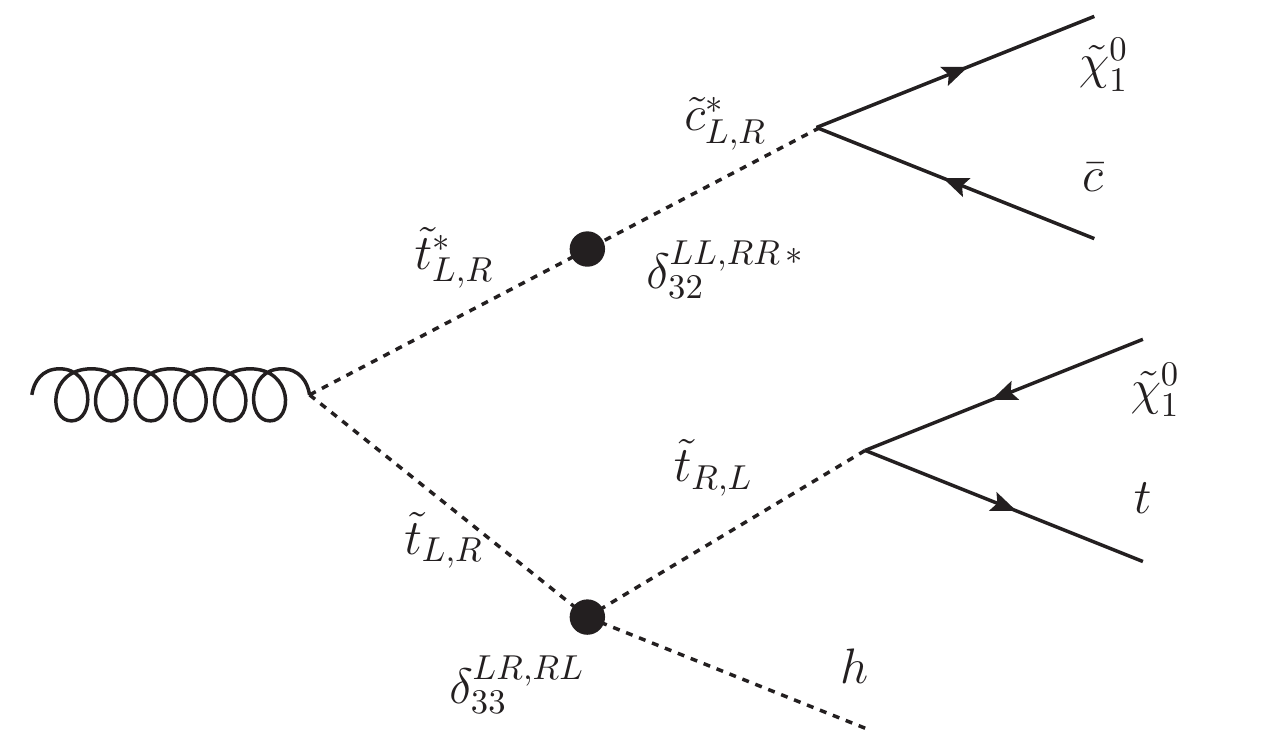}}
\put(0,0){\includegraphics[width=8cm]{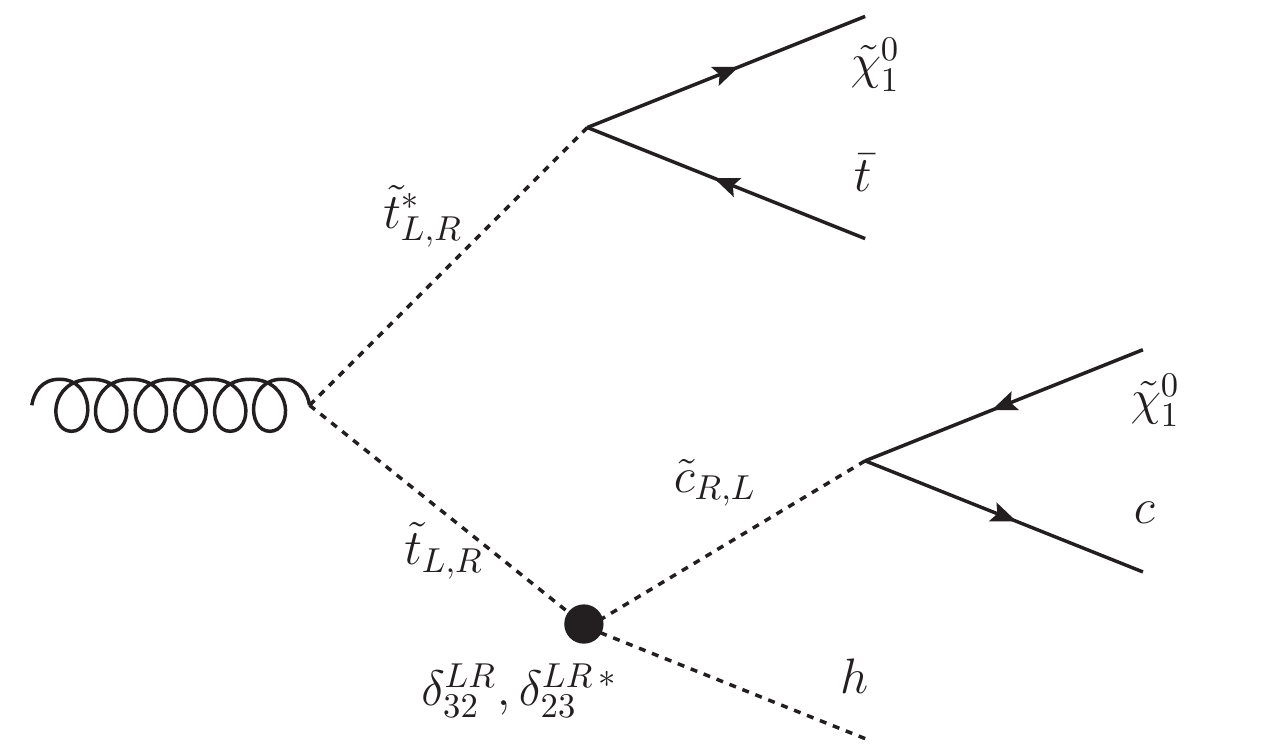}}
\put(150,160){$(N_{htc})$}
\end{picture}	
\vspace*{0.5cm}
	   	\caption{Cascade decays in case of a nearly degenerate top-charm spectrum. The dots represent mass-insertions. First row: cascade decays used for the $SU(5)$ test defined in	Eq.\ \eqref{eq:higgs_test_degen}. Second row: other processes leading to the same final state. 	   	
	   	}
    \label{fig:processes_higgs_tag_degen} 
\end{figure}

The LHC SUSY processes of interest are thus stop and scharm pair production, followed by a flavour-violating decay into a squark and a Higgs-boson in one of the decay chains. These processes are depicted  in Fig.\ \ref{fig:processes_higgs_tag_degen}. We further assume that the squarks decay into the bino. These processes can be identified requiring a single top, a hard jet (from a charm quark), a Higgs, and large missing transverse energy ($E_{\rm T}$). Higgs production through up-squark flavour-violating decays have been also studied in Refs.\ \cite{QFV2}. Note that in the degenerate case, not all particles can be on-shell in the decay chain producing the Higgs.

As in previous cases, this test has to rely on a distinction between the chiralities, which is possible only for the top quark. Reconstructing the events is necessary in order to select the ones where the Higgs comes from $\tilde{c}_{L,R}\rightarrow h\, \tilde{t}_{R,L}$ and reject the ones from $\tilde{t}_{L,R}\rightarrow h\, \tilde{c}_{R,L} $. There is in principle enough information from the decay chain to carry out this distinction. The former of these processes is shown in first row of Fig.\ \ref{fig:processes_higgs_tag_degen}. The latter is shown in second row. Other processes leading to the same final states are also possible but are suppressed by extra mass-insertions (see second row of Fig.\ \ref{fig:processes_higgs_tag_degen}). 

Provided that the cascade decay with $\tilde{c}_{L}\rightarrow h\, \tilde{t}_{R}$ can be isolated, top polarimetry then readily provides a $SU(5)$-test, as ${\rm BR}(\tilde{c}_{L}\rightarrow h\, \tilde{t}_{R})\propto |\delta^{LR}_{23}|^2 $ and ${\rm BR}(\tilde{c}_{R}\rightarrow h\, \tilde{t}_{L})\propto |\delta^{LR}_{32}|^2$. Denoting the event number from the relevant  flavour-changing Higgs decay chain as $N_{htc}$, top polarimetry provides a splitting of the events into $N_{htc}=N_{htc,+}+N_{htc,-}$ (see Subsec. \ref{se:top_polarization}). The $SU(5)$ test then takes the form 
\be
R=\frac{|N_{htc,+}-N_{htc,-}|}{N_{htc,+}+N_{htc,-}}\,. \label{eq:higgs_test_degen}
\ee
This situation is similar to the one of Sec. \ref{se:Heavy_SUSY} and will not be further discussed. 

Instead we focus on a somewhat different spectrum, where stop mixing is large while the scharms are nearly degenerate. 
The MIA applies to the scharm sector, but not inside the stop sector. Instead, we rotate exactly the stops into their mass eigenbasis. After rotating the stops (see Sec.\ \ref{se:natural} for the definition of the stop angle), the scharm-stop mass matrix takes  the form
\begin{equation}
	\begin{pmatrix}
		m^2_{\tilde c}~~ &~~ 0~~ & ~0~ & ~0~ \\
						 &~~ m^2_{\tilde c}~~ &~ 0~  & ~0~ \\
						 & & ~ m^2_{\tilde t_{1}}~ &  ~0~  \\
						 & & 0 & ~ m^2_{\tilde t_{2}}~     \\
	\end{pmatrix}
	+ m_{\tilde c}^2
	\begin{pmatrix}
		0 &~~ \delta^{LR}_{22}~~ & ~\delta^{LL}_{23}c_\theta-\delta^{LR}_{23}s_\theta~ & 		~\delta^{LL}_{23}s_\theta+\delta^{LR}_{23}c_\theta~ \\
		& ~~0~~ &~ \delta^{LR}_{32}c_\theta-\delta^{RR}_{23}s_\theta~  & ~ 		\delta^{LR}_{32}s_\theta+\delta^{RR}_{23}c_\theta~ \\
		& & ~ 0~ &  0  \\
		& & 0 & ~ 0~     \\
	\end{pmatrix}\,.
\label{eq:tc_matrix}
\end{equation}
Following the MIA approach, the first matrix above corresponds to the squark mass eigenvalues, while the second matrix is treated as a mass insertion. By definition, $m^2_{\tilde t_{2}} > m^2_{\tilde t_{1}}$. 
Note that the MIA is expected to be valid to a good precision for $m^2_{\tilde t_{2}}+m^2_{\tilde t_{1}} \sim 2m^2_{\tilde{c}}$ \cite{Raz:2002zx,Hiller:2008sv}. For the rest of this Section we focus on the case $m^2_{\tilde t_{2}}>m^2_{\tilde{c}}>m^2_{\tilde t_{1}}$. Note that this ordering would happen with degenerate stop-scharm soft masses and a hierarchical $a_u$ with large  $(3,3)$ element. Again, this scenario can naturally happen in the five-dimensional GUTs discussed in Subsec.\ \ref{se:5Dtopcharm}.

\begin{figure}
\begin{picture}(400,270)
\put(240,0){\includegraphics[width=6.5cm]{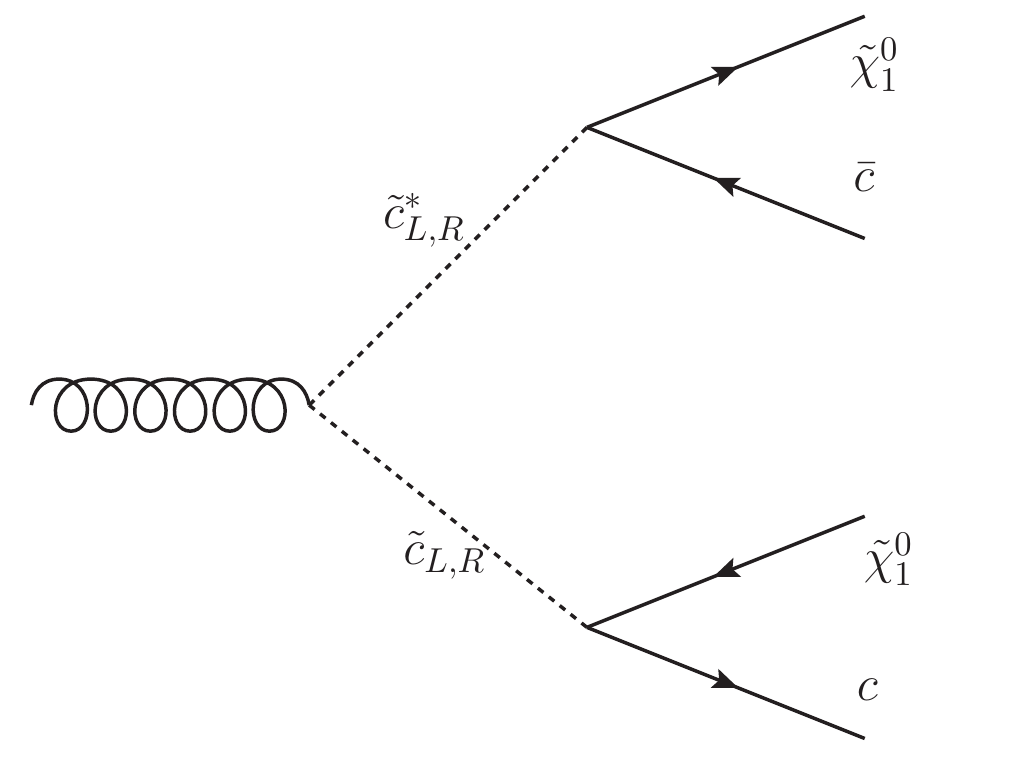}}
\put(0,0){\includegraphics[width=8cm]{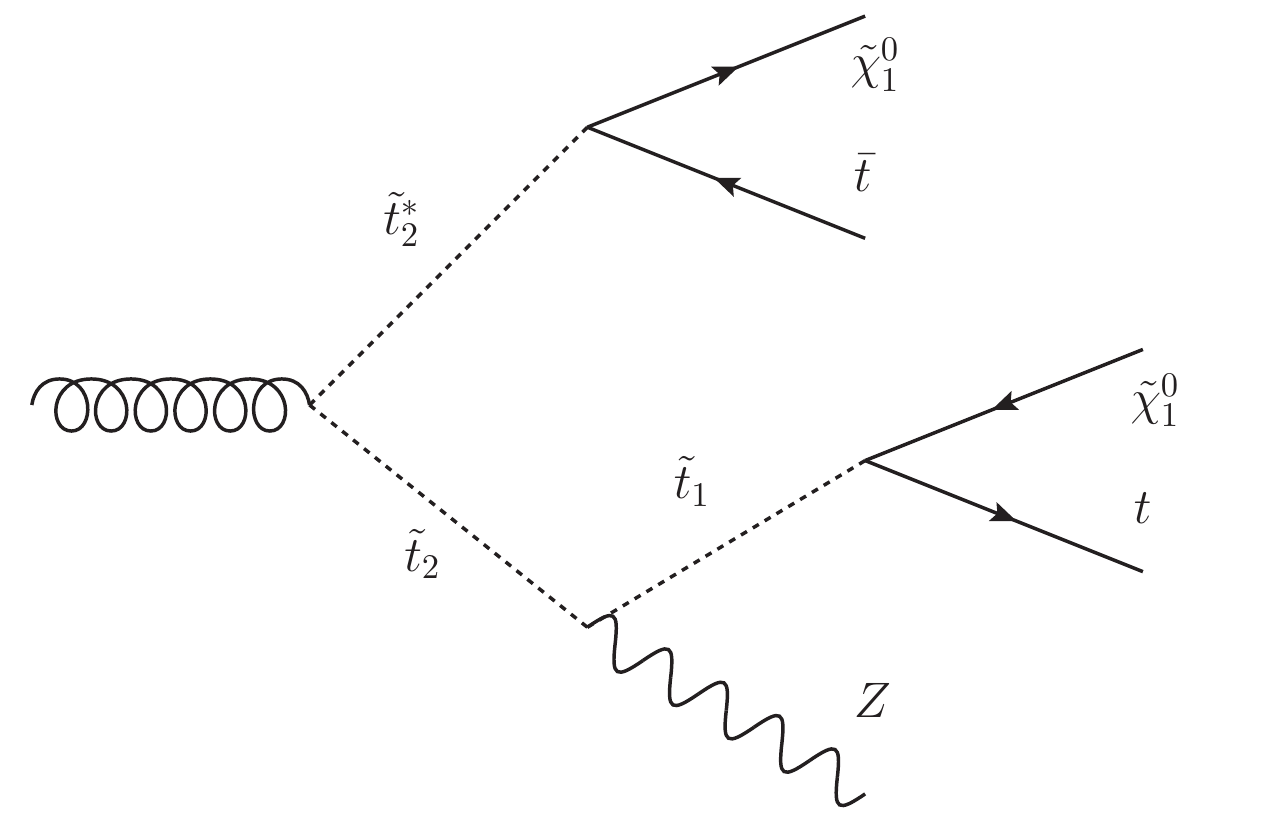}}
\put(0,170){\includegraphics[width=8cm]{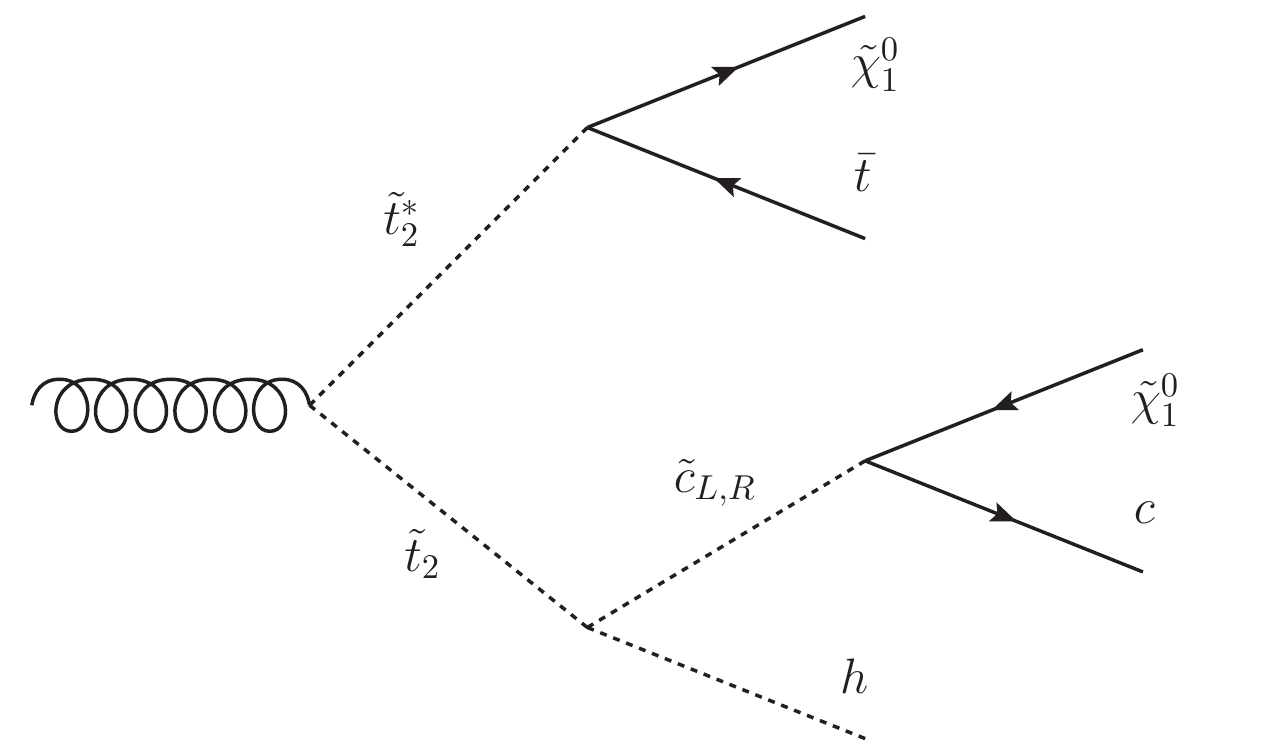}}
\put(240,170){\includegraphics[width=8cm]{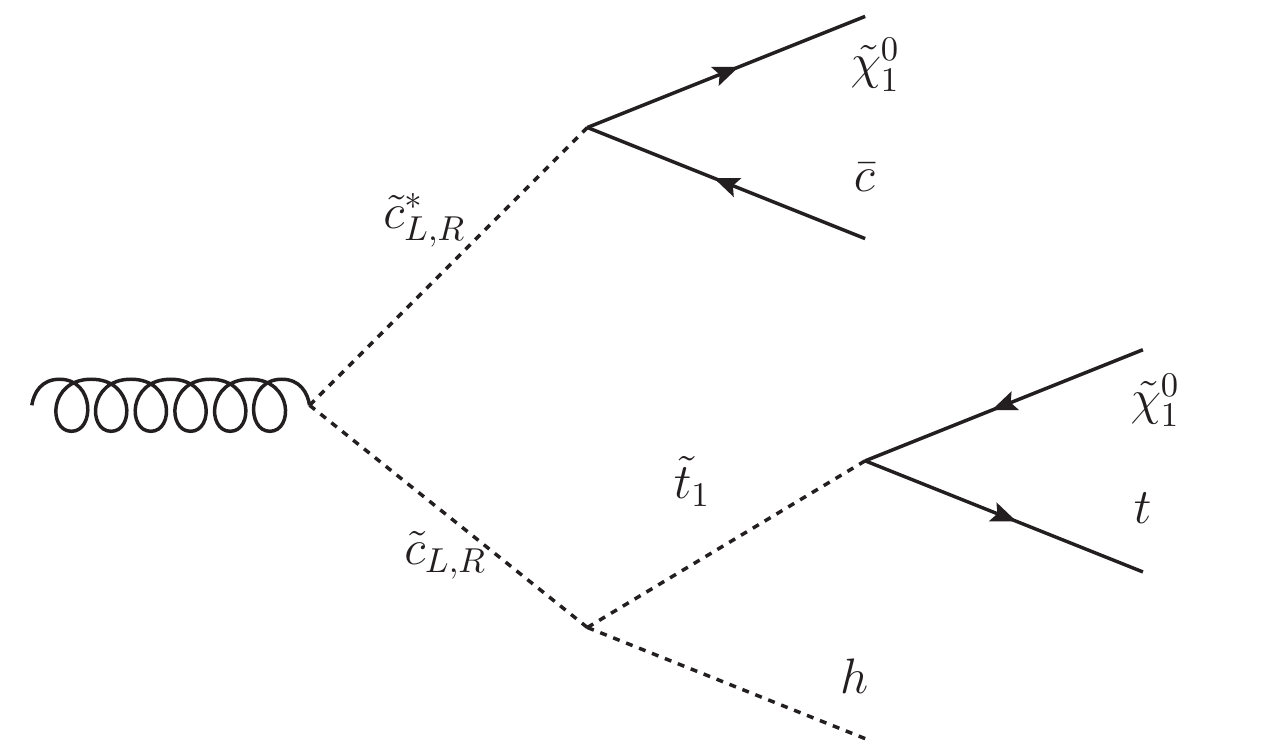}}
\put(80,-5){$(N_{Zt})$}
\put(320,-5){$(N_{jj})$}
\put(80,165){$(N_{hj})$}
\put(320,165){$(N_{ht})$}
\end{picture}	
\vspace*{0.5cm}
	   	\caption{Cascade decays used for the $SU(5)$ test based on Higgs detection defined in Eq.\ \eqref{eq:higgs_test}.}
    \label{fig:processes_higgs_tag} 
\end{figure}
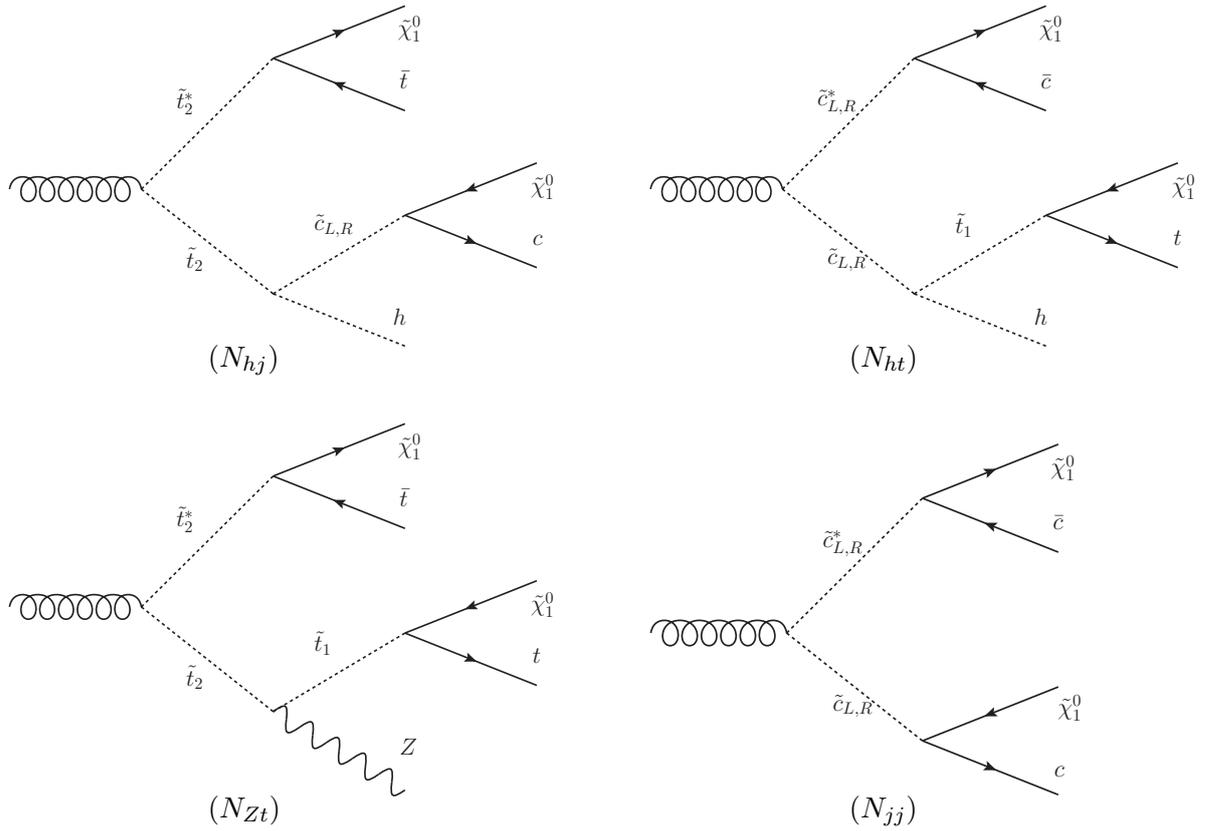

The physical Higgs couples only to the left-right mixing terms $\delta^{LR}$. The vertices $h \tilde{c}_L \tilde{t}_1$, $h \tilde{c}_L \tilde{t}_2$ , $h \tilde{c}_R \tilde{t}_1$, $h \tilde{c}_R \tilde{t}_2$ are respectively proportional to $|\delta^{LR}_{23} s_\theta|^2$, $|\delta^{LR}_{23} c_\theta|^2$, $|\delta^{LR}_{32} c_\theta|^2$, $|\delta^{LR}_{32} s_\theta|^2$. Within the given mass ordering, the SUSY cascade decays are rather different than from the degenerate case  discussed above. Flavour changing scharm decays going through $\tilde{t}_2$ are now suppressed because of $m^2_{\tilde t_{2}}>m^2_{\tilde c}$. As a consequence, contrary to the degenerate spectrum discussed above, top polarimetry is not useful anymore. On the other hand, real decays $\tilde{t}_2\rightarrow h \tilde{c}$, and $\tilde{c}\rightarrow h \tilde t_1 $ are now opened. 

We require again a single top, a hard jet, a Higgs, and large missing $E_{\rm T}$ from both sides of the decay chains. Two event topologies lead to this final state: the Higgs can either come from $\tilde{t}_{2}\rightarrow h\, \tilde{c}_{R,L} $ or from $\tilde{c}_{L,R} \rightarrow h\, \tilde{t}_{1}$. These processes are shown in first row of Fig.~\ref{fig:processes_higgs_tag}. The first of these diagrams is proportional to $|\delta^{LR}_{23}|^2c_\theta^2 + |\delta^{LR}_{32}|^2s_\theta^2$, while the second is proportional to $|\delta^{LR}_{23}|^2s_\theta^2 + |\delta^{LR}_{32}|^2c_\theta^2$. These two types of events can be disentangled using the topology of the decay chain. We denote the event rates associated with these two diagrams $N_{hj}$ and $N_{ht}$, respectively. For maximal stop mixing ($c_{\theta}=s_{\theta}$), the two quantities become equal so that the power of the test is expected to vanish. 

\begin{figure}
  	\centering
	\includegraphics[width=8cm]{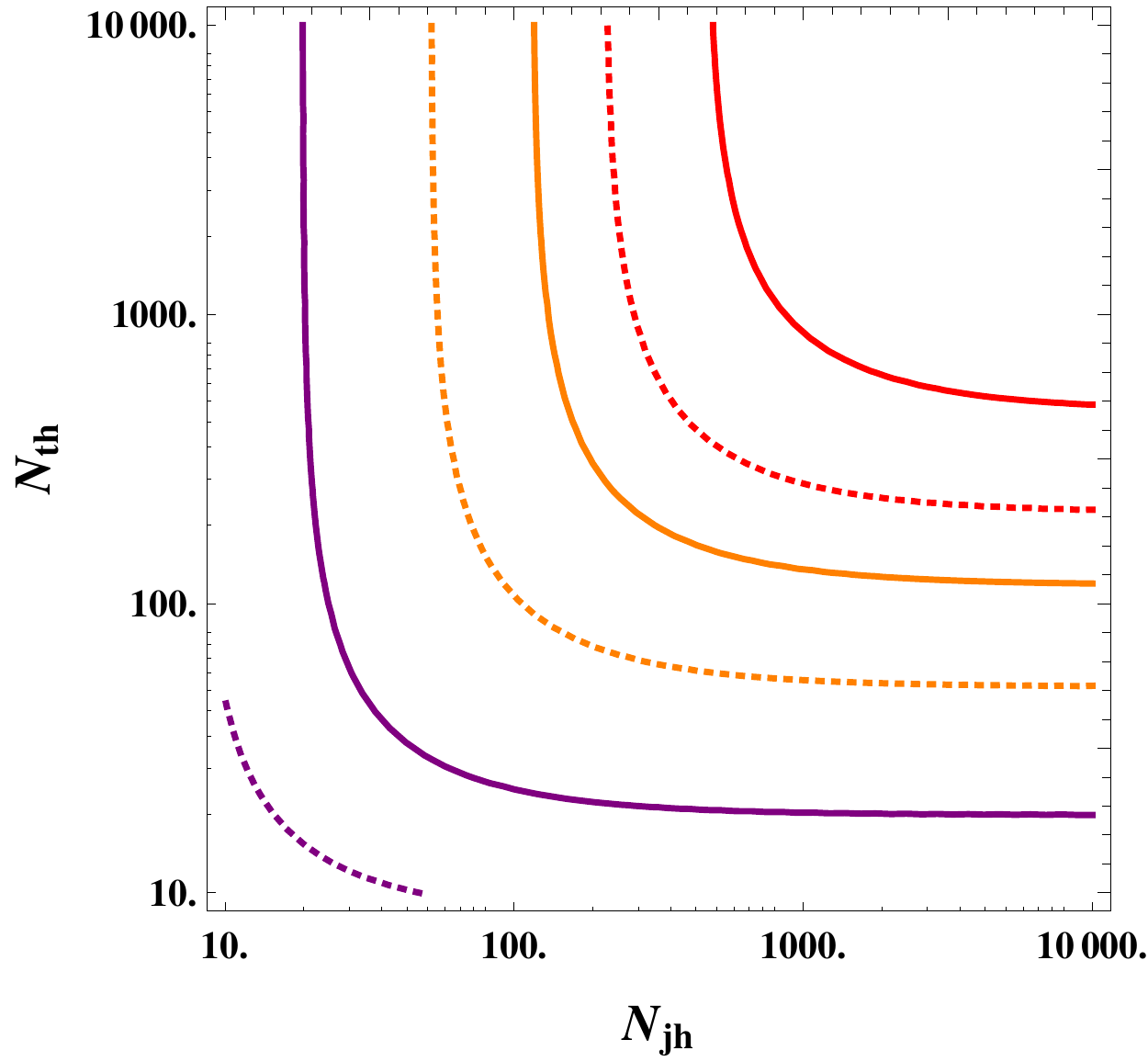}
   	\caption{ Expected precision $P_Z$ for the $SU(5)$ test based on Higgs detection in SUSY cascade decays Eq.\ \eqref{eq:higgs_test}. $N_{hj}$ and $N_{ht}$ are the number of observed cascade decays from $\tilde t_2$ and $\tilde c_{L,R}$ pair production, respectively. One fixed the stop angle to $\theta_t=0.4$. Plain (dotted) lines denote $2\sigma$ ($3\sigma$) significance, respectively. Purple, orange and red lines respectively show $P_{2,3}=(50\%,20\%,10\%)$ isolines of expected precision.}
     \label{fig:higgs_tag}
\end{figure}

Contrary to the degenerate case, the stops and scharms have different production rates. Moreover the theoretical predictions suffer from a large QCD error. One way to proceed is to normalize $N_{hj}$, $N_{ht}$ using observed data. In order to normalize $N_{ht}$, we ask for the measurement of flavour-conserving decay chains of $\tilde c$-pairs into two jets plus large missing $E_{\rm T}$. The corresponding event rate is noted $N_{jj}$. Because of stop mixing, the same process cannot be used to normalize $N_{hj}$. Instead we ask for one of the two $\tilde t_2$ to decay into $Z \tilde t_1$. This event rate is noted $N_{Zt}$. These processes are depicted in the second row of Fig.\ \ref{fig:processes_higgs_tag}. 

Normalizing $N_{hj}$ by $N_{Zt}$ and $N_{ht}$ by $N_{jj}$   cancels the cross-sections, leaving only the ratio of partial decay widths,
\begin{equation}
	{\rm E}\bigg[\frac{N_{hj}}{N_{Zt}} \bigg] ~=~ 
	\frac{\Gamma(\tilde{t}_{2}\rightarrow h\, \tilde{c}_{L,R})}{\Gamma(\tilde{t}_{2}\rightarrow Z \tilde{t}_{1})} 	\,,\qquad\quad
	{\rm E}\bigg[\frac{N_{ht}}{N_{jj}} \bigg]~=~ 
	\frac{\Gamma(\tilde{c}_{L,R}\rightarrow h\, \tilde{t}_{1})}{\Gamma(\tilde{c}_{L,R}\rightarrow c\, \tilde{B})}\, .
\end{equation}
We now consider the quantity
\begin{equation}
	\eta ~\equiv~ 
	\frac{\Gamma(\tilde{c}_{L,R}\rightarrow h\, \tilde{t}_{1})}{\Gamma(\tilde{t}_{2}\rightarrow h\,  \tilde{c}_{L,R})}
	\frac{\Gamma(\tilde{t}_{2}\rightarrow Z \tilde{t}_{1})}{\Gamma(\tilde{c}_{L,R}\rightarrow c\, \tilde{B})}
	~\approx~
	\frac{m^2_{\tilde{c}}- m^2_{\tilde{t}_1}}{m^2_{\tilde{t}_2}- m^2_{\tilde{c}}}
	\frac{m^2_{\tilde{t}_2}- m^2_{\tilde{t}_1}}{m^2_{\tilde{c}}- m^2_{\tilde{B}}} \,,
\end{equation}
which is obtained when neglecting the SM masses \footnote{Neglecting the SM masses is made only for illustration purpose, this approximation is not needed for the analysis.}. This factor may be evaluated using extra information from kinematic analyses, for example using the kinematic edges of the $ht$, $hj$, $Zt$ invariant masses. 

The normalized $SU(5)$ test then reads
\begin{equation}
	R ~=~ \frac{1}{c_{2\theta}}\bigg|\frac{N_{hj}}{N_{jj}} - 		\eta\frac{N_{ht}}{N_{Zt}}\bigg|\bigg/\bigg(\frac{N_{hj}}{N_{jj}}+\eta\frac{N_{ht}}{N_{Zt}}\bigg)\,,
	\label{eq:higgs_test}
\end{equation}
which satisfies
\begin{equation}
	{\rm E}[R] ~=~ \frac{|\delta_{23}^{LR}-\delta_{32}^{LR}|}{\delta_{23}^{LR}+\delta_{32}^{LR}}\,.
	\label{eq:higgs_test_E}
\end{equation}
The associated expected precision reads 
\begin{equation}
	P_Z ~\approx~ \frac{Z}{2 c_{2\theta}}\bigg(\frac{1}{N_{hj}}+\frac{1}{N_{ht}}\bigg)^{1/2}\,.
	\label{eq:higgs_test_PZ}
\end{equation}
Note that $P_Z\rightarrow \infty$ when $\theta_t\rightarrow \pi/4$, i.e.\ the power of the test vanishes in the limit of maximal stop mixing as expected. We display only the leading statistical uncertainty that comes from the small flavour-changing event rates.

The expected precision is shown in Fig.\ \ref{fig:higgs_tag} for the intermediate value $\theta_t=0.4$. Because of the mass ordering of this scenario, one expects $N_{hj}< N_{ht}$ because the $\tilde{t}_2$ is heavier than the scharms and thus produced less abundantly. Assuming $N_{hj} \ll N_{ht}$ and $\theta_t=0.4$, testing the relation with $50\%$, $20\%$, or $10\%$ precision at $3\sigma$ requires respectively $ N_{hj} \gtrsim 19$, $116$, or $464$ events. Roughly twice less events are needed if $\theta_t=0$. For comparison, assuming flavour-violating branching ratios of $0.05$,  and $300$ fb$^{-1}$ of integrated luminosity, one expects about $1340$, $130$, and $11$ events for a squark  with respectively $m_{\tilde q}=700$, $1000$, and $1400$ GeV. 

Other possibilities of normalization of the $N_{hj}$, $N_{ht}$ rates are in principle possible -- using either  observed or theoretical event rates. In any case the approach relies on evaluating the appropriate $\eta$ parameter such that Eq.\ \eqref{eq:higgs_test_E} is satisfied. The expected precision Eq.\ \eqref{eq:higgs_test_PZ} is valid as long as the statistical uncertainty from $N_{hj}$ and $N_{ht}$ dominates.

Finally, the analysis of the specific mass orderings carried out above can be taken as a guideline to build more global $SU(5)$ tests. In particular the flavour-changing squark decays with Higgs production are expected to always carry relevant information regarding $a_u\approx a_u^t$, for any mass ordering. Assuming an arbitrary mass ordering, a global $SU(5)$ test can be setup by putting  into a likelihood function the $N_{hj}$, $N_{ht}$ event rates plus some information to normalize them (i.e.\ all the potentially useful information found above). This global test constitutes a natural extension to the cases analysed above and is subject to future work.


\section{Summary and conclusion \label{Sec:Conclusions}}

A fairly intriguing feature of the Standard Model (SM) is that quarks and leptons fit naturally into complete representations of the $SU(5)$ gauge group. This might be taken as a hint that the SM is the low-energy effective theory of a Grand-Unified Theory (GUT), either $SU(5)$-symmetric or  symmetric under a group broken to $SU(5)$ at an intermediate energy scale. Many realizations of $SU(5)$-like GUTs are possible, and imply a variety of low-energy features. Moreover, the large quantum corrections occurring when flowing down the SM effective theory from the GUT to the weak scale potentially wipe out a lot of precious information. Testing whether the world is microscopically $SU(5)$-like is thus a rather challenging task. The large model-dependence coming from both the GUT realization and from quantum corrections constitute a irreducible theoretical uncertainty, from the point of view  of testing the $SU(5)$ hypothesis. 

Low-energy supersymmetry is a natural ingredient of $SU(5)$-like GUTs, as it typically favours the high-energy unification of gauge couplings. As a happy coincidence, the superpartners potentially carry information about the possible GUT group. Low-energy SUSY is therefore both favoured by $SU(5)$-like GUTs and useful to reveal their nature. Still, testing the $SU(5)$ hypothesis rises the double challenge of taming the model-dependence and finding observations that are doable in the real world. While such a program may seem at first view insurmountable, it happens that a new $SU(5)$ relation between the up-squark trilinear couplings  $a_u=a_u^t$, first identified in Ref.\  \cite{Fichet:2014vha}, opens a number of possibilities. This relation is exact at the GUT scale, and is immune to any GUT threshold corrections.

Moreover, the relation $a_u\approx a_u^t$ is found to survive through the MSSM renormalization group flow, such that it is spoiled by typically $O(1\%)$ of relative error at the TeV scale. This relation can be thus taken as a window on the GUT physics, that can be  exploited to test the $SU(5)$ hypothesis. Because of the peculiar structure of $a_u\approx a_u^t$, all the $SU(5)$-tests coming from this relation have to rely on either flavour-violation or chirality flip  in the sector of up-squarks and quarks. The discussed $SU(5)$-tests are set up for the Large Hadron Collider (LHC). 

Even though the $SU(5)$ relation $a_u\approx a_u^t$ is immune to many corrections, a model-dependence remains in the up-squark spectrum. Indeed, although certain patterns of the up-squark mass matrix can be justified in a number of ways, overall a lot of freedom remains. We consider therefore various typical scenarios for the up-squark spectrum.
The scenarios that we have investigated are ``Heavy supersymmetry'' (i.e.\ $M_{\rm SUSY}\gg $ TeV, Sec.\ \ref{se:Heavy_SUSY}), ``Natural supersymmetry'' (i.e.\ real stops are accessible at the LHC, Sec.\ \ref{se:natural}), and ``Top-charm supersymmetry'' (i.e.\ both scharm and stops are accessible at the LHC, Sec.\ \ref{se:Top_charm}).  

Depending on the up-squark spectrum, the $SU(5)$ tests involve only virtual squarks, only real squarks, or both real and virtual squarks. Whenever the virtual squarks are heavier than the TeV scale, they can be integrated out to form a TeV-scale effective Lagrangian. The effect of virtual squarks is then enclosed into effective operators, either generated at tree-level or at one-loop. A complementary expansion is the mass insertion approximation, that we use both in the broken and unbroken EW phase. 

In many cases, the $SU(5)$ tests that we find consist in determining whether a relation among certain observables is satisfied or not. In order to quantify the feasibility of a test, we introduce a systematic procedure relying on the frequentist $p$-value. The associated expected precision tells, for a given amount of data, up to which magnitude a violation of the $SU(5)$ relation can be assessed within a given statistical significance. 

In the ``Heavy SUSY'' case, it appears that the $SU(5)$ hypothesis can be tested using  flavour-changing dipole operators in the up-sector. These dipoles from SUSY are loop-generated.  A top quark has to be present in the dipole in order to use top-polarimetry. The expected sensitivity as a function of signal and background event rates  is evaluated. The common LHC searches from either anomalous single top production or anomalous flavour changing top decays seem however not sensitive enough to be able to discover SUSY dipoles. 

We propose thus an alternative kind of precision measurements relying on ultraperipheral collisions between proton and/or heavy ions. A set of exclusive processes is proposed in order to probe the SUSY dipoles operators. We also point out that, contrary to a seemingly widespread belief, the $W$/$Z$-boson fluxes from proton and heavy ions are coherent in the UPC regime. The vector boson fusion cross-sections can thus be expected to be rather large.

In the ``Natural SUSY'' scenario, already investigated in Ref.\ \cite{Fichet:2014vha}, the two mass orderings $m_{\tilde t_{1,2}} > m_{\tilde B, \tilde W}$ and $m_{\tilde W} > m_{\tilde t_{1,2}}>m_{\tilde B}$ are considered. A test associated to the former relies on flavour-changing stop decays and on charm-tagging, while no knowledge of stop masses nor the stop angle is needed. We evaluate the expected precision as a function of charm-tagging efficiency and event rates. 

For the second mass ordering, the $SU(5)$ test relies on top polarimetry. Its potential greatly depends on the stop mixing angle, and vanishes in the limit of no stop mixing and maximal stop mixing. In contrast, the potential for arbitrary stop mixing is more promising. We evaluate the expected precision as a function of the stop mixing angle, event rates and spin-analyzing power. A few events from the mainly left-handed stop are sufficient, while a more copious production is needed for the mainly right-handed stop. This test is favoured by a SUSY spectrum with a mainly right-handed  lightest stop.

We finally introduce the ``Top-charm SUSY'' scenario, featuring a heavy first generation of up-squarks, and demonstrate that it can arise in a variety of five-dimensional GUT realizations. $SU(5)$ tests can be found using the SM Higgs boson to probe the $LR$ components of the squarks. The tests therefore rely on cascade decays involving Higgs production and flavour-violation. For degenerate scharms and stops, a test using top polarimetry is presented. For the case of $m_{\tilde t_2}>m_{\tilde c_{L,R}}>m_{\tilde t_1}$, a test relying on the reconstruction of the decay chains involving the Higgs is discussed. The expected precision as a function of the stop angle and event rates is evaluated. 
 
The $SU(5)$ tests that appear in the various SUSY scenarios considered are summarized in Tab. \ref{tab:summary}. The typical amount of events needed to reach an expected precision of $50\%$ at 3$\sigma$ is also shown for each of the tests. For these numbers the experimental conditions  are chosen to be the most favorable, e.g $\kappa=1$. The number of needed event  ranges roughly from about 10 to about 100. For the required precision, the tests involving top polarimetry require typically $O(100)$ events, because they rely on the shape of kinematic distributions. All these tests rely on the existence of a given relation among certain observables. In cases where no such relation is available, a more global hypothesis testing has to be done. This will be the subject of a further work \cite{TODO}.

\begin{table}
\centering
\begin{tabular}{|c|c|c|c|c|c|} 
\hline 
  & Heavy & \multicolumn{2}{|c|}{Natural SUSY} & \multicolumn{2}{|c|}{Top-charm SUSY} \\ 
 & SUSY & $m_{\tilde t_{1,2}} \!>\! m_{\tilde B, \tilde W}$  & 
\specialcell{$m_{\tilde W} \!>\! m_{\tilde t_{1,2}}$\\$>m_{\tilde B}$} & 
 $m_{\tilde t_{L,R}} \!\sim\! m_{\tilde c_{L,R}}$ & \specialcell{ $m_{\tilde t_{2}} \!>\! m_{\tilde c_{L,R}}$ \\ $>m_{\tilde t_{1}}$}
  \\ 
  \hline
  Squarks involved & virtual & \multicolumn{2}{|c|}{virtual/real} & \multicolumn{2}{|c|}{real} \\
\hline 
Top polarimetry & yes & no & yes & yes & no \\ 
\hline 
Charm-tagging & no & yes & no & no & no \\ 
\hline 
Higgs detection & no & no & no & yes & yes \\ 
\hline 
$\theta_t$-dependence & no & no & yes & no & yes \\ 
\hline 
$P_3=50\%$ & $144$ & $72$ & $108$ & $144$ & $10$ \\
\hline
\end{tabular} 
\caption{Summary of the $SU(5)$-tests appearing in the various SUSY scenarios considered.
The last line shows the typical number of events needed to reach a $50\%$ precision at 3$\sigma$.}
 \label{tab:summary}
\end{table}

\acknowledgments

The authors would like to thank Rohini Godbole for helpful discussions. S.F.\ thanks Felix Br\"ummer for fruitful discussions during an earlier collaboration, and Christophe Royon, Matthias Saimpert, Gero von Gersdorff, Suchita Kulkarni and David d'Enterria for providing useful comments and discussions. Y. S. thanks Werner Porod, Florian Staub for discussions about SPheno and Diego Guadagnoli for discussions about flavour observables.
 This work is supported by Campus France, PHC PROCOPE, project no.\ 26794YC. The work of Y.S.\ is supported by a Ph.D.\ grant of the French Ministery for Education and Research. Y. S acknowledges  J\"urgen Reuter and DESY for hospitality. S.F.\ acknowledges the Brazilian Ministry of Science, Technology and Innovation for financial support, the Institute of Theoretical Physics of Sao Paulo (ICTP/SAIFR) and the LAPTh (Annecy-le-Vieux, France) for hospitality and support. 

\appendix

\section{The dipole form factors}
\label{App:dipole_ff}

The form factors from penguin diagrams are \cite{Altmannshofer:2010pqa,Gabbiani:1996hi}
\begin{align}
F^{(1)}_{s}  &~=~ \frac{10\,(19+172x+x^2)}{21(1-x)^4}+\frac{20\, x(18+15x -x^2)}{7(1-x)^5}\log x , \\
F^{(2)}_{s}  &~=~ -\frac{12\,(11+x)}{5(1-x)^3}-\frac{6\,(9+16x-x^2)}{5(1-x)^4}\log x , \\
F^{(1)}_{EW} &~=~ \frac{10\,(1-8x-17x^2)}{3(1-x)^4}-\frac{20\,x^2(3+x)}{(1-x)^5}\log x , \\
F^{(2)}_{EW} &~=~ \frac{6\,(1+5x)}{(1-x)^3} +\frac{12\,x(2+x)}{(1-x)^4}\log x .
\end{align}

\section{Event rates for $\tilde{t}_{a,b}\rightarrow t \tilde{B}$ }
\label{app:stop_decay}
The expected event rates have the form
\be
N_{a+}= N_{aL}\frac{1-\kappa/2}{2}+N_{aR}\frac{1+\kappa/2}{2}\,,
\ee
\be
N_{a-}= N_{aL}\frac{1+\kappa/2}{2}+N_{aR}\frac{1-\kappa/2}{2}\,,
\ee
\be
N_{b+}= N_{bL}\frac{1-\kappa/2}{2}+N_{bR}\frac{1+\kappa/2}{2}\,,
\ee
\be
N_{b-}= N_{bL}\frac{1+\kappa/2}{2}+N_{bR}\frac{1-\kappa/2}{2}\,,
\ee
with 
\be
N_{aL}=N_a\,\frac{(c_\theta-x s_\theta)^2}{(c_\theta-x s_\theta)^2+16(s_\theta-x c_\theta)^2}\,,
\ee
\be
N_{aR}=N_a\,\frac{16(s_\theta-x c_\theta)^2}{(c_\theta-x s_\theta)^2+16(s_\theta-x c_\theta)^2}\,,
\ee
\be
N_{bL}=N_b\,\frac{(s_\theta+x c_\theta)^2}{(s_\theta+x c_\theta)^2+16(c_\theta+x s_\theta)^2}\,,
\ee
\be
N_{bR}=N_b\,\frac{16(c_\theta+x s_\theta)^2}{(s_\theta+x c_\theta)^2+16(c_\theta+x s_\theta)^2}\,.
\ee
Here $N_{a,b}$ is the production rate of $\tilde{t}_{a,b}$.
The expected precision associated with the test of Eq.\ (\ref{eq:R_stop_decay}) reads
\be
\begin{split}
P_Z=&\frac{1}{255\sqrt{2}\, \kappa\, (3+c_{ 4\theta})}
\bigg[
\frac{(17-15c_{2\theta})^2 
(3212-739\kappa^2+1020(\kappa^2-4)c_{2\theta}
+(900-289\kappa^2)c_{4\theta})
}{N_a}\\
&+\frac{(17+15c_{2\theta})^2 
(3212-739\kappa^2-1020(\kappa^2-4)c_{2\theta}
+(900-289\kappa^2)c_{4\theta})
}{N_b}
\bigg]^{1/2} .
\end{split}
\ee

\bibliographystyle{JHEP} 

\end{document}